\documentclass[12pt]{article}

\usepackage[utf8]{inputenc}
\usepackage{graphicx}
\usepackage{float}
\usepackage{bbm}
\usepackage{amsmath}
\usepackage{braket}
\usepackage{amsfonts}
\usepackage{amssymb}
\usepackage{appendix}
\begin{document}

\begin{center}

\textbf{\Large 
New sources of leptonic CP violation at the DUNE neutrino experiment
%Hints of new physics at DUNE using CP asymmetries
} 

\vspace{30pt}
 A. Giarnetti$^a$ and D. Meloni$^a$
\vspace{16pt}

\textit{$^a$Dipartimento di Matematica e Fisica, 
Universit\`a di Roma Tre\\Via della Vasca Navale 84, 00146 Rome, Italy}\\
\vspace{16pt}

\end{center} 

% \maketitle
\abstract

We check the capability of the DUNE neutrino experiment to detect new sources of leptonic CP violation beside the single phase expected in the Standard Model. We illustrate our strategy based on the measurement of CP asymmetries in the case New Physics will show up as Non-Standard neutrino  Interactions and sterile neutrino states and show that the most promising one, once the experimental errors are taken into account in both scenarios, is the one related to the $\nu_\mu \to \nu_e$ transition.

\section{Introduction}
Neutrino flavor oscillations have been one of the most important particle physics discoveries of the last decades \cite{Fukuda:1998mi}. Experiments using natural and artificial neutrino sources were able to measure mixing parameters with a good precision \cite{Esteban:2020cvm,deSalas:2020pgw}. In particular, the three mixing angles, as well as the absolute values of the mass splittings,  have been determined with a few percent uncertainty. However, some degeneracies still exist and future experiments aim to solve them. In particular, the atmospheric mass splitting sign (hierarchy problem) and the octant in which the atmospheric angle $\theta_{23}$ lies are still unknown. Moreover, the phase of the PMNS matrix, which is the only source of CP violation in neutrino oscillation\footnote{The other two phases appearing in the most general neutrino Lagrangian, in which both Majorana and Dirac mass terms are taken into account, do not affect the oscillation probabilities. Thus, they could only be tested using other phenomenon (i.e. neutrinoless double beta decay).}, has not been measured with a satisfactory precision \cite{Kolupaeva:2020pug, Abe:2021gky}.  

Even though most of the experimental results are in agreement with the three neutrino paradigm, the long-standing short baseline \cite{Athanassopoulos:1995iw}-\cite{Aguilar-Arevalo:2018gpe} and reactor \cite{Mention:2011rk}-\cite{Berryman:2020agd} anomalies gave some hint of New Physics (NP) phenomena such as the presence of additional sterile neutrino states \cite{Agarwalla:2016mrc}-\cite{Abe:2019fyx}. Moreover, uncertainties on the oscillation parameters leave rooms for the presence of additional effects in neutrino interaction \cite{Roulet:1991sm}-\cite{Guzzo:1991hi} and propagation in matter not contemplated in the Standard Model (SM) and that can be described in a model independent way by four fermions effective operators, namely the Non Standard Interactions (NSI) operators \cite{Farzan:2017xzy}-\cite{Verma:2018gwi}.

Both types of NP models introduce new sources of CP violation in neutrino oscillations. In particular, when one sterile state is contemplated, the $4\times 4$ PMNS matrix contains two more phases than in the SM case. On the other hand, if we take into account the possibility that during their travel through matter neutrinos can undergo NSI, three new phases in the state evolution equations emerge. 

The NP phases inevitably affect CP violation in neutrino oscillations, whose presence can be highlighted from CP-odd observables; among them, 
asymmetries of the type $A_{\alpha\beta} \sim P(\nu_\alpha\to \nu_\beta)-P(\bar{\nu}_\alpha \to \bar{\nu}_\beta)$ can be directly measured at future neutrino experiments capable of distinguishing neutrino from antineutrino events. Such quantities are generally dependent on the CP phases, and become a smoking gun for new CP violation as soon as their values deviate from the SM predictions, including matter effects \cite{Altarelli:2008yr}.

The goal of the present paper is to test whether neutrino experiments will perform sufficiently well as to establish that other phases beside the single one expected in the SM are responsible for CP violation in the lepton sector. To achieve our aim, we adopt the following strategy:
\begin{itemize}
\item compute the leptonic asymmetries in the SM;
\item evaluate the experimental uncertainty on them;
\item re-compute the asymmetries including the effects of NP;
\item check whether the new results are sufficiently away from the                                                        SM predictions.
\end{itemize}
The outcome of our procedure is the identification, if any, of the most promising asymmetry capable to produce an experimental signature well beyond the SM expectations. 
%Motivated by the fact that asymmetries can be computed very easily, without requiring energy spectra reconstructions or fits on the oscillation parameters, we study in this paper how measuring $A_{\alpha\beta}$ using different transition channels, can suggest the presence of NP effects, like sterile neutrinos or NSI.

We validate our strategy studying the asymmetries as they can be measured at the Deep Underground Neutrino Experiment (DUNE)  \cite{Acciarri:2016crz}-\cite{Abi:2020evt}. 
%The performances of this long-baseline accelerator experiment have been widely studied in recent years. 
An unprecedented feature of DUNE will be the possibility to study three different transition channels (namely $\nu_\mu\to \nu_\mu$, $\nu_\mu\to \nu_e$, $\nu_\mu\to\nu_\tau$) and the Neutral Current (NC) neutrino interactions using both neutrino and antineutrino fluxes; thus, this experiment will be able to measure four independent asymmetries. 
%In this manuscript we will discuss which of the asymmetries in DUNE are the most useful in searching hints of NSI or sterile neutrinos, taking into account different energy spectra for the initial fluxes.

The manuscript is organized as follows: in Sect.\ref{CPasySM}, we discuss the analytic structure of the relevant asymmetries in the SM, by means of the perturbation theory in the small parameters (to be defined later in the manuscript) and in the regime of small matter effects (a perfectly viable approximation for DUNE); we then repeat the same calculation in the NSI scenario (Sect.\ref{CPasyNSI}) and in the $3+1$ sterile neutrino model (Sect.\ref{CPasysterile}).  Sect.\ref{DUNE} is devoted to a description of the DUNE experiment and the impact of NP of the neutrino energy spectra while in Sect.\ref{num_asy} we show the foreseen  experimental values on the asymmetries built in terms of number of events (integrated asymmetries) and their related uncertainties, both in the SM and in the NP scenarios. Eventually, in Sect.\ref{concl} we draw our conclusions. Expressions of the probabilities in the NP models analyzed here are provided in the Appendices A and B.

\section{CP asymmetries in the Standard Model} \label{CPasySM}

The neutrino state evolution equations when they travel through matter can be written as:
\begin{eqnarray} \label{evol}
i \frac{d}{dt} \left( \begin{array}{c} 
                   \nu_e \\ \nu_\mu \\ \nu_\tau 
                   \end{array}  \right)
 = \left[\frac{1}{2 E_\nu} U\left( \begin{array}{ccc}
                   0   & 0          & 0   \\
                   0   & \Delta m^2_{21}  & 0  \\
                   0   & 0           &  \Delta m^2_{31}  
                   \end{array} \right) U^{\dagger} +  
                  A_{CC} \left( \begin{array}{ccc}
            1      & 0 & 0 \\
            0  & 0  & 0 \\
            0 & 0 & 0
                   \end{array} 
                   \right) \right] ~
\left( \begin{array}{c} 
                   \nu_e \\ \nu_\mu \\ \nu_\tau 
                   \end{array}  \right)\, ,
\label{eq:matter}
\end{eqnarray}
where $U$ is the usual neutrino mixing matrix and $A_{CC}\equiv \sqrt 2 G_F n_e$, with $n_e$ being the electron density in the Earth crust. Defining $P(\nu_\alpha \to \nu_\beta)$ as the transition probability from a flavor $\alpha$ to a flavor $\beta$, one can construct the CP-odd asymmetries as:
\begin{equation}
A_{\alpha \beta} \equiv \frac{
P(\nu_\alpha\to \nu_\beta)-P(\bar{\nu}_\alpha \to \bar{\nu}_\beta)}
{P(\nu_\alpha \to \nu_\beta)+P(\bar{\nu}_\alpha \to \bar{\nu}_\beta)} \ .
\label{CPodd}
\end{equation}

It is well known that 
matter effects modify the behaviour of the asymmetries as a function of the Standard Model CP phase $\delta$ (see, e.g., \cite{Donini:1999jc}): in fact, the passage of neutrinos through matter introduces fake CP-violating effects which allows $A_{\alpha\beta}\ne 0$ even when $\sin \delta=0$. In principle, to extract genuine CP violating effects, one could defines the  subtracted asymmetries as  $A^{\rm sub}_{\alpha\beta}(\delta)=A_{\alpha\beta}(\delta)-A_{\alpha\beta}(\delta=0)$. However, we prefer to deal with more directly measurable quantities and we will use eq.(\ref{CPodd}) which, for non negligible matter effects, are non vanishing when $\delta=0,\pm \pi$.

To derive the analytic expressions for the asymmetries, 
we use perturbation theory in the small $\alpha=\Delta m^2_{21}/\Delta m^2_{31}$ ratio \cite{Kopp:2007ne} and expand the mixing angles according to:
\begin{equation}
\label{expansion}
s_{13} = \frac{r}{\sqrt{2}}\,, \qquad s_{12} = \frac{1}{\sqrt{3}}(1+s)\,,\qquad s_{23} =  \frac{1}{\sqrt{2}}(1+a)\,,
\end{equation}
where $r, s$ and $a$ represent the deviation from the tri-bimaximal mixing values of the neutrino mixing parameters, namely $\sin \theta_{13}=0, \sin \theta_{23} = 1/\sqrt{2}, \sin \theta_{12} = 1/\sqrt{3}$ \cite{King:2007pr,Pakvasa:2007zj}. It turns out that, given the recent fit to neutrino oscillation experiments, $r,s,a \sim {\cal O}(0.1)$. To simplify the notation, we further introduce $\Delta_{21} = \Delta m_{21}^2 L/ 4E_\nu$, $\Delta_{31} = \Delta m_{31}^2 L/ 4E_\nu$ and $V_{CC} = A_{CC} L/ 2 \Delta_{31} = 2 A_{CC} E_\nu / \Delta m^2_{31}$; at the DUNE peak energy, namely 2.5 GeV,  we estimate $V_{CC} \sim 0.2$ and we can further expand in the small $V_{CC}$. 

To start with, let us consider the {\it vacuum} case; for the $\nu_\mu \to \nu_e$ channel, the leading term of the asymmetry is the following:
\begin{eqnarray} 
\label{amueSM}
 A_{\mu e}^{SM_0} &=& -\frac{12}{f_1}\, r\, \alpha \Delta_{31}\sin \delta \sin^2 \Delta_{31} \,, \\ \nonumber 
 \end{eqnarray}
 where
 \begin{eqnarray}
 f_1 &=& 9 r^2\sin^2 \Delta_{31}
 +4 \alpha \Delta_{31}\left(\alpha \Delta_{31} + 3 r \cos \delta \cos \Delta_{31}\sin \Delta_{31}\right).
\end{eqnarray}
Being the numerator and the denominator of eq.(\ref{amueSM}) doubly suppressed by small quantities, we expect  $A_{\mu e}^{SM_0} \sim {\cal O}(1)$.\\
%(in fact, two small parameters appear both at the numerator and the denominator), the experimental fact that  $r$ is a bit larger than $\alpha$ brings $A_{\mu e}$ in vacuum to be of $\mathcal{O}(0.1)$. 
For the $\nu_\mu\to\nu_\tau$ channel, on the other hand, we find that the leading contribution to the asymmetry is given by a simpler expression:
\begin{eqnarray}\label{amutauSM}
 A_{\mu\tau}^{SM_0} &=& \frac{4}{3}r \alpha \Delta_{31}\sin\delta\,,
\end{eqnarray}
which is clearly smaller than $A_{\mu e}$.
%because of the presence of the factor $r\alpha$ only in the numerator. 
Notice also that, differently from  $A_{\mu e}$, this asymmetry becomes negative if $\delta>180^\circ$, as emerging from fits to neutrino oscillation data \cite{Esteban:2020cvm,deSalas:2020pgw}.
\\
A third possible asymmetry, namely $A_{\mu\mu}$, is obviously vanishing in vacuum because of CPT conservation but can assume a relevant role when matter effects are taken into account (as we will discuss later on). 

As it is well known, the inclusion of matter effects complicates the analytic expressions of the transition probabilities and, more importantly, that of the asymmetries. In order to deal with readable formulae, we can work in the regime of week matter potential $V_{CC} \ll 1$ which, as outline before, is a good approximation in the case of DUNE. Thus, we can organize our perturbative expansion as follows: 
\begin{equation}\label{orgpert}
A_{\alpha\beta} =A_{\alpha\beta}^{SM_0} +V_{CC} \, A_{\alpha\beta}^{SM_1} + {\cal O}(V_{CC}^2) \,,
\end{equation}
where $A_{\alpha\beta}^{SM_1}$ represents the first order correction to  the vacuum case $V_{CC} =0$. Thus, the asymmetries considered in this study acquire the following corrections:
\begin{eqnarray}\nonumber
A_{\mu e}^{SM_1} &=&-\frac{6}{f_1} r \nonumber
\left(\Delta_{31} \cos \Delta_{31}-\sin \Delta_{31}\right)\left[2\alpha\Delta_{31}\cos\delta\cos\Delta_{31}+3 r \sin\Delta_{31}+\right. \\&-& \left.   
  \frac{24}{f_1} r \alpha^2\sin^2\delta \,\Delta_{31}^2\sin^3 \Delta_{31}
\right] \,, \\
%\end{eqnarray}
%\begin{equation}\nonumber
A_{\mu \tau}^{SM_1} &=& -2 r^2 \left(1-\Delta_{31} \cot\Delta_{31}\right) + \frac{8}{27}\alpha^2\Delta_{31}^3\cot\Delta_{31} \,,\\
%\end{equation}
%\begin{eqnarray}
\label{amumuSM}
 A_{\mu\mu}^{SM_1} &=& \frac{4}{3}r \alpha \Delta_{31}\cos\delta\left(\Delta_{31}-\tan \Delta_{31}\right)-\frac{8}{27}\alpha^2 \Delta_{31}^3 \tan \Delta_{31}\,.
\end{eqnarray}
It is evident that $A_{\mu e}$ increases because a term proportional to $r^2/f_1$ appears, which is of $\mathcal{O}(1)$. Since at the atmospheric peak $\sin{\Delta_{31}}\gg \cos{\Delta_{31}}$, the $r^2/f_1$ correction is positive and adds an $\mathcal{O}(V_{CC})$ contribution to the total $A_{\mu e}$, that at the DUNE peak energy becomes roughly $1/2$. \\
On the other hand, $A_{\mu \tau}^{SM_1}$ contains only terms proportional to $V_{CC} r^2$ and $V_{CC} \alpha^2$ which are not balanced by any small denominator. Thus, both contributions set a correction to the vacuum asymmetry. \\
A similar situation arises for $A_{\mu\mu}$, where only terms proportional to $V_{CC} r \alpha$ and $V_{CC} \alpha^2$ appear. %Notice the lack of any  dependence on the standard CP phase in both $A_{\mu\tau}$ and $A_{\mu\mu}$.
%Finally, the asymmetry in the disappearance channel $A_{\mu\mu}$ is, in principle, the smallest of the three, since it does not contain any $r^2$ terms. However, when we are close to the atmospheric peak, $\Delta_{31}\sim \pi/2$. Thus the terms in $\tan{\Delta_{31}}$ can be very big and the expansion is no longer valid. This happens because in the definition of the CP-odd asymmetry, we have at the denominator $P(\nu_\mu\to\nu_\mu)+P(\bar{\nu}_\mu\to\bar{\nu}_\mu)$ which is very close to zero at the appearance peak. {\bf questa cosa non mi e' chiara; anche se la tangente e' molto grande, i termini non inclusi nella 8 saranno sempre piu' piccoli di erre*alpha e alpha quadrato; a cosa ti riferivi esattamente?}However, far from this special case, this expansion clearly shows that also the matter corrections to the disappearance asymmetry are very small.

\section{NSI and CP asymmetries}\label{CPasyNSI}

As mentioned in the Introduction, the uncertainties on the mixing parameters leave room for the possibility of the presence of Non Standard Interactions between neutrinos and the particles they meet travelling through the Earth. The strength of such new interactions can be parameterized in terms of the complex couplings 
$\varepsilon_{\alpha \beta}= |\epsilon_{\alpha\beta}| e^{i \phi_{\alpha\beta}}$, which modify the matter potential of eq.(\ref{evol}) to: 
% of which $\epsilon_{e\mu}$, $\epsilon_{e\tau}$ 
% and $\epsilon_{e\tau}$ are independent.
% 
\begin{eqnarray} 
A_{CC} \left( \begin{array}{ccc}
            1 + \varepsilon_{ee}     & \varepsilon_{e\mu} & \varepsilon_{e\tau} \\
            \varepsilon_{e \mu }^*  & \varepsilon_{\mu\mu}  & \varepsilon_{\mu\tau} \\
            \varepsilon_{e \tau}^* & \varepsilon_{\mu \tau }^* & \varepsilon_{\tau\tau} 
                   \end{array} 
                   \right)\, .
\label{eq:matterNSI}\nonumber
\end{eqnarray}
%\begin{eqnarray} 
%i \frac{d}{dt} \left( \begin{array}{c} 
%                   \nu_e \\ \nu_\mu \\ \nu_\tau 
%                   \end{array}  \right)
% = \left[\frac{1}{2 E_\nu} U_{PMNS} \left( \begin{array}{ccc}
%                   0   & 0          & 0   \\
%                   0   & \Delta m^2_{21}  & 0  \\
%                   0   & 0           &  \Delta m^2_{31}  
%                   \end{array} \right) U_{PMNS}^{\dagger} +  
%                  A_{CC} \left( \begin{array}{ccc}
 %           1 + \epsilon_{ee}     & \epsilon_{e\mu} & \epsilon_{e\tau} \\
%            \epsilon_{e \mu }^*  & \epsilon_{\mu\mu}  & \epsilon_{\mu\tau} \\
%            \epsilon_{e \tau}^* & \epsilon_{\mu \tau }^* & \epsilon_{\tau\tau} 
%                   \end{array} 
%                   \right) \right] ~
%\left( \begin{array}{c} 
%                   \nu_e \\ \nu_\mu \\ \nu_\tau 
%                   \end{array}  \right)\, ,
%\label{eq:matter}\nonumber
%\end{eqnarray}
%
Since the Hamiltionian has to be Hermitian, the three diagonal couplings $\varepsilon_{\alpha\alpha}$ must be real. Moreover, we can always subtract a matrix proportional to the identity without changing the transition probabilities. If we choose to subtract $\varepsilon_{\mu\mu} \mathbb{I}$, only two independent diagonal parameters ($\varepsilon_{ee}'=\varepsilon_{ee}-\varepsilon_{\mu\mu}$ and $\varepsilon_{\tau\tau}'=\varepsilon_{\tau\tau}-\varepsilon_{\mu\mu}$) will appear in the NSI matrix\footnote{Notice that, since from non-oscillation experiments bounds on $\varepsilon_{\mu\mu}$ are very stringent, $\varepsilon_{ee}'\sim\varepsilon_{ee}$ and $\varepsilon_{\tau\tau}'\sim\varepsilon_{\tau\tau}$.}. Thus, beside the standard oscillation angles and phases, the parameter space is enriched by five more moduli $|\varepsilon_{\alpha\beta}|$ and three more phases $\phi_{\alpha\beta}$, which could provide new sources of CP violation in the lepton sector. 

\subsection{Asymmetries in the NSI framework}

Since NSI effects are strongly intertwined with standard matter effects driven by $V_{CC}$, the asymmetries can be cast in a form which generalizes eq.(\ref{orgpert}):
\begin{equation}\label{genform}
A_{\alpha\beta} =A_{\alpha\beta}^{SM_0} +V_{CC}(A_{\alpha\beta}^{SM_1} +  A_{\alpha\beta}^{NSI}) + {\cal O}(V_{CC}^2) \,,
\end{equation}
where $A_{\alpha\beta}^{SM_{0,1}}$ refers to the pure Standard Model results and all the effects of the NSI are included in the $A_{\alpha\beta}^{NSI}$ term. \\
Bounds on the magnitude of the NSI couplings have been widely discussed \cite{Dev:2019anc}; even though some of them could in principle be of $\mathcal{O}(1)$ and give rise, for example, to degeneracies leading to  the so-called  LMA-Dark  solution \cite{Miranda:2004nb},  we decided nonetheless to consider all $\varepsilon_{\alpha\beta}$'s on the same footing and of the same order of magnitude as the other small standard parameters $a,s,r,\alpha$ and $V_{CC}$. In this way, we are able to catch the leading dependence on NP carried on by the CP asymmetries.

For the $\nu_\mu \to \nu_e$ channel, the leading order NSI contributions can be arranged as follows:
\begin{equation}
A_{\mu e}^{NSI} =\varepsilon_{e \mu} a_{\mu e}^{ \varepsilon_{e\mu}} +
\varepsilon_{e \tau} a_{\mu e}^{\varepsilon_{e\tau}}\,,
\label{mue_1}
\end{equation}
where the $a's$ functions are given by:
\begin{eqnarray}\nonumber
 a_{\mu e}^{\varepsilon_{e\mu}} &=& \frac{3}{f_1} \left[6r\cos(\delta-\delta_{e\mu}) \sin \Delta_{31} \left(\Delta_{31} \cos \Delta_{31}+\sin \Delta_{31}\right) \right. + \nonumber \\
 &&\left. 4 \alpha\Delta_{31}\cos\delta_{e\mu} \left(\Delta_{31}+ \cos \Delta_{31}\sin \Delta_{31}\right)\right] \label{mue_2}
 \\ &-& \frac{72}{f_1^2} r \alpha  \sin\delta\Delta_{31}^2 \sin^4\Delta_{31}\left[3 r \sin(\delta-\delta_{e\mu})+2\alpha \sin\delta_{e\mu})\right]\nonumber\,,
\end{eqnarray}

\begin{eqnarray}\nonumber
 a_{\mu e}^{\varepsilon_{e\tau}} &=& \frac{3}{f_1} \left[6r\cos(\delta-\delta_{e\tau}) \sin \Delta_{31} \left(-\Delta_{31} \cos \Delta_{31}+\sin \Delta_{31}\right) \right. + \nonumber \\ 
 &&\left. 2 \alpha\Delta_{31}\cos\delta_{e\tau} \left(-2\Delta_{31}+ \sin 2\Delta_{31}\right)\right] \label{mue_3_last}
 \\ &+& \frac{72}{f_1^2} r \alpha  \sin\delta\Delta_{31}^2 \sin^4\Delta_{31}\left[3 r \sin(\delta-\delta_{e\tau})-2\alpha \sin\delta_{e\tau}\right]\nonumber\,.
\end{eqnarray}

The only NP parameters appearing at the considered perturbative level are $\varepsilon_{e\mu}$
and $\varepsilon_{e\tau}$ which, in turn, carry the dependence on the CP phases $\delta_{e\mu},\delta_{e\tau}$. %Eqs.(\ref{mue_2},\ref{mue_3_last}) show that the new contributions are dominated by the new parameters
%$\varepsilon_{e \mu}$ and $\varepsilon_{e \tau}$,
%with the related coefficients $a_{\mu e}\sim {\cal O}(1/r)$ (consider that the $f_1$ function is doubly suppressed itself by $r^2,\alpha^2$ and $r \alpha$).  
All in all, the NSI contributions set an ${\cal O} (V_{CC})$ correction to $A_{\mu e}^{SM_0}$.
We also notice that the largest of the considered terms, namely the ones linear in $r$ in the numerator, have similar expressions in both $a_{\mu e}^{\varepsilon_{e \mu}}$ and in $a_{\mu e}^{\varepsilon_{e\tau}}$, apart from the sign in front of $\cos{\Delta_{31}}$. This means that, around the atmospheric peak, the phases $\delta_{e\mu}$ and $\delta_{e\tau}$ are  equally important even though the magnitude of their impact strongly depends on the value of the standard CP phase $\delta$.

For the asymmetry in the $\mu\tau$-channel, we found the following structure:
\begin{eqnarray}\nonumber
A_{\mu \tau}^{NSI} &=& 8 \varepsilon_{\mu\tau}\cos\delta_{\mu\tau}\Delta_{31}\cot\Delta_{31} + \\ 
&-&\frac{4}{3}\alpha\Delta_{31}^2\left(\varepsilon_{e\mu}\cos\delta_{e\mu}-\varepsilon_{e\tau}\cos\delta_{e\tau}-4 \varepsilon_{\mu\tau}\cos\delta_{\mu\tau}\csc^2\Delta_{31}\right) + \label{amtcorr}\\
&-&2 r \left[\varepsilon_{e\mu}\cos(\delta-\delta_{e\mu})+\varepsilon_{e\tau}\cos(\delta-\delta_{e\tau})\right]\left(1-\Delta_{31}\cot\Delta_{31}\right) \nonumber
\\
&+& 4 a \varepsilon_{\tau\tau}'\left(1-\Delta_{31}\cot\Delta_{31}\right)\nonumber
\,.
\label{mutau_formul}
\end{eqnarray}
In this case, four different NSI parameters enter the leading order corrections, namely $\varepsilon_{\mu\tau},\varepsilon_{e\tau},\varepsilon_{e\mu}$ (together with their phases) and $\varepsilon^\prime_{\tau\tau}$.
Contrary to the $\mu e$ case, the largest correction to the vacuum expression is given by the first order term $\varepsilon_{\mu\tau}$ in the first line of eq.(\ref{amtcorr}), which is not suppressed by any of the standard small parameters $a,r,s$ and $\alpha$. Considering that  $A_{\mu \tau}^{SM_0} \sim {\cal O}(r\alpha)$, this makes the $\mu \tau$-channel very promising for searching for NP, at least at the probability level where possible complications due to small $\tau$ statistics do not enter. 
%However it is worth to mention that the presence of $\cot{\Delta_{31}}$ suppress the first term of $A_{\mu \tau}^{NSI}$ at the atmospheric peak.}

Finally,  for the $\mu\mu$ channel, matter effects generate a substantial difference in the propagation of neutrinos versus antineutrinos, which results in the following NSI contributions:

\begin{eqnarray}\nonumber
 A_{\mu\mu}^{NSI} &=& -8 \varepsilon_{\mu\tau} \Delta_{31}  \cos\delta_{\mu\tau}\tan\Delta_{31} - 4 r \varepsilon_{e\mu} \Delta_{31}\cos \left(\delta-\delta_{e\mu}\right)\tan \Delta_{31} + \\
 %\\ && \label{amumuNSI} \\ 
 &+& 4 a \varepsilon_{\tau\tau}' \left(\Delta_{31}-\tan \Delta_{31}\right) \tan\Delta_{31} - \frac{4}{3}\alpha \Delta_{31}\times \\ \nonumber &&\left[\varepsilon_{e\mu}\cos\delta_{e\mu}\left(\Delta_{31}+\tan\Delta_{31}\right)-\varepsilon_{e\tau}\cos\delta_{e\tau}\left(\Delta_{31}-\tan\Delta_{31}\right)- \right.\\ && \left.  4 \Delta_{31}\varepsilon_{\mu\tau}\cos\delta_{\mu\tau}\sec^2\Delta_{31}\right]\nonumber \,.
\end{eqnarray}

As expected from unitarity relations, we get an opposite linear dependence on $\varepsilon_{\mu\tau}$ but with a coefficient proportional to $\tan{\Delta_{31}}$ which, close to the atmospheric peak, gives an important correction to $A_{\mu \mu}^{SM_1}$. 
%Thus, we expect that these correction can be very large, but again, as discussed in the previous section, near the peak the expansion is no longer valid {\bf insomma questa cosa del picco che invalida l'espansione e' un terreno minato}. 
% Also the other terms of the corrections, in particular the ones in $\varepsilon_{\tau\tau}' \tan^2{\Delta_{31}}$ and in $\varepsilon_{\mu\tau} \sec^2{\Delta_{31}}$ are strongly amplified at the atmospheric peak.

\section{Sterile neutrinos and CP asymmetries}\label{CPasysterile}

The next NP scenario under discussion is the so-called {\it 3+1} model, in which a sterile neutrino state supplements the three standard active neutrinos.
Even though the new state cannot interact with the ordinary matter, it can have a role in neutrino oscillations thanks to the mixing with the active partners. The long-standing reactor, gallium and short-baseline anomalies \cite{Boser:2019rta} suggested that, if present, the fourth mass eigenstate $m_4$ should have a mass such that $\Delta m_{41}^2 = m_4^2 - m_1^2\sim {\cal O}(1)$ eV$^2$, that is orders of magnitude larger than the solar and the atmospheric mass splittings, thus capable to drive very fast oscillations visible at accordingly small $L/E$. In addition to the new mass splitting $\Delta m_{41}^2$, the PMNS matrix becomes a $4\times 4$ matrix which can be parametrized in terms of 6 angles and 3 phases. In this manuscript, we adopt the following multiplication order of the rotation matrices $R(\theta_{ij})$ \cite{Maltoni:2007zf}-\cite{Meloni:2010zr}:
\begin{eqnarray}
 U=R(\theta_{34})R(\theta_{24})R(\theta_{23},\delta_3)R(\theta_{14})R(\theta_{13},\delta_2)R(\theta_{12},\delta_1).
 \label{parameterization}
\end{eqnarray}
Apart from $\delta_2$, which becomes the standard CP phase for $m_4 \to 0$, we have two potential new sources of CP violation, encoded in two phases $\delta_1$ and $\delta_3$.
In the description of neutrino propagation in matter, we cannot disregard the role of the NC interactions because the sterile state does not feel at all the presence of matter; this results in the following evolution equations:
\begin{eqnarray} 
&& i\frac{d}{dt} \left( \begin{array}{c} 
                   \nu_e \\ \nu_\mu \\ \nu_\tau  \\ \nu_s
                   \end{array}  \right) =\\ 
 &=& \left[\frac{1}{2 E_\nu} U \left( \begin{array}{cccc}
                   0   & 0          & 0  & 0 \\
                   0   & \Delta m^2_{21}  & 0 & 0 \\
                   0   & 0           &  \Delta m^2_{31} & 0 \\
                   0 & 0 & 0 & \Delta m^2_{41}
                   \end{array} \right) U^{\dagger} +  
                   \left( \begin{array}{cccc}
            A_{CC} + A_{NC}     & 0 & 0 & 0 \\
            0  & A_{NC}  & 0 & 0 \\
            0 & 0 & A_{NC} & 0 \\
            0 & 0 & 0 & 0
                   \end{array} 
                   \right) \right] ~
\left( \begin{array}{c} 
                   \nu_e \\ \nu_\mu \\ \nu_\tau \\
                   \nu_s
                   \end{array}  \right) ,
\label{eq:matter3p1}\nonumber
\end{eqnarray}
where $\nu_s$ is the new sterile state, $A_{CC}$ is the usual matter charged current potential and $A_{NC}$ is the matter NC potential, $A_{NC}\equiv 1/ \sqrt 2 G_F n_n$, with $n_n$ being the neutron density in the Earth crust. 

\subsection{Asymmetries in 3+1 framework}

The  parameter space of the 3+1 model is enlarged compared to the SM case by three new mixing angles $\theta_{i4}$, two more CP phases $\delta_{1,3}$ and the mass-squared difference  $\Delta m^2_{41}$. Thus, in addition to the expansion parameters used in the previous sections ($r,s,a$), we also expand in the small $s_{14},s_{24}$ and $s_{34}$ (where $s_{i4} =\sin \theta_{i4}$) that we can still assume of ${\cal O}(0.1)$. To further simplify the analytic expressions of the asymmetries, we also introduce  $V_{NC} = A_{NC} L/ 2 \Delta_{31}$. It is useful to present the results
 in a form similar to eq.(\ref{genform}):
\begin{equation}
A_{\alpha\beta} =A_{\alpha\beta}^{SM}  +  A_{\alpha\beta}^{3+1} + {\cal O}(\lambda^n) \,,
\end{equation}
where $A_{\alpha\beta}^{SM}$ are the SM asymmetries and the symbol $\lambda$ represents a common order of magnitude of all small quantities used in our perturbation theory, including $V_{CC}$ (but not $V_{NC}$, whose dependence in $A_{\alpha\beta}$ is exact). 
The exponent amount to $n=3$ for $A_{\mu \tau}, A_{\mu \mu}$ and $n=2$ for $A_{\mu e}$.
Notice that, due to the parametrization adopted in this manuscript, the SM phase $\delta$ of eqs.(\ref{amueSM})-(\ref{amumuSM}) must be replaced by the combination $\delta_2-\delta_1-\delta_3$. 
Averaging out all the fast oscillations driven by $\Delta m_{41}^2$, the various $A_{\alpha\beta}^{3+1}$ have the following expressions:
\begin{eqnarray}\nonumber
A_{\mu e}^{3+1} &\sim&\frac{s_{14} s_{24}}{f_1}\{-6 \left[2 \alpha \Delta_{31} \sin\delta_1 +
 3 r \cos\Delta_{31} \sin(\delta_2 - \delta_3) \sin\Delta_{31}\right]\} + \nonumber \\ &&\frac{s_{14}s_{24}}{f_1^2}\{
 216 r^2 \alpha \Delta_{31} \cos(\delta_2 - \
\delta_3) \sin(\delta_1 - \delta_2 + \delta_3) \
\sin^4\Delta_{31}\nonumber\}\,,\\
 A_{\mu \tau}^{3+1} &=& 2 s_{24} s_{34} \cot\Delta_{31} (\sin\delta_3-2 V_{NC} \Delta_{31} \cos\delta_3) \,,\label{3+1asy}\\
A_{\mu \mu}^{3+1} &=& 4 s_{24} s_{34} V_{NC} \Delta_{31} \cos\delta_3 \tan\Delta_{31}\nonumber\,.
\end{eqnarray}
To avoid large expressions, for $A_{\mu e}$ we only quote the corrections due to the new mixing angles.

%\begin{eqnarray}\nonumber
% g(s_{14}s_{24}\alpha)+ g^\prime(r,s,\alpha)&=&-6 s_{14} s_{24}\left[2 \alpha \Delta_{31} \sin\delta_1 +
% 3 r \cos\Delta_{31} \sin(\delta_2 - \delta_3) \sin\Delta_{31}\right]\\
% &&-6 r s \alpha \Delta_{31}  \sin(\delta_1 - \delta_2 + \delta_3)\sin\Delta_{31}^2\nonumber 
%\end{eqnarray}

First of all, we notice that the corrections to the $\mu e$ asymmetry are only linearly  suppressed compared to the leading order results; thus, we expect such an asymmetry to be quite sensitive to new sources of CP violation. 
Then, both corrections to the $\mu\tau$ and the $\mu\mu$ asymmetries are linear in the combination $s_{24}s_{34}$. Since the angle $\theta_{34}$ has weak constraints (values of $20-30^\circ$ are still allowed), these corrections can be relatively large. Notice also that, since $V_{NC}$ is roughly of the same order of magnitude as $V_{CC}$, $A_{\mu\tau}^{3+1}$ is expected to provide a large correction to the standard model asymmetries, making the $\nu_\tau$ appearance channel, at least in principle, very sensitive to NP effects. 
%However, as discussed in the previous sections, at the atmospheric peak $\cot\Delta_{31}\ll 1$ and $\tan\Delta_{31}\gg 1$; this effect can amplify the $\mu\mu$ asymmetry and suppress the $\mu\tau$ one. \\
As for the PMNS phases, all leading order corrections depend only on the new phase $\delta_3$. This means that a long baseline experiment is mostly sensitive only to the combination $\delta_2-\delta_1-\delta_3$ and to the single phase $\delta_3$.

Beside the results of eq.(\ref{3+1asy}), it is worth considering a new asymmetry corresponding to the $\nu_\mu\to\nu_s$ transition. Even though sterile neutrinos cannot be directly detected, the probability $P(\nu_\mu\to\nu_s)$ is a measure of  the NC events in the detector. Indeed, being the NC interactions flavor independent, the number of events is proportional to the sum of the transition probabilities from the starting flavor ($\nu_\mu$) to the three active final flavors ($\nu_{e,\tau,\mu})$ because of the unitarity relation $P(\nu_\mu\to\nu_s)=1-P(\nu_\mu\to\nu_{e,\mu,\tau})$. The new asymmetry has vanishing matter corrections and, at the leading non-vanishing order, reads:
\begin{equation}
    A_{\mu s}^{3+1}=-\frac{2 s_{24}s_{34} \sin\delta_3 \sin{\Delta_{31}} \cos{\Delta_{31}}}{2s_{24}^2+(s_{34}^2-s_{24}^2) \sin^2{\Delta_{31}}}\,.
\end{equation}
This is clearly an $\mathcal{O}(1)$ result since both numerator and denominator are of $\mathcal{O}(\lambda^2)$. 
%However, at the atmospheric peak, also this asymmetry is suppressed by the small  $\cos{\Delta_{31}}$.{\bf riflessione generale da fare su queste considerazioni del picco}
In Tab.(\ref{summary}) we summarize the outcome of our analytic considerations on the magnitude of the NP corrections to the asymmetries discussed in this paper. 

\begin{table}[]
\centering
\begin{tabular}{|c|c|c|c|}
\hline
\textbf{Asymmetry}   & \textbf{SM}  & \textbf{NSI}  & \textbf{3+1} \\ \hline
$A_{\mu e}$       &  $1$ & $\lambda^2$&  $\lambda$     \\ \hline
$A_{\mu \mu}$ & $\lambda^3$       & $\lambda^2$ & $\lambda^2$  \\ \hline
$A_{\mu \tau}$   & $\lambda^2$      &  $\lambda^2$ &  $\lambda^2$   \\ \hline
$A_{\mu s}$    & -    & - & 1      \\ \hline
\end{tabular}
\caption{\label{summary}\it  Order of magnitude estimates of the various contributions to the asymmetries discussed in this paper. $\lambda$ is a common order parameter such that: $r,s,a,\Delta_{21}, V_{CC}, \varepsilon_{\alpha\beta},\theta_{i4}\sim {\cal O} (\lambda)$.}
\end{table}

\section{The DUNE experiment}\label{DUNE}

The DUNE (Deep Underground Neutrino Experiment) experiment is a proposed long-baseline experiment based in the USA \cite{Acciarri:2016crz}-\cite{Abi:2020evt}. The accelerator facility and the Near Detector will be located at Fermilab, while the Far Detector is going to be built at the SURF (Sanford Underground Research Facility) laboratories in South Dakota, 1300 km away from the neutrino source. \\
Both far and near detectors will be LAr-TPCs, namely detectors with very good imaging capabilities, which are expected to collect a huge number of different neutrino interaction events. \\
The neutrino beam will be a $\nu_\mu$ beam with a small $\nu_e$ contamination. Focusing horns, which are able to select particles with a given electric charge before the decay tunnel, will produce particle and antiparticles beams, allowing the experiment to run in two different modes, namely the $\nu$-mode and the $\bar{\nu}$-mode. The neutrino flux energy spectra should be peaked at $E_{peak}=2.5$ GeV, however different proposal have been promoted for higher energy fluxes.
Indeed, even though $E_{peak}$ is the energy of the atmospheric peak of the oscillation probabilities at 1300 km baseline, we are below the $\tau$ production threshold ($E_{thr}=3.1$ GeV). Thus, with such a flux,  CC interactions of the huge number of $\nu_\tau$-s arriving at the far detector are forbidden. A broader and more energetic flux would overcome this problem, allowing the tau neutrinos to be energetic enough to produce $\tau$ leptons. This {\it $\tau$-optimized} flux \cite{DUNEwebsite,Bi} would be less performing in constraining oscillation parameters from $\nu_e$ appearance channel (due to the increased number of background events such as misidentified $\nu_\tau$-s) but at the same time very useful for NP searches thanks to the increased number of $\nu_\tau$ events. \\
In order to simulate the DUNE experiment, we used the GLoBES package \cite{Huber:2004ka,Huber:2007ji}. Initial fluxes, far detector efficiencies and energy resolutions, backgrounds and systematic uncertainties have been provided by the DUNE collaboration for the $\nu_e$ appearance and $\nu_\mu$ disappearance channels \cite{Alion:2016uaj, Abi:2021arg}. For the former, backgrounds are misidentified $\nu_\mu$, $\nu_\tau$ and NC events, as well as $\nu_e$-s from the flux contamination; for the latter, the main background source are NC events. Systematic normalization uncertainties at the far detector are 2\% for the $\nu_e$ appearance signal and 5\% for the $\nu_\mu$ disappearance signal, numbers proposed by the DUNE collaboration from the foreseen performance of the Near Detector. \\
In the last years, the possibility to study also the $\nu_\tau$ appearance and the NC channels as signals have been taken into account. For the first one, the use of the hadronic and electronic decays of $\tau$ leptons to identify the event topology have been discussed in \cite{deGouvea:2019ozk,Ghoshal:2019pab}. According to the cited literature, we used an efficiency of 30\% for both electronic and hadronic decay events, 20\% systematic normalization uncertainty and misidentified $\nu_e$ and NC events as backgrounds. For the NC channel, 90\% effieciency, 10\% systematic uncertainty and 10\% of the $\nu_\mu$ events as a background have been proposed in \cite{Coloma:2017ptb} and adopted here. 

\subsection{Effects of NSI and sterile neutrinos on DUNE spectra}\label{spectrasect}
\begin{figure}
    \centering
    \includegraphics[width=5cm, height=5cm]{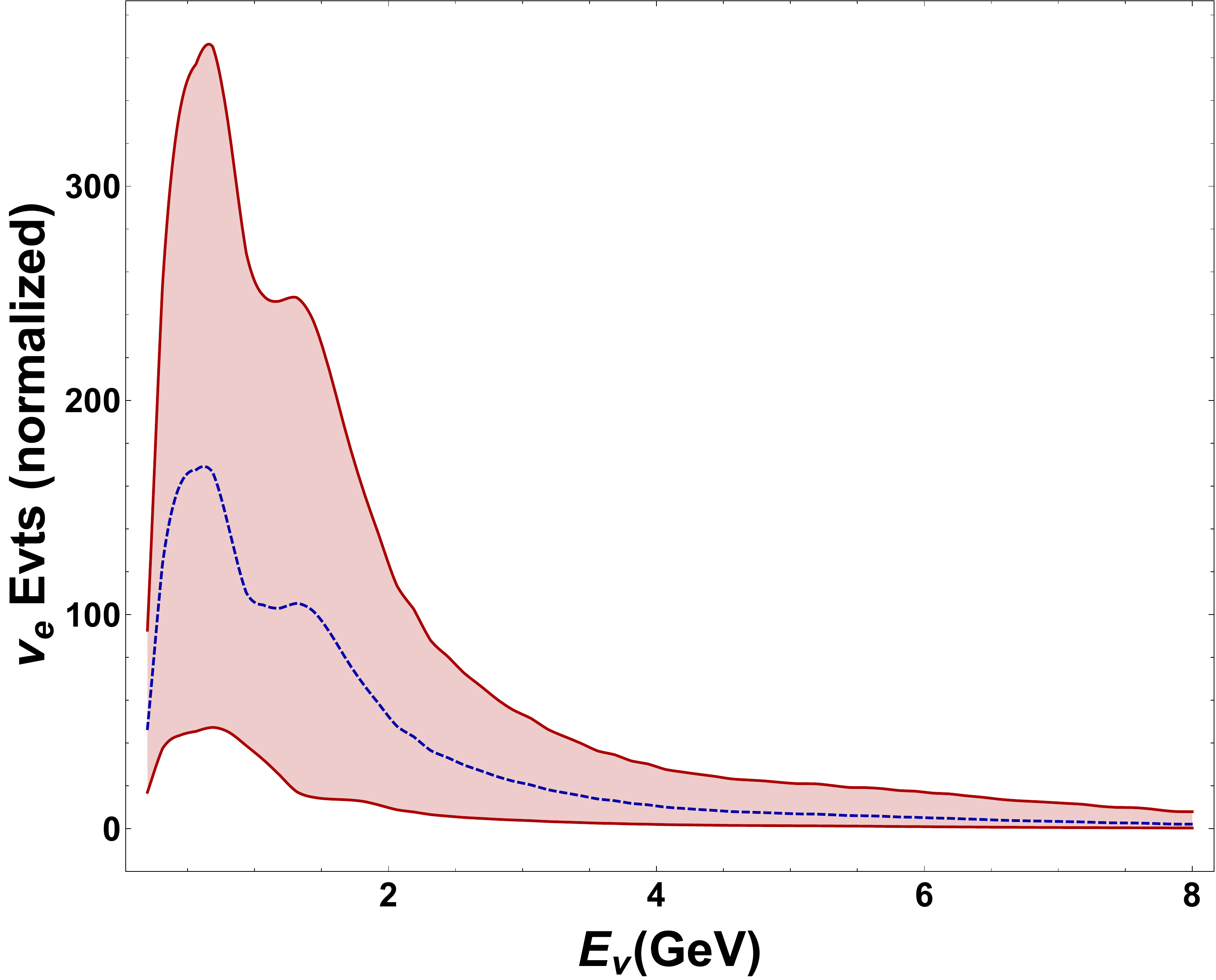}
    \includegraphics[width=5cm, height=5cm]{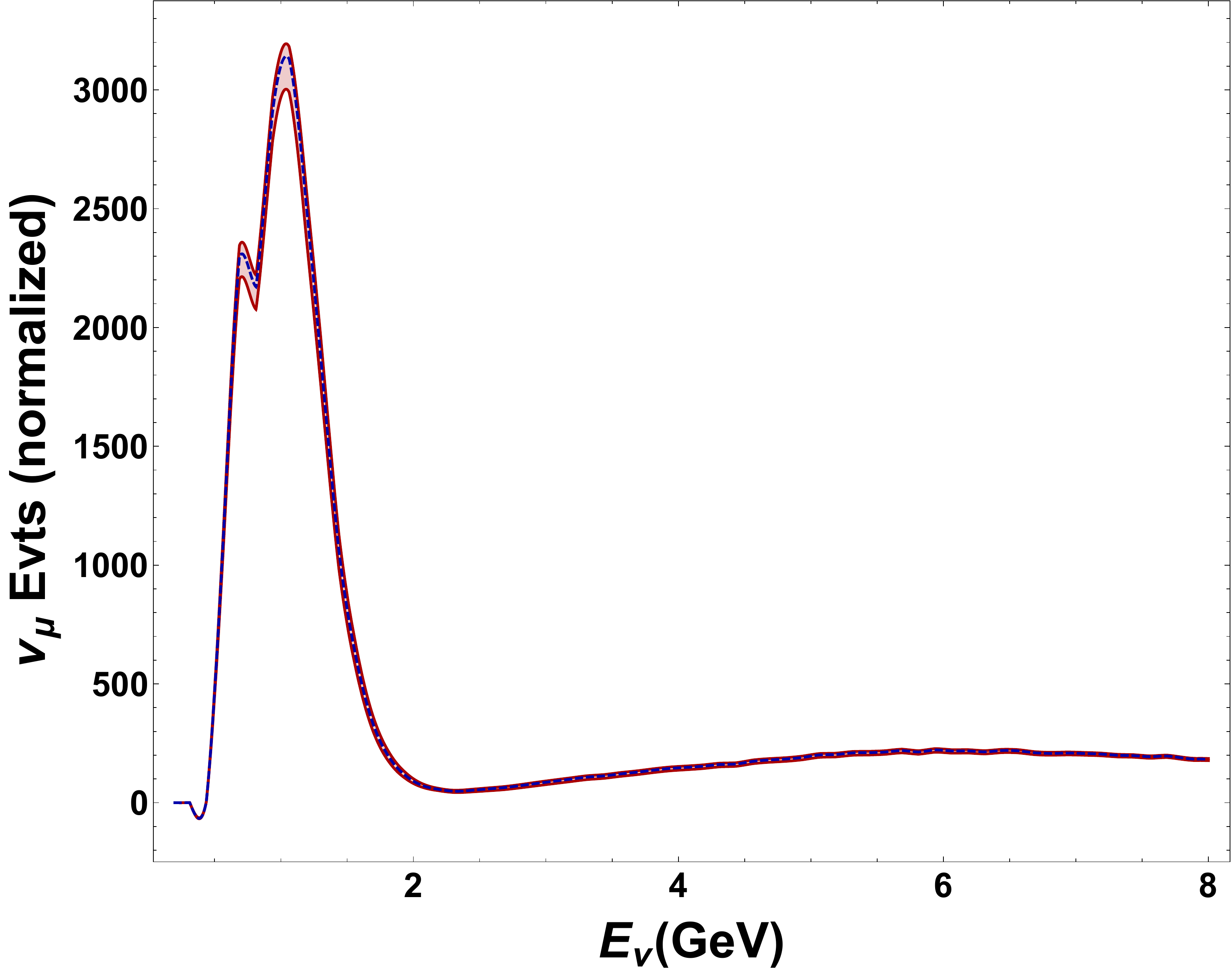}
    \includegraphics[width=5cm, height=5cm]{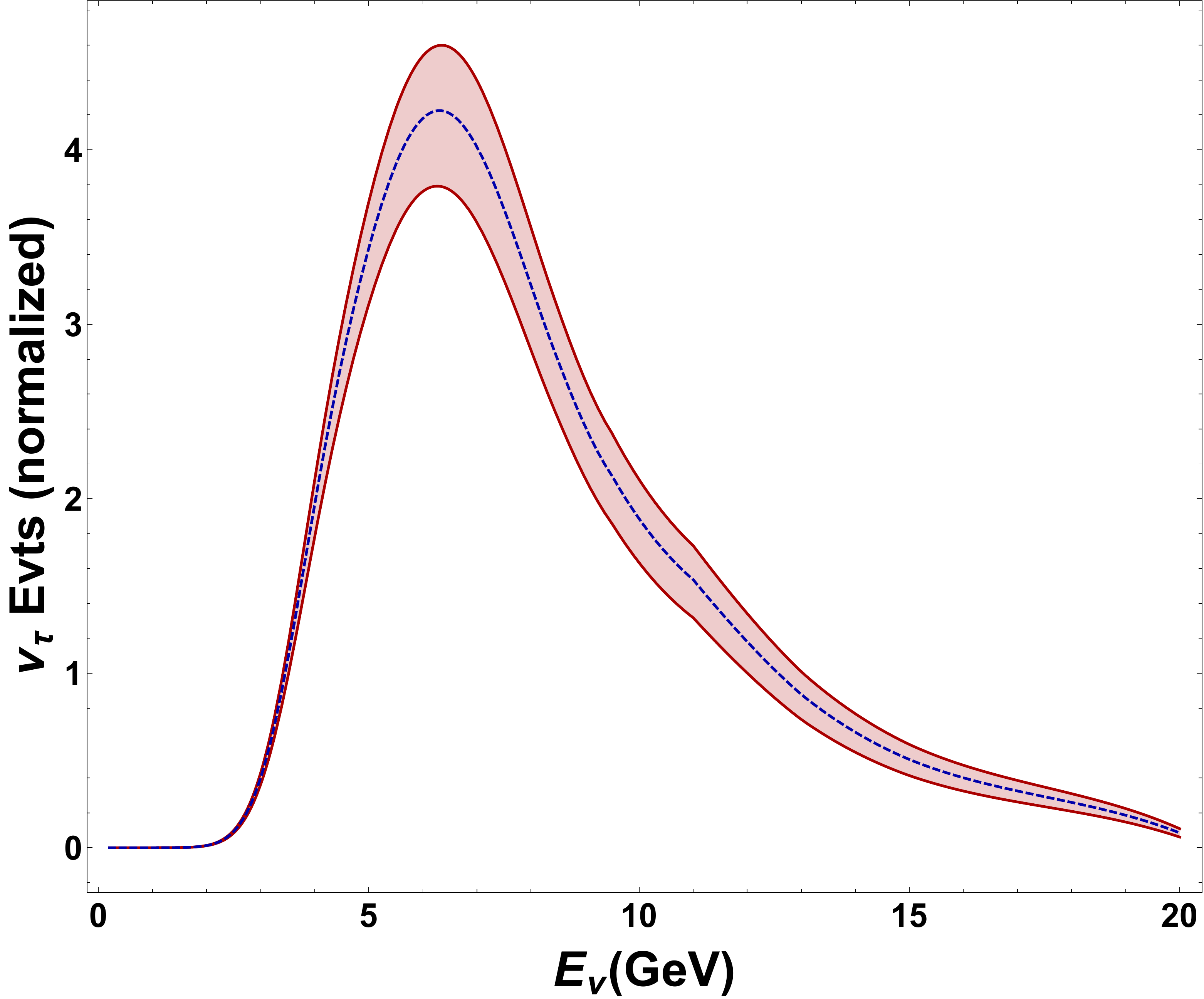}
    \includegraphics[width=5cm, height=5cm]{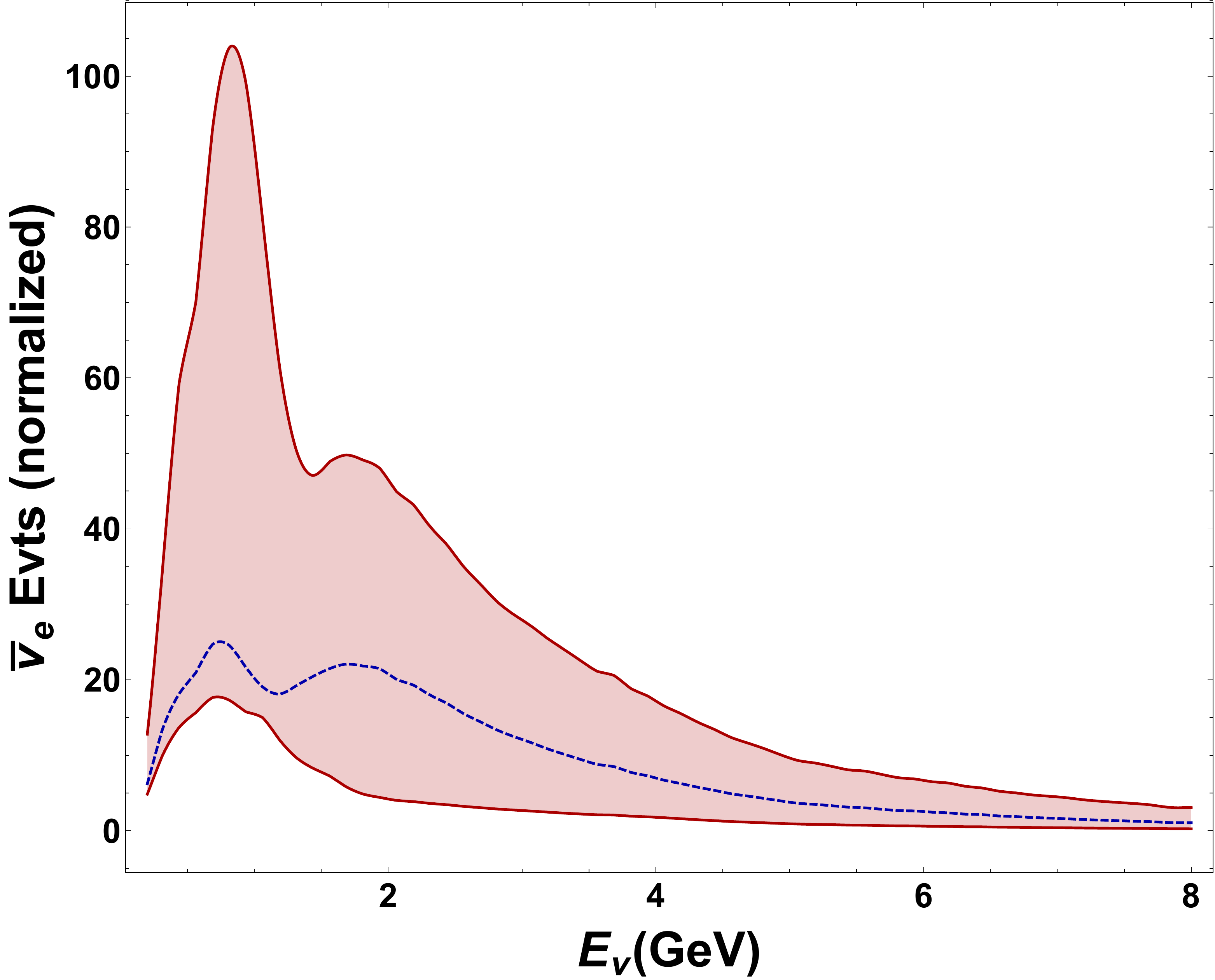}
    \includegraphics[width=5cm, height=5cm]{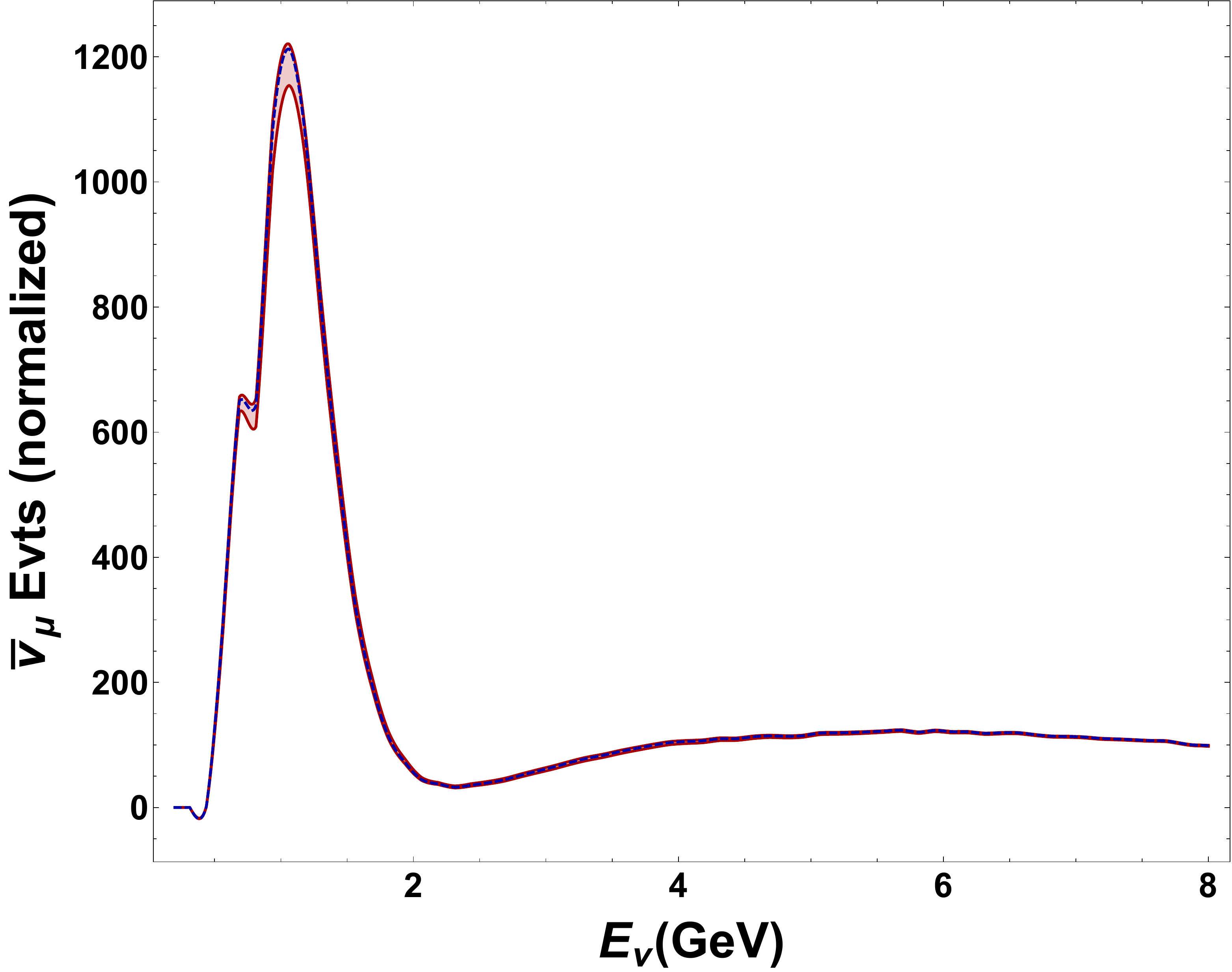}
    \includegraphics[width=5cm, height=5cm]{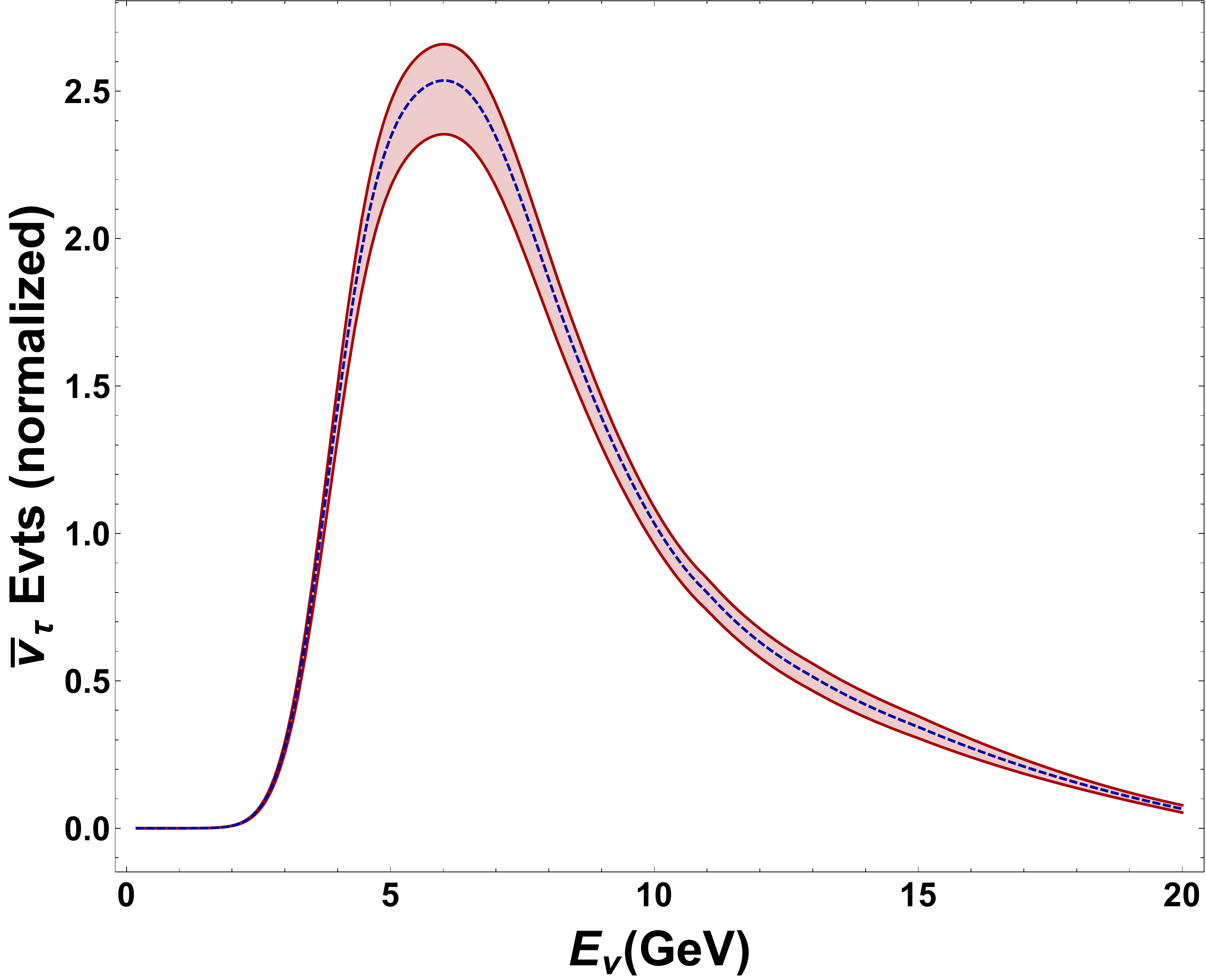}
    \caption{\it Expected number of neutrino (upper panels) and antineutrino (lower panels) events at the DUNE far detector as a function of the neutrino energy. The (orange) bands show the effects of including the NSI parameters into the transition probabilities while the blue dashed line is referred to the spectra obtained in the standard model. Best fits for the standard oscillation angles, phase and mass differences are taken from \cite{Esteban:2020cvm} and reported in Tab.\ref{SMparam}. Standard fluxes for neutrinos have been used.} 
    \label{NSIspectra}
\end{figure}

Before discussing the sensitivity on the CP asymmetries, it is useful to  have a look at the effects of New Physics on the neutrino spectra, which will help in the interpretation of our numerical results. \\
For the NSI case, different global analyses on oscillation experiments have been done \cite{Esteban:2019lfo,Huitu:2016bmb,Dev:2019anc}. The 2$\sigma$ current limits on the various $\varepsilon_{\alpha\beta}$ from ref.\cite{Esteban:2019lfo} have been summarized in Tab.\ref{tabNSI}.
\begin{table}[]
\centering
\begin{tabular}{|c|c|}
\hline
\textbf{NSI parameters}   & \textbf{$2\sigma$ bounds} \\ \hline
$\varepsilon_{ee}'$       & (-0.2 , 0.45)            \\ \hline
$\varepsilon_{\tau\tau}'$ & (-0.02 , 0.175)          \\ \hline
$|\varepsilon_{e \mu}|$   & \textless{}0.1           \\ \hline
$|\varepsilon_{e \tau}|$    & \textless{}0.3           \\ \hline
$|\varepsilon_{\mu\tau}|$   & \textless{}0.03          \\ \hline
\end{tabular}
\caption{\label{tabNSI}\it  2$\sigma$ bounds on the moduli of the NSI parameters, from \cite{Esteban:2019lfo}.}
\end{table}
Scanning the NSI parameters in the allowed ranges and taking the new phases $\delta_{e\mu}$,$\delta_{e\tau}$ and $\delta_{\mu\tau}$ in the range $[0,2\pi]$ (both sets of parameters extracted randomly flat), we get the  neutrino and antineutrino spectra at the far detector, as shown in Fig.\ref{NSIspectra}. The number of events is normalized by the bin width and the exposure for both $\nu$-mode and $\bar{\nu}$-mode is 3.5 years.\\
In all panels, the blue dashed lines refer to the spectra obtained in the standard oscillation framework with normal hierarchy (NH), for which the  best fit values are summarized in Tab.\ref{SMparam}.
\begin{table}[]
\centering
\renewcommand{\arraystretch}{1.6}
\begin{tabular}{|c|c|c|}
\hline
\textbf{Oscillation parameters}              & \textbf{Best fits (NH)}   & \textbf{Best fits (IH)}    \\ \hline
$\theta_{12}/^\circ$                         & $33.44^{+0.78}_{-0.75}$   & $33.45^{+0.78}_{-0.75}$    \\ \hline
$\theta_{13}/^\circ$                         & $8.57^{+0.13}_{-0.12}$    & $8.61^{+0.12}_{-0.12}$     \\ \hline
$\theta_{23}/^\circ$                         & $49^{+1.1}_{-1.4}$        & $49.3^{+1.0}_{-1.3}$       \\ \hline
$\delta/^\circ$                              & $195^{+51}_{-25}$         & $286^{+27}_{-32}$          \\ \hline
$\frac{\Delta m_{21}^2}{10^{-5} \, \, eV^2}$ & $7.42^{+0.21}_{-0.20}$    & $7.42^{+0.21}_{-0.20}$     \\ \hline
$\frac{\Delta m_{31}^2}{10^{-3} \, \, eV^2}$ & $2.514^{+0.028}_{-0.027}$ & $-2.497^{+0.028}_{-0.028}$ \\ \hline
\end{tabular}
\caption{\label{SMparam}\it Best fits for oscillation parameters obtained by the global analysis in \cite{Esteban:2020cvm}.}
\end{table}
The figures clearly show that the $\nu_e$ and $\bar{\nu}_e$ spectra are the ones affected the most by the NSI parameters. This is because at the DUNE energies and baseline, the SM $\nu_e$ appearance probability is suppressed and the effect of the NSI parameters, in particular that of $\varepsilon_{e\tau}$ which has weaker bounds compared to the others, results more evident. 
%We also observe that the modifications of the spectra due to New Physics are different for the neutrinos and antineutrinos: this suggests that the $\mu e$ asymmetry should be quite sensitive to NSI. {\bf anche qui bisogna fare attenzione, dato che dal punto di vista analitico la correzione e' di ordine $\lambda^2$. Forse potremmo cancellare questa frase dato che al LO  $A_{\mu e}\sim 1$ vuol dire che una probabilita' domina sull'altra anche nel MS, non solo sull'NSI}\\
Conversely, in the $\nu_\mu$ spectra the NSI have a very small impact: in fact, in the disappearance probability the first term is of $\mathcal{O}(1)$, and the largest NSI corrections is driven by $\varepsilon_{\mu\tau}$ which has very strong bounds $\sim \mathcal{O}(10^{-3})$.\\
Finally, in the $\nu_\tau$ appearance channel the modifications in the spectra due the NSI's are evident but, due to the small number of expected events, the changes with respect to the SM case are difficult to observe. 

\begin{figure}
    \centering
    \includegraphics[width=5cm, height=5cm]{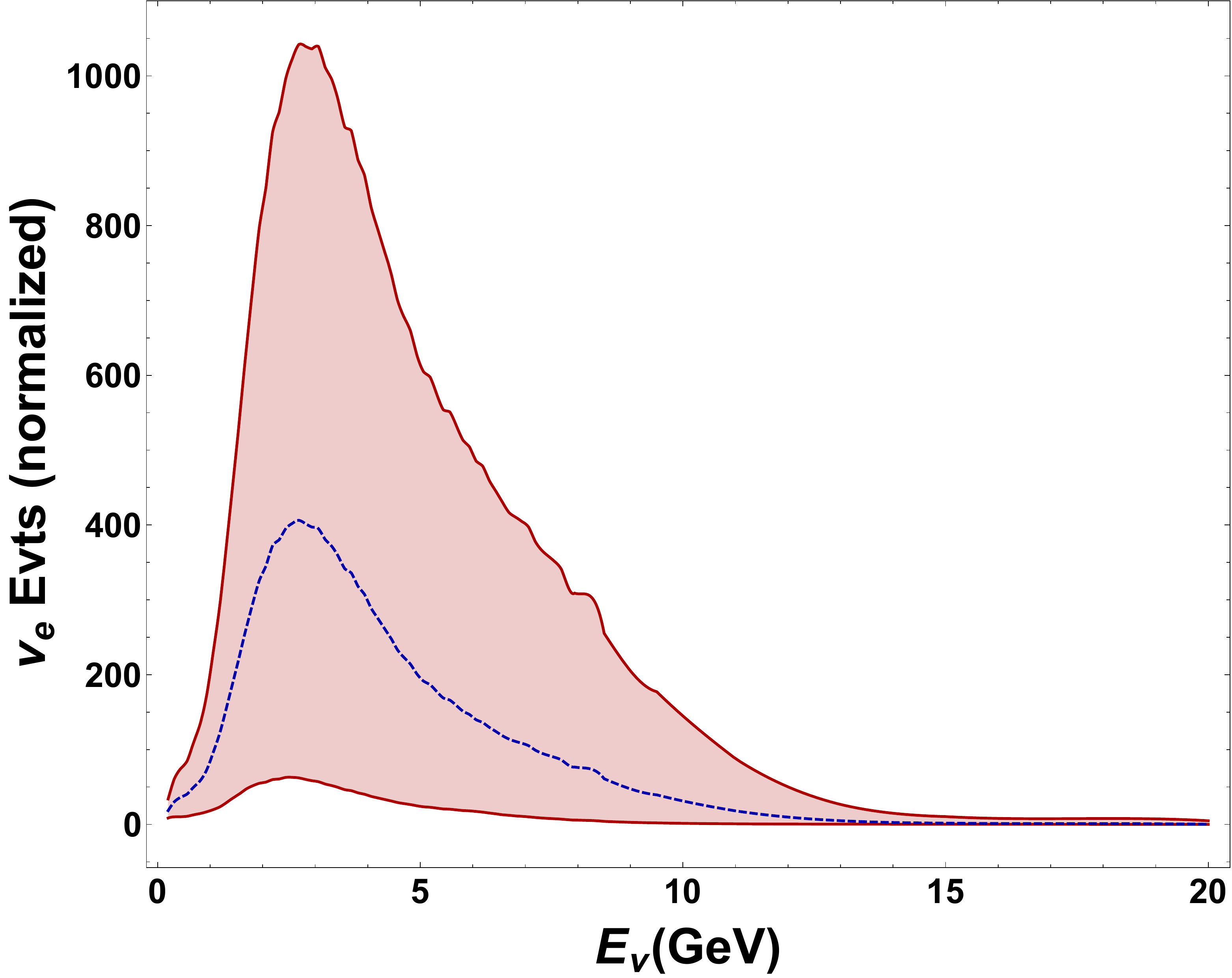}
    \includegraphics[width=5cm, height=5cm]{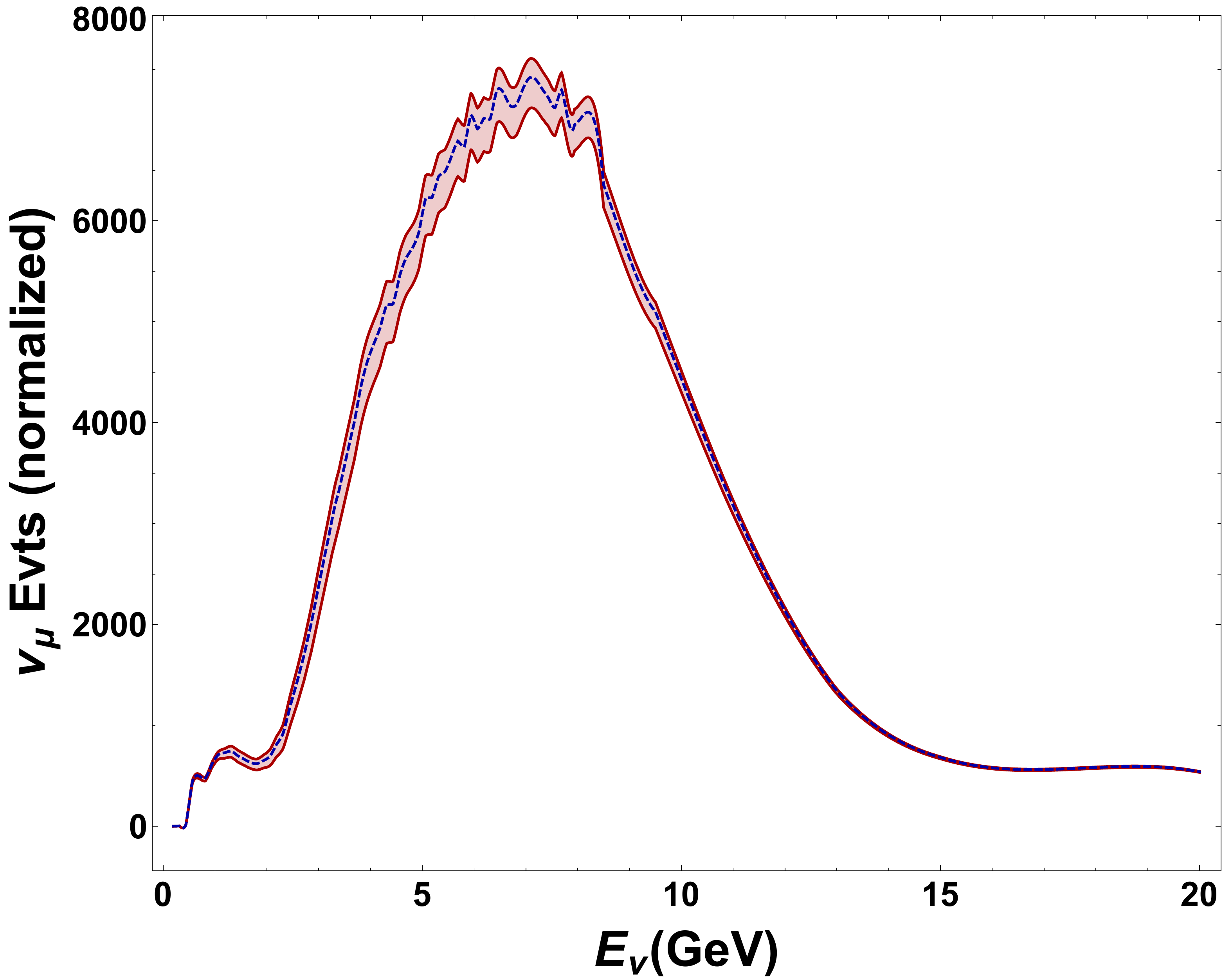}
    \includegraphics[width=5cm, height=5cm]{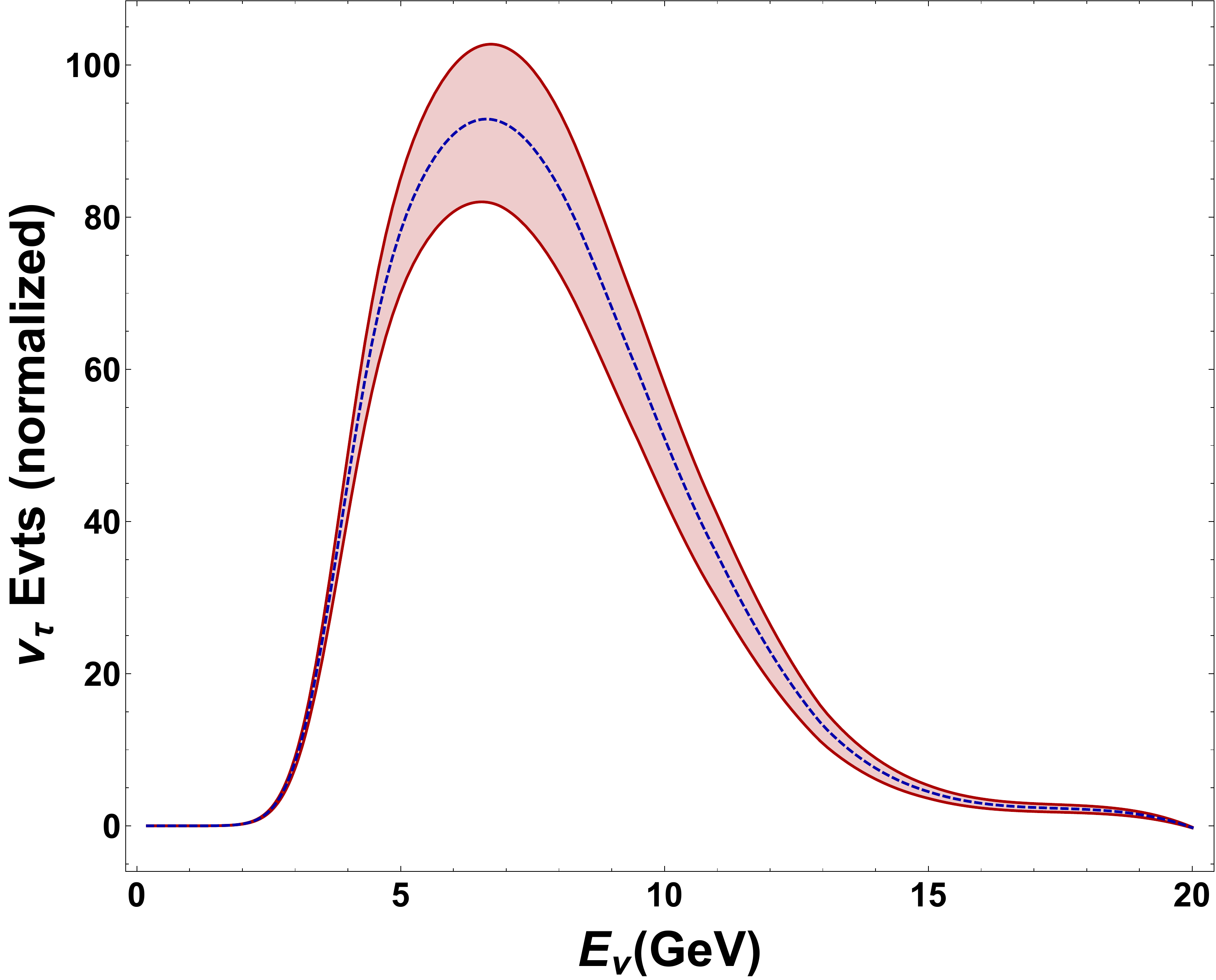}
    \includegraphics[width=5cm, height=5cm]{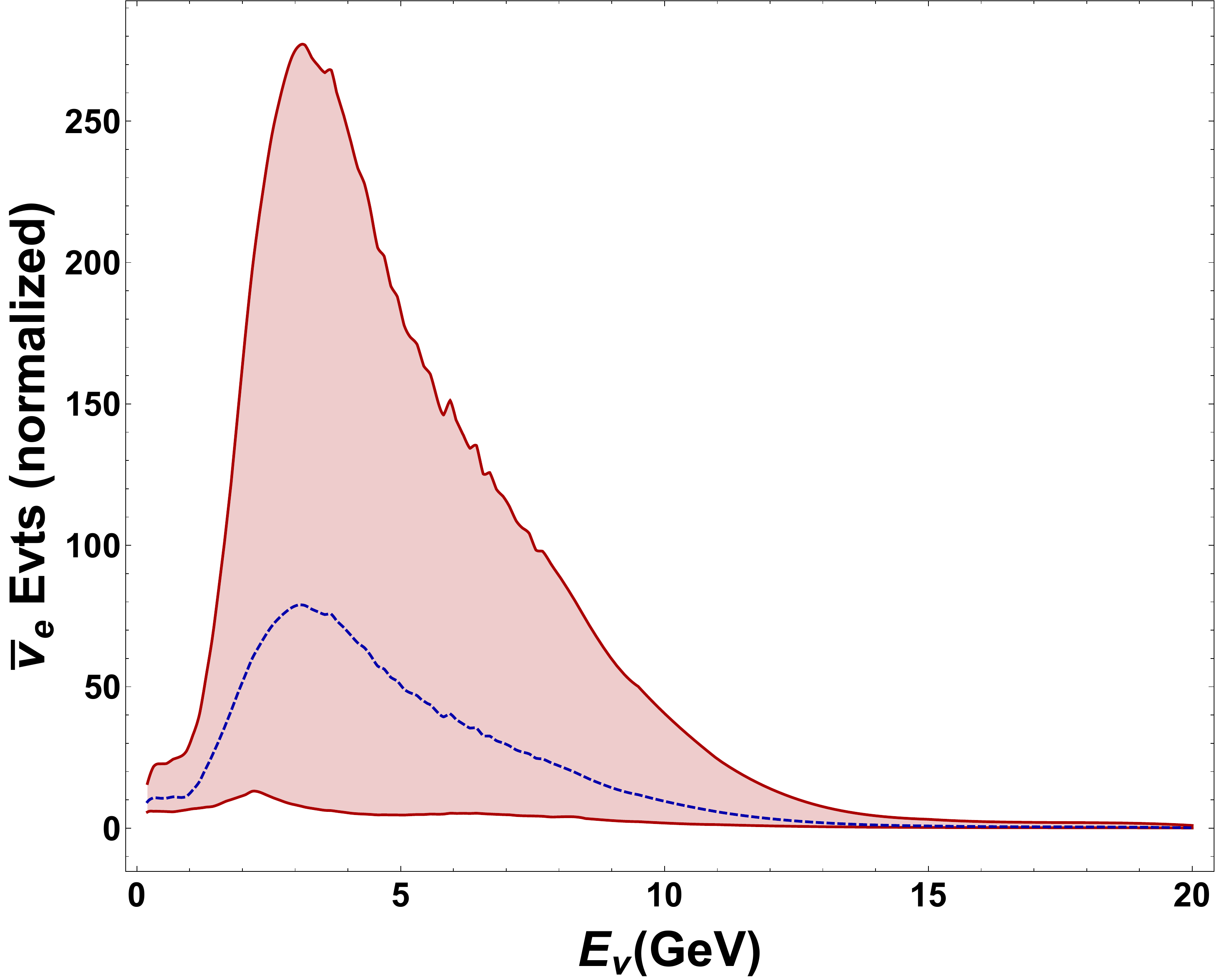}
    \includegraphics[width=5cm, height=5cm]{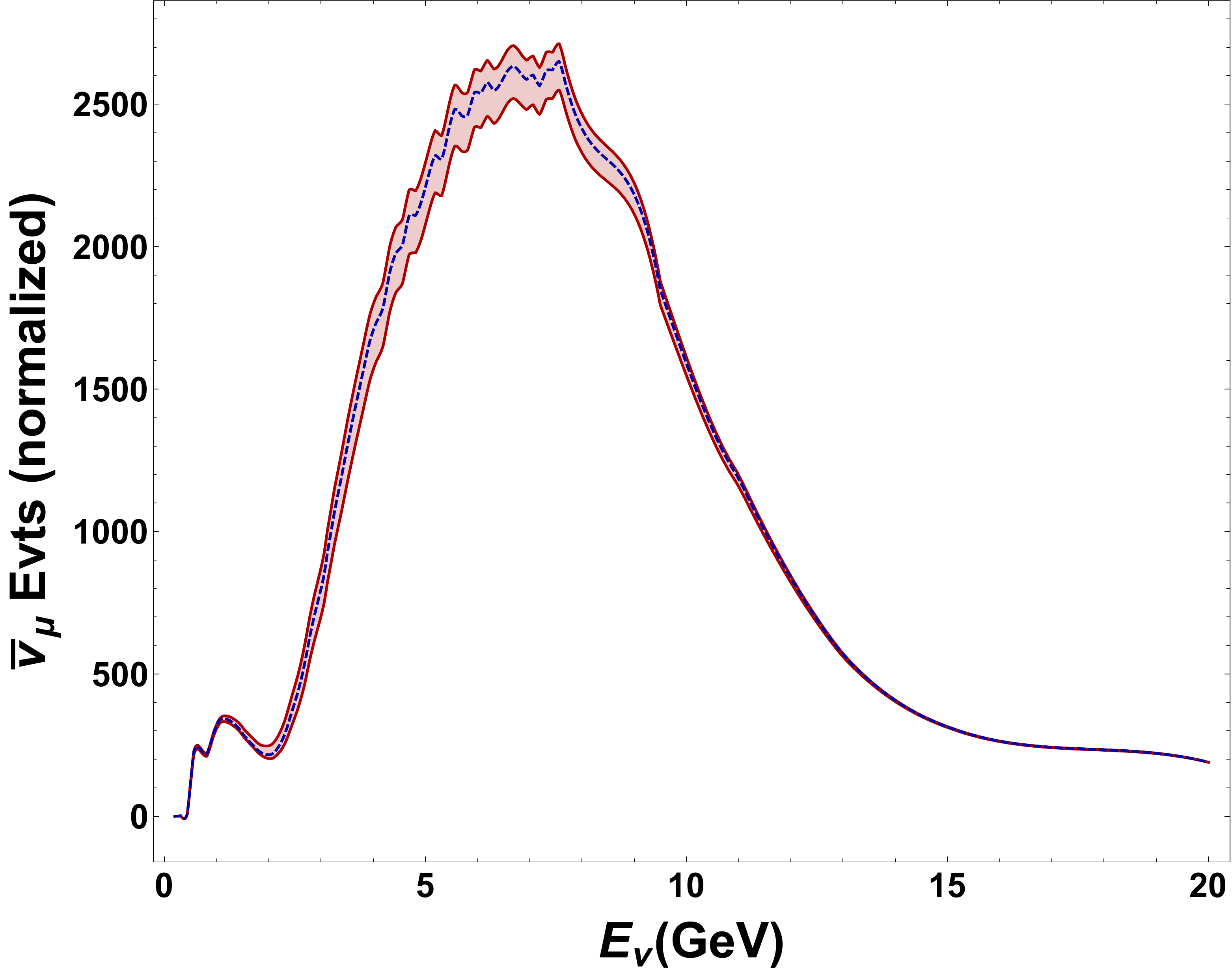}
    \includegraphics[width=5cm, height=5cm]{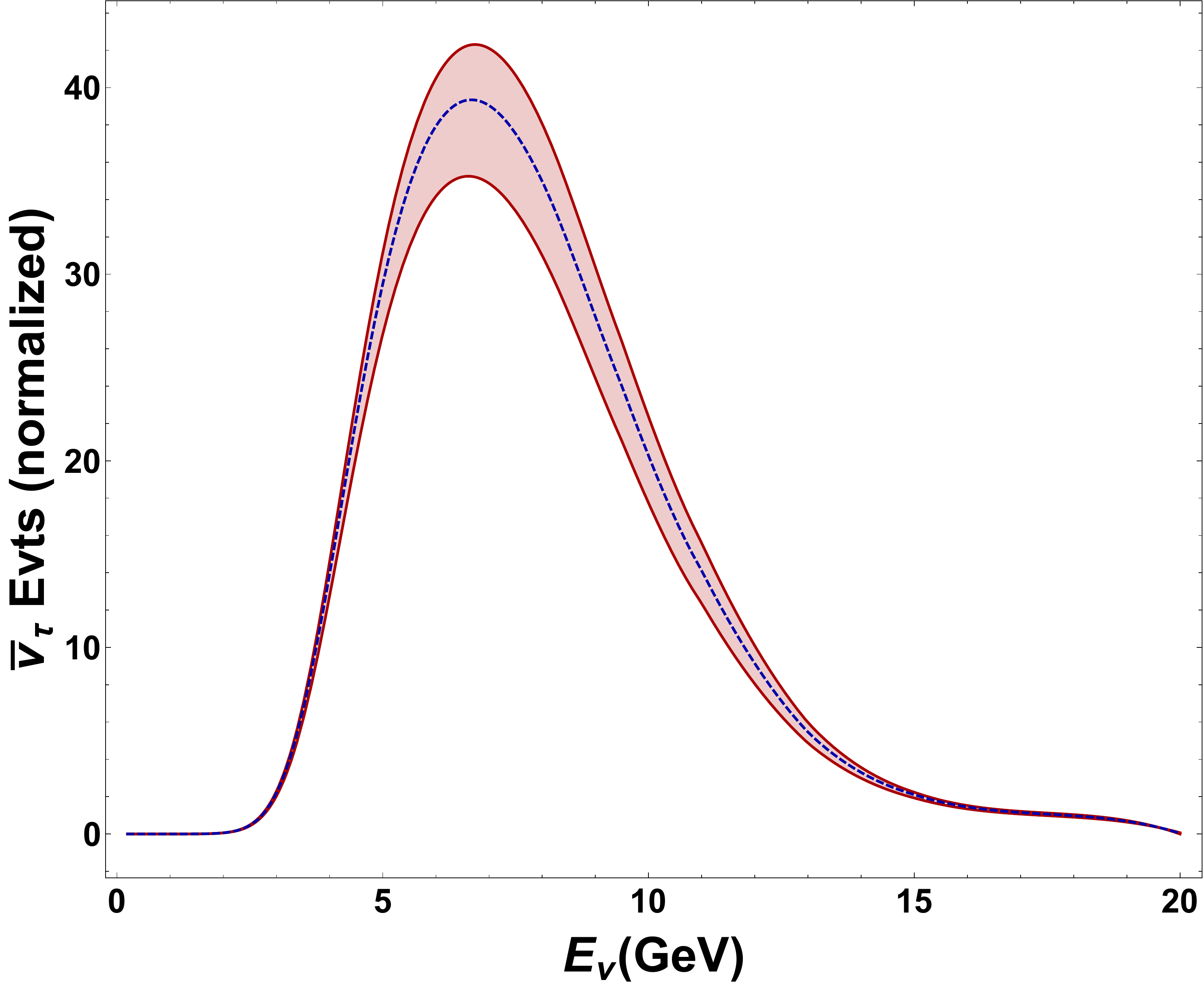}
    \caption{\it Same as Fig.\ref{NSIspectra} but  using the $\tau$-optimized flux.}
    \label{NSIspectra_opt}
\end{figure}

If we repeat the same study using the high energy $\tau$-optimized flux, (see Fig.\ref{NSIspectra_opt}\footnote{Wiggles near the disappearance spectra peaks are an artifact of extrapolating the smearing matrices from $\mathcal{O}(1)$ GeV to high energies.}), we get very similar features as before but for the $\nu_\tau$ events, which are obviously much larger. %Thus, the variation in the number of events due to NSI is more evident, making the $\nu_\tau$ appearance more useful in new physics studies than in the standard flux case {\bf a me sembra che la variazione percentuale al picco dovuto a nsi sia del 10\% con entrambi i flussi. Non mi aspetto che la nuova fisica sia piu' evidente con l'ottimizzato eccetto per il fatto che l'errore statistico e' minore}.

\begin{figure}
    \centering
    \includegraphics[width=5cm, height=5cm]{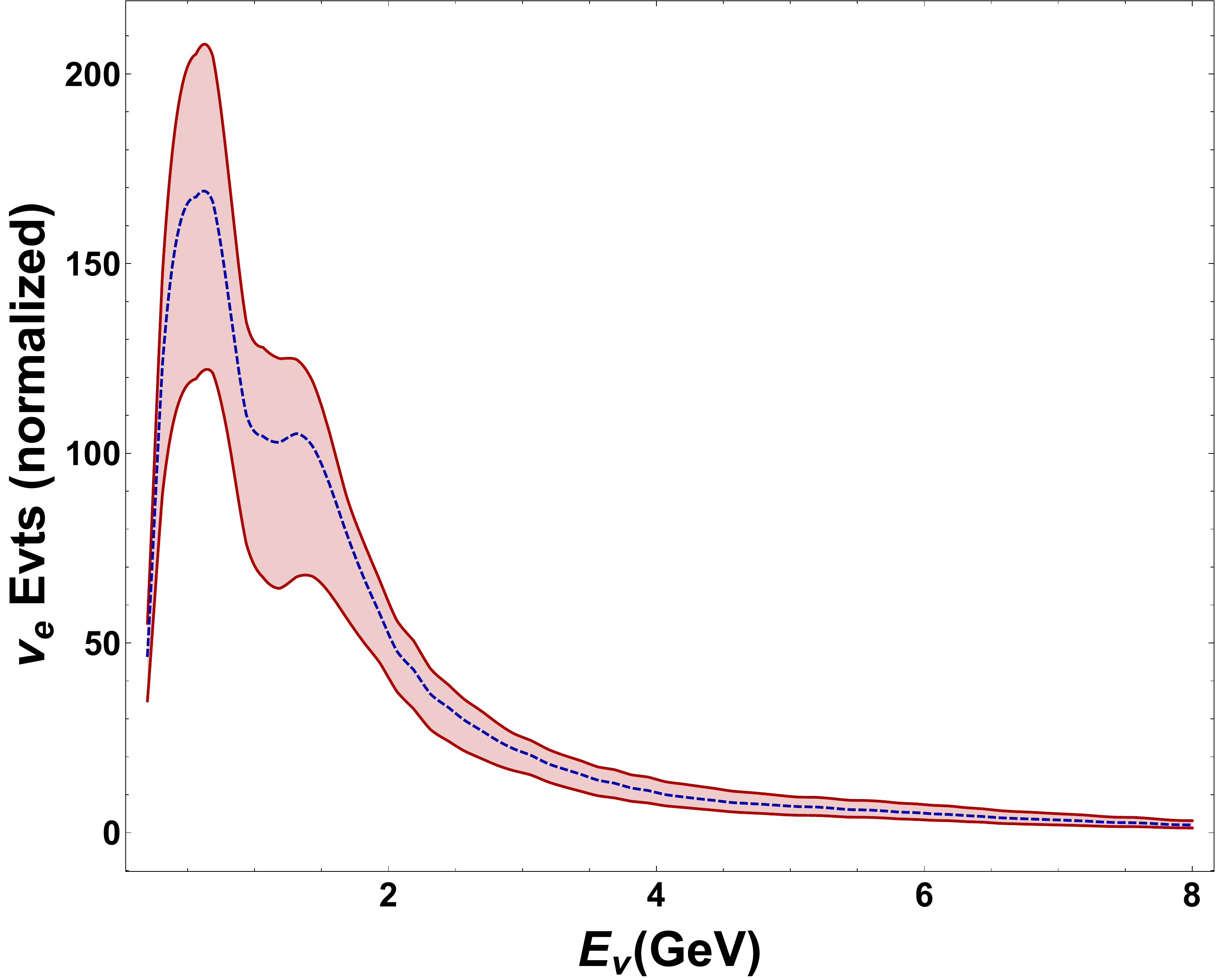}
    \includegraphics[width=5cm, height=5cm]{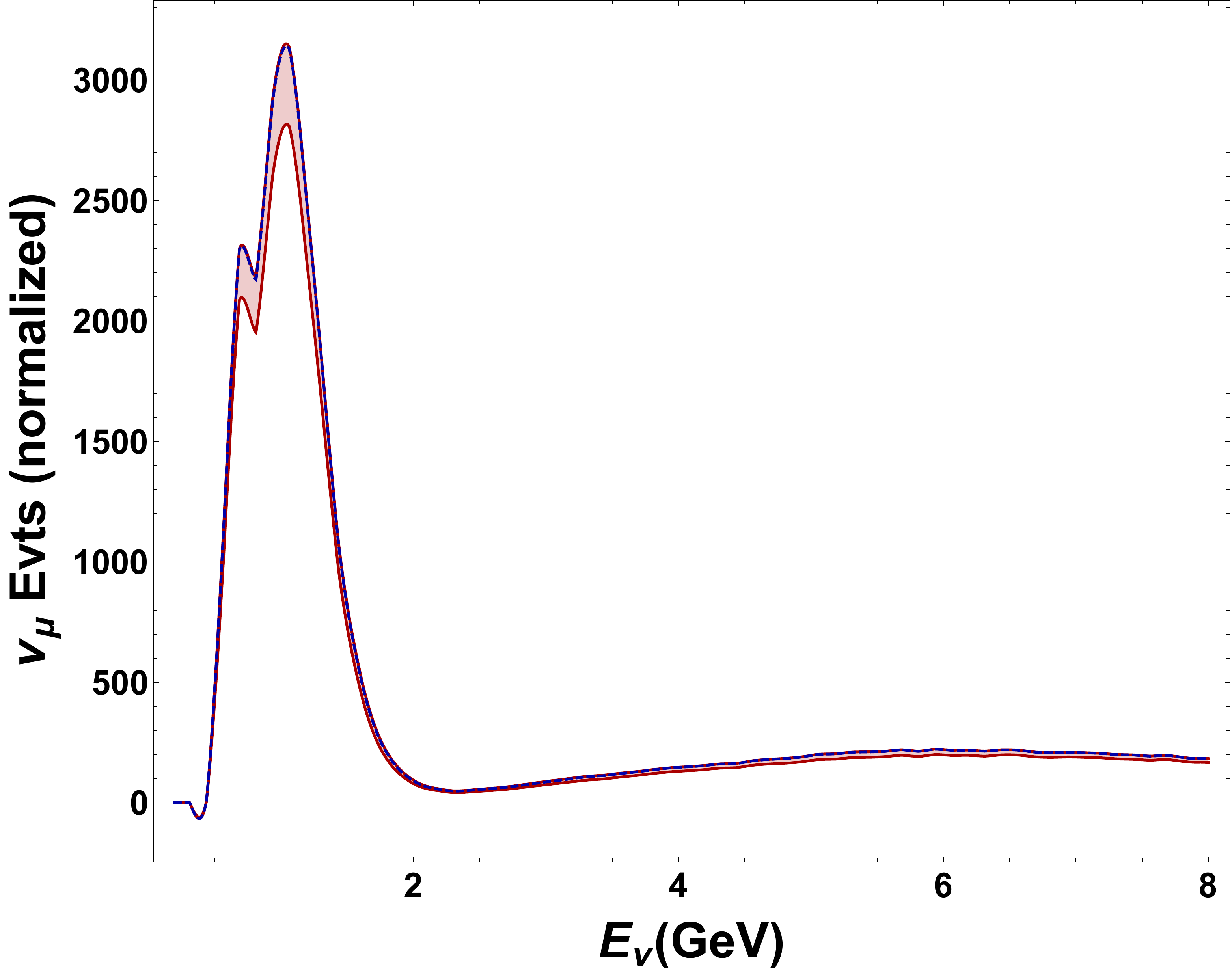}
    \includegraphics[width=5cm, height=5cm]{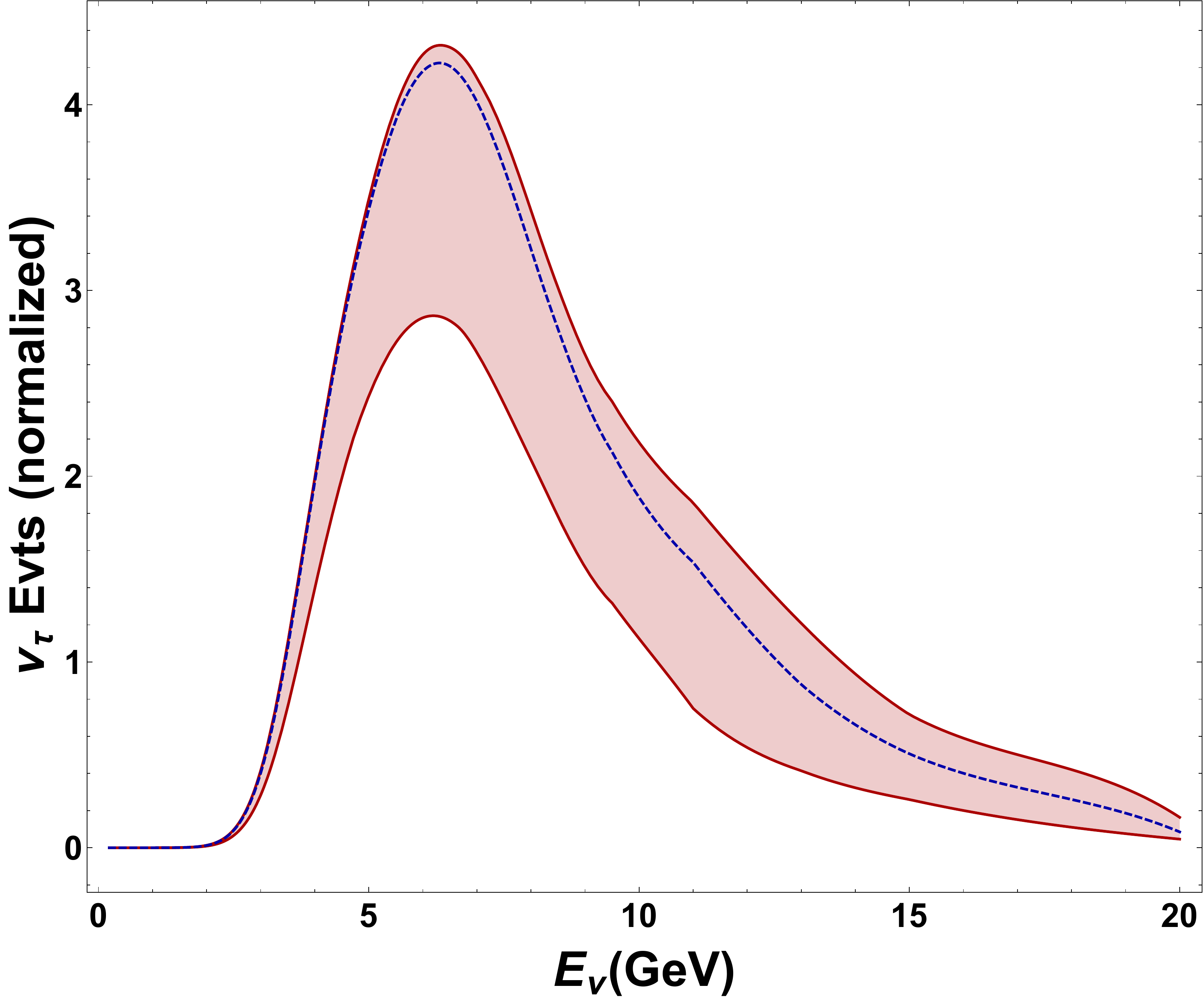}
    \includegraphics[width=5cm, height=5cm]{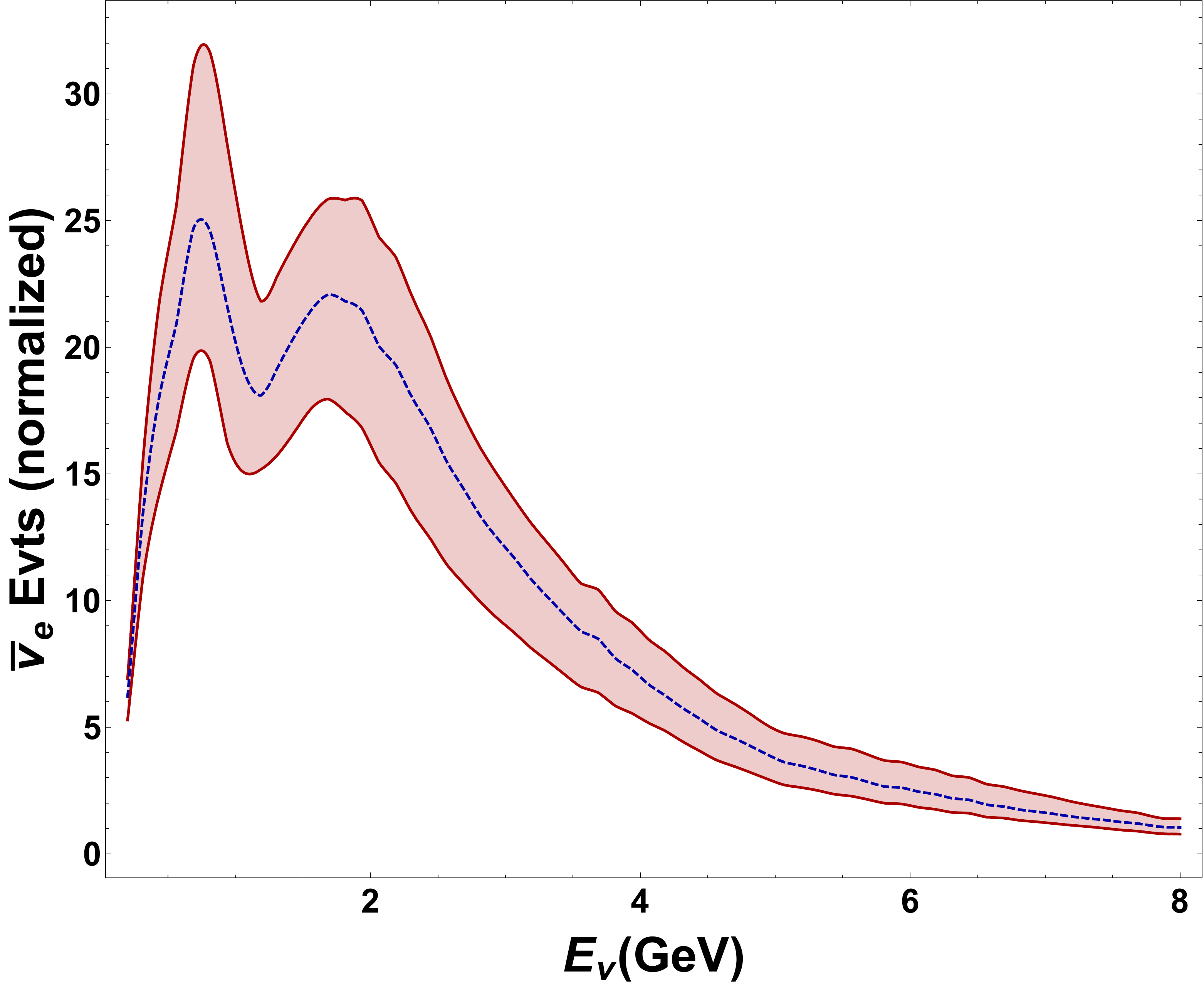}
    \includegraphics[width=5cm, height=5cm]{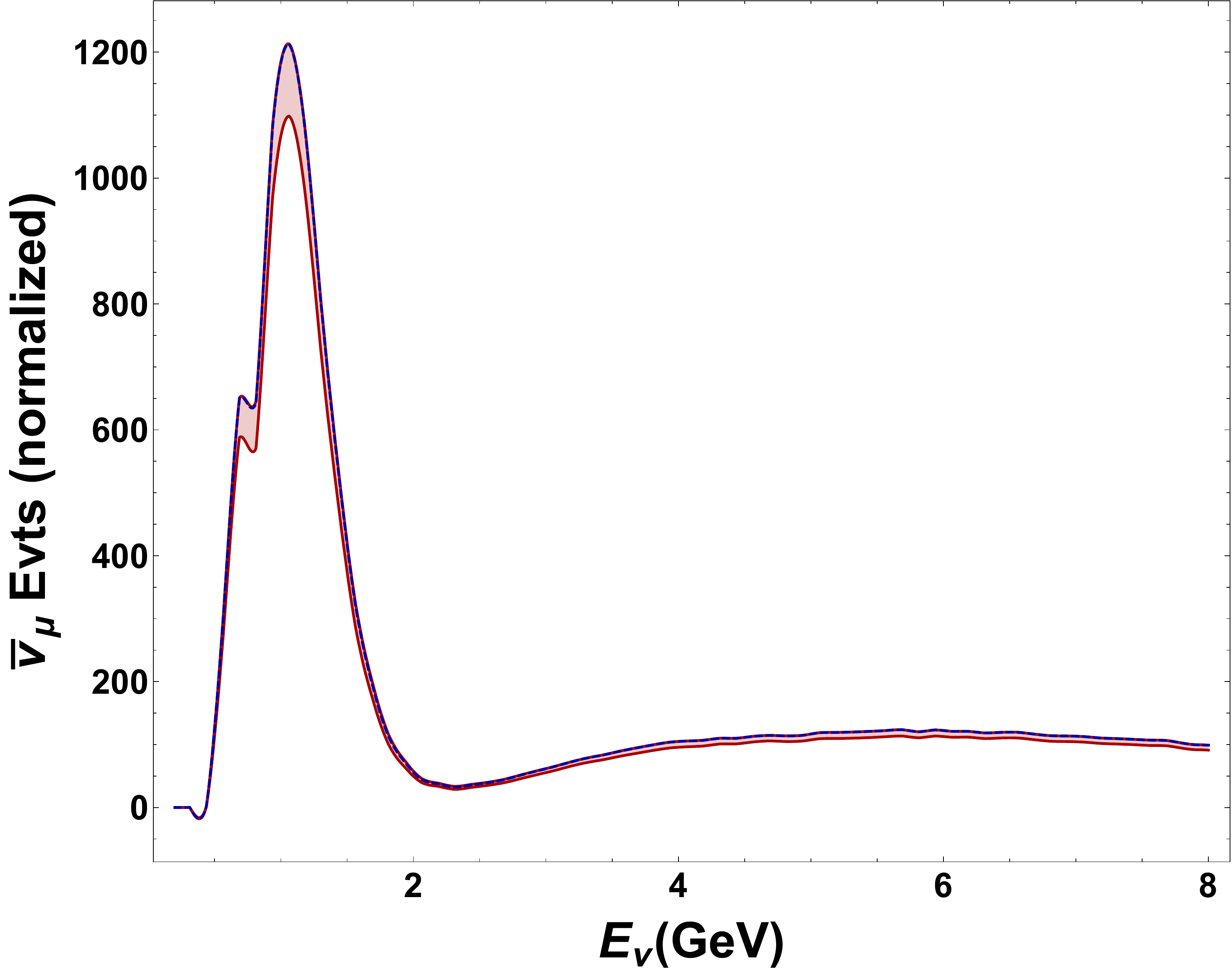}
    \includegraphics[width=5cm, height=5cm]{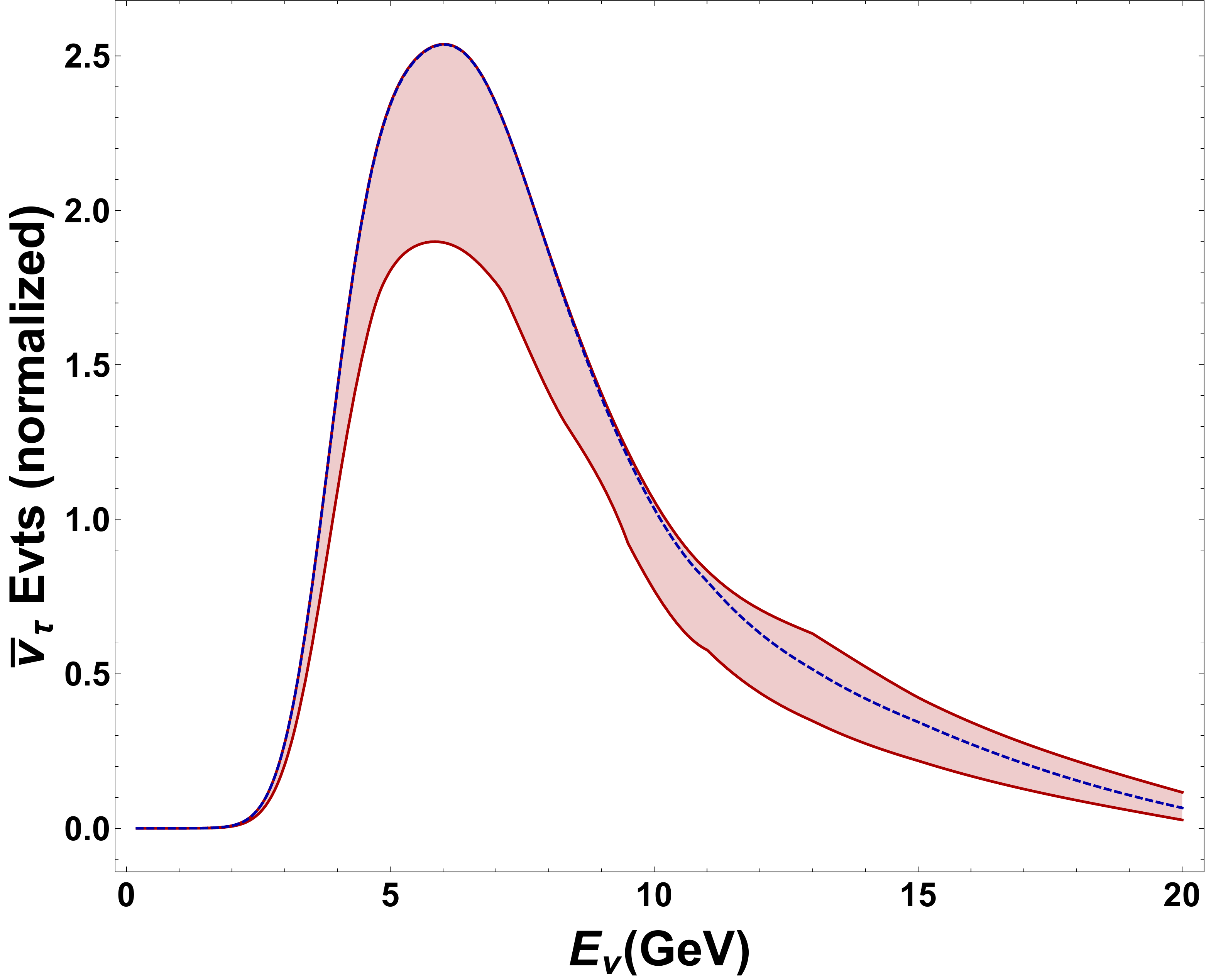}
    \caption{\it Same as Fig.\ref{NSIspectra} but  for the $3+1$ neutrino mass model.}
    \label{sterilespectra}
\end{figure}

\begin{figure}
    \centering
    \includegraphics[width=5cm, height=5cm]{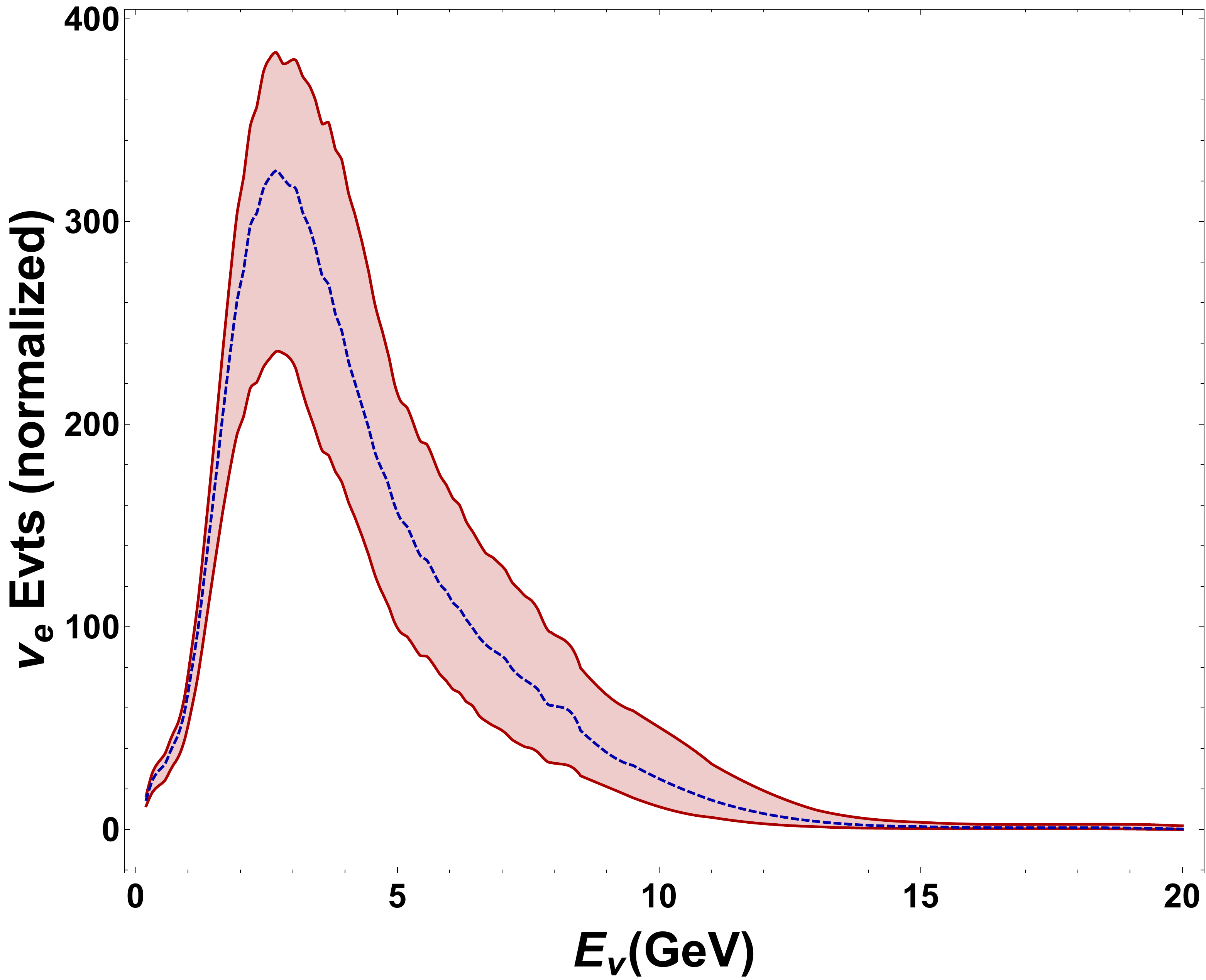}
    \includegraphics[width=5cm, height=5cm]{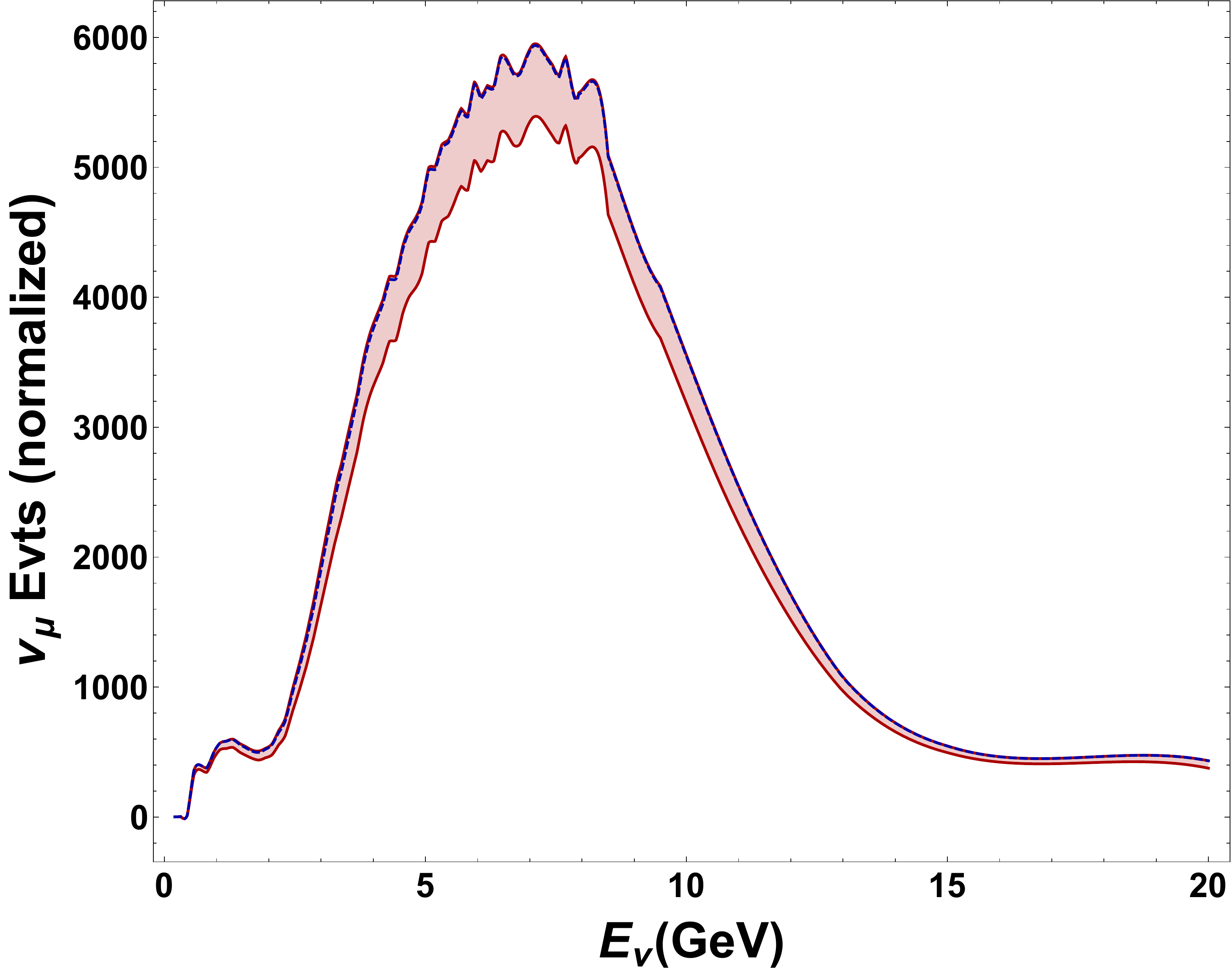}
    \includegraphics[width=5cm, height=5cm]{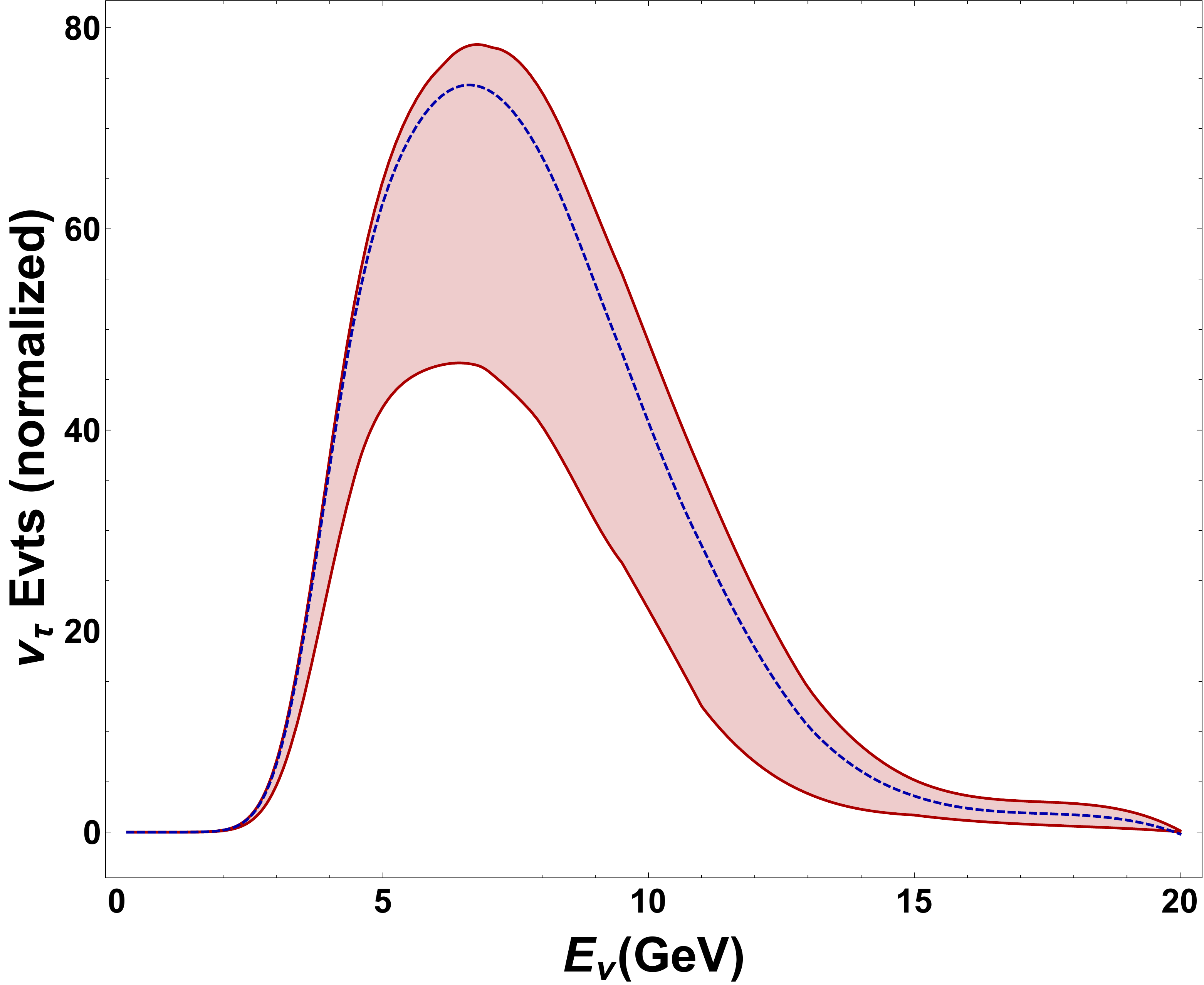}
    \includegraphics[width=5cm, height=5cm]{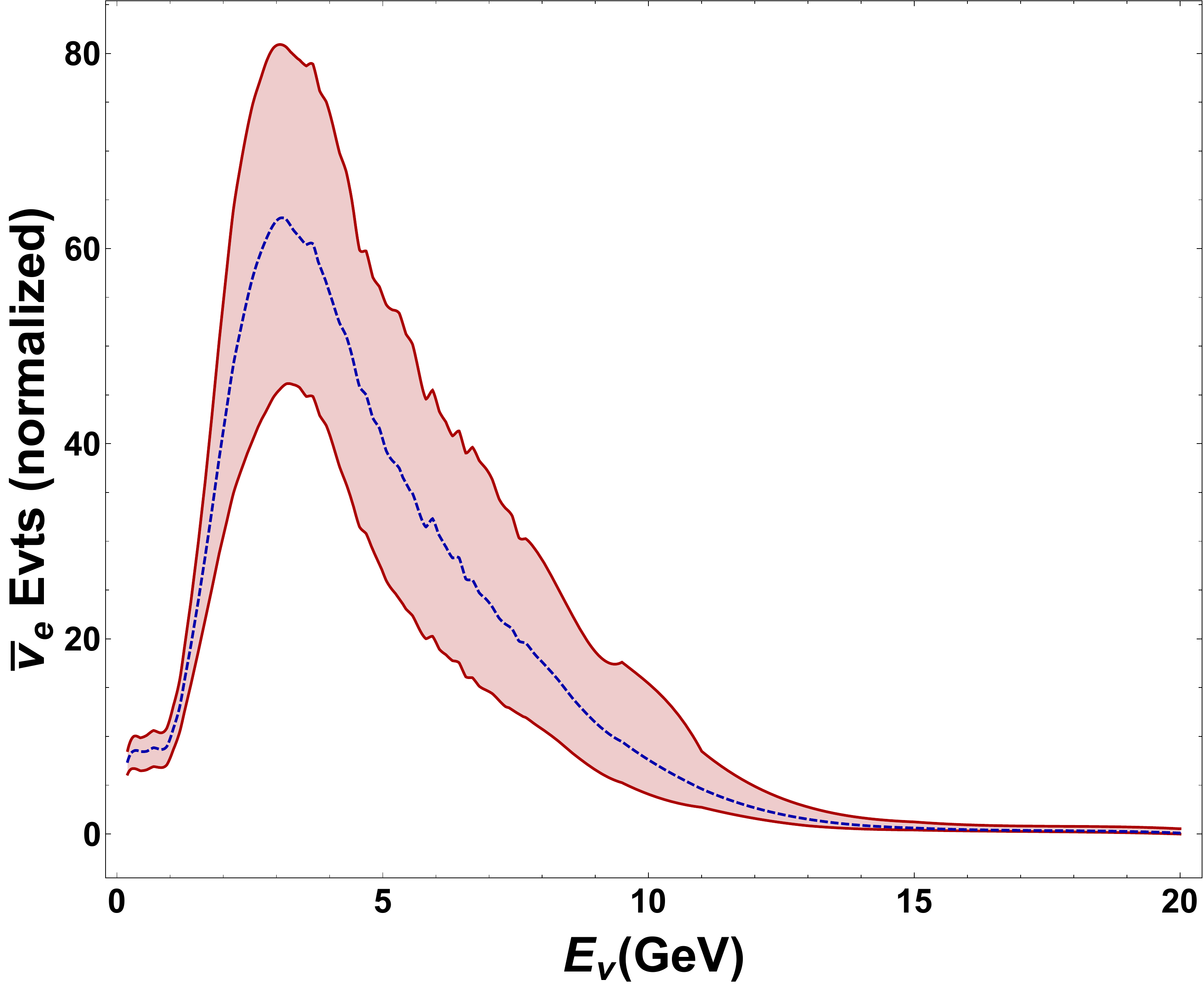}
    \includegraphics[width=5cm, height=5cm]{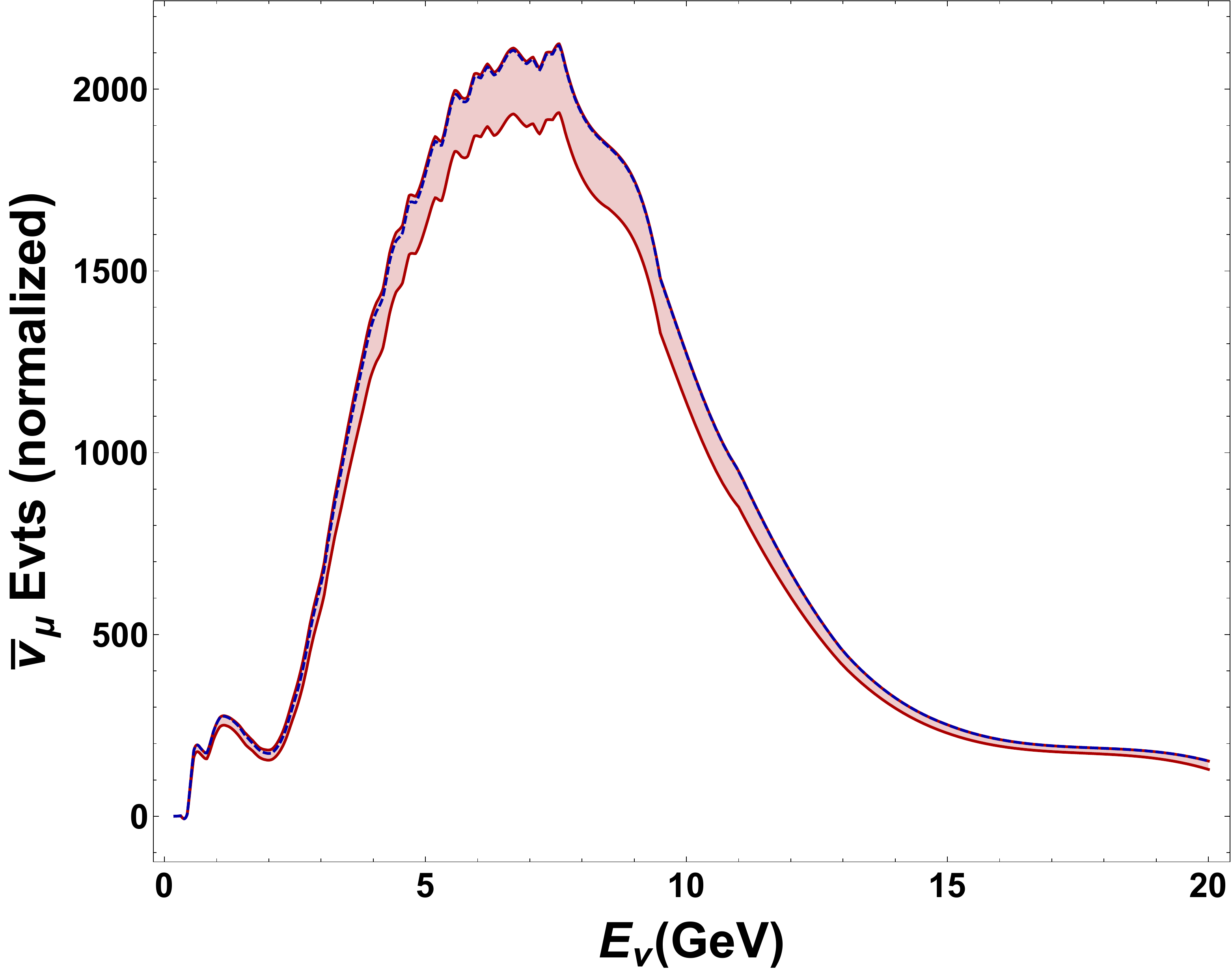}
    \includegraphics[width=5cm, height=5cm]{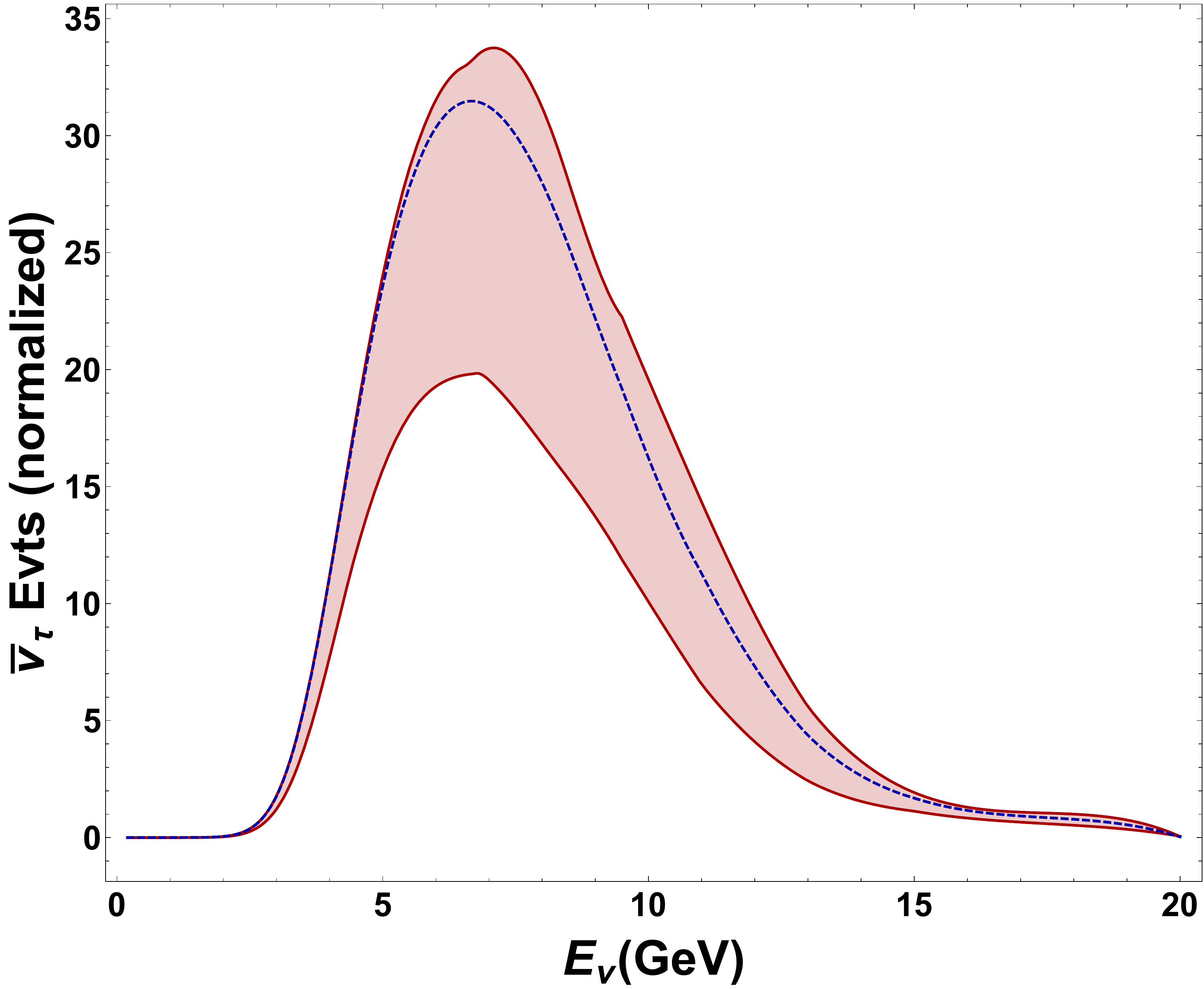}
    \caption{\it Same as Fig.\ref{NSIspectra_opt}  but  for the $3+1$ neutrino mass model.}
    \label{sterilespectra_opt}
\end{figure}

In the 3+1 framework, bounds on the parameters strongly depend on the adopted parametrization of the $4\times 4$ mixing matrix $U$. In addition, transition probabilities are affected by many degeneracies among standard and non-standard mixing angles that makes the extraction of the allowed/excluded ranges more complicated. However, in the parametrization of eq.(\ref{parameterization}), if we allow the standard parameters to vary only in their allowed ranges (Tab.\ref{SMparam}) and we fix the new mass splitting to be $\Delta m^2_{41} \sim$ 1 eV$^2$, different studies \cite{Boser:2019rta,Dentler:2018sju} suggest that $\theta_{14}$ and $\theta_{24}$ can be taken in the range $[0-10]^\circ$ while $\theta_{34}$ in the range $[0-30]^\circ$. The variations of the neutrino spectra for the 3+1 model are shown in Figs.\ref{sterilespectra} and \ref{sterilespectra_opt}.
As for the NSI case,  the largest deviation from the SM results  are found in the $\nu_e$ spectra, since, as showed in eqs.(\ref{3+1asy}), the 3+1 corrections to the standard asymmetry are at the first order of our expansion.

As for the other flavors, the $\nu_\mu$ spectra are not really affected by NP, while the deviation of the $\nu_\tau$ spectra from the SM predictions can be mostly ascribed to $\theta_{34}$ and its relatively large allowed range. We also see that the contributions to NSI is mailny negative, as it should be because 
 the corrections to the probabilities driven by $\theta_{34}$ are negative indeed. \\
Computing the same spectra using the $\tau$-optimized flux (Fig.\ref{sterilespectra_opt}), the changes in the $\nu_\tau$ appearance channel, as in the NSI case, are amplified.

\section{Numerical evaluation of asymmetries}
\label{num_asy}

\begin{figure}
    \centering
    \includegraphics[width=7cm,height=7cm]{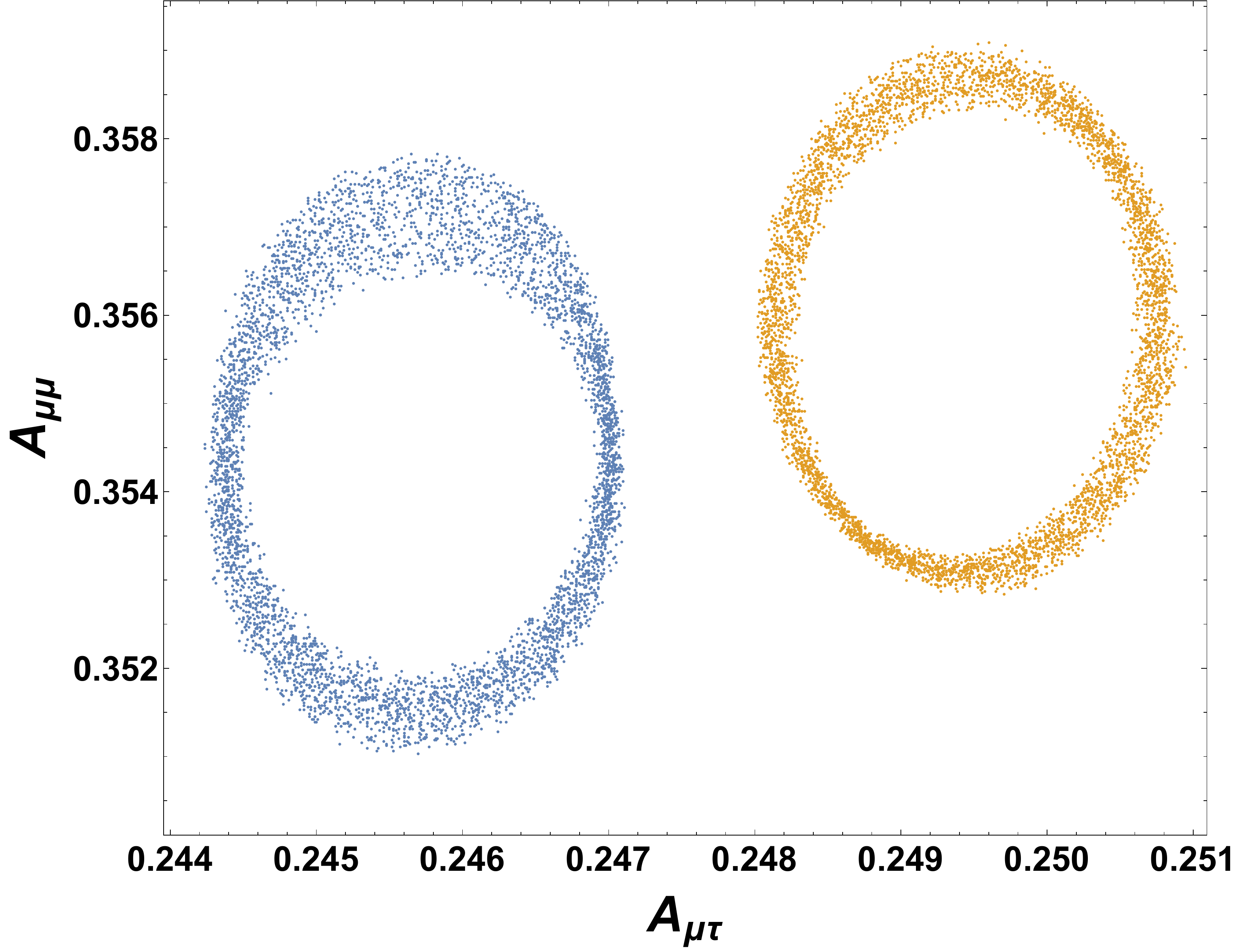}
    \includegraphics[width=7cm,height=7cm]{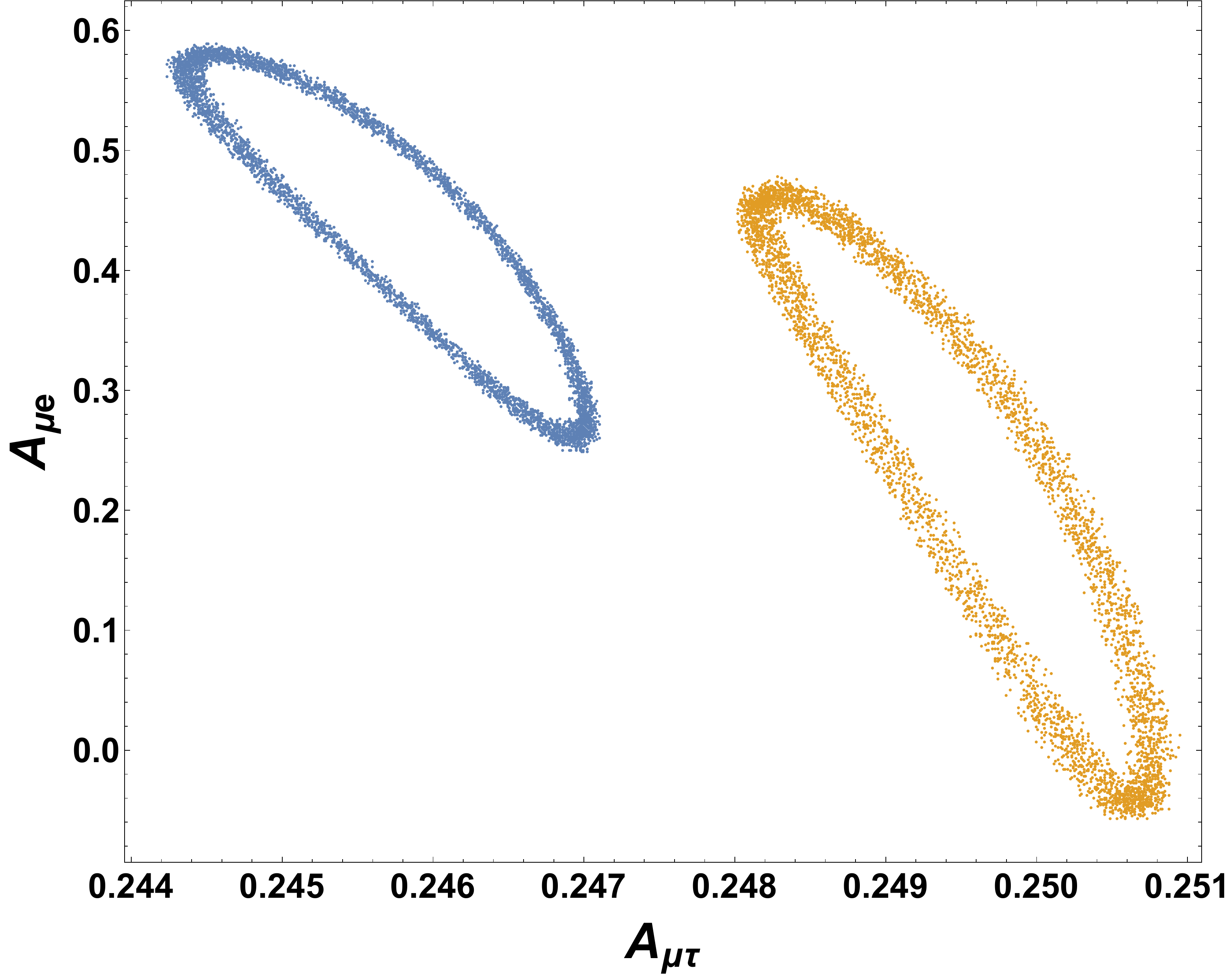}
    \caption{\it Numerical evaluation of the SM asymmetries of eq.(\ref{eq:asirate}) at DUNE, with standard flux. The SM parameters have been allowed to vary in their 1$\sigma$ range, while all possible values for the CP phase have been taken into account. The blue dots represent asymmetries in the normal hierarchy hypothesis, while the orange ones represent asymmetries in the inverted hierarchy hypothesis. }
    \label{asystd}
\end{figure}

\begin{figure}
    \centering
    \includegraphics[width=7cm,height=7cm]{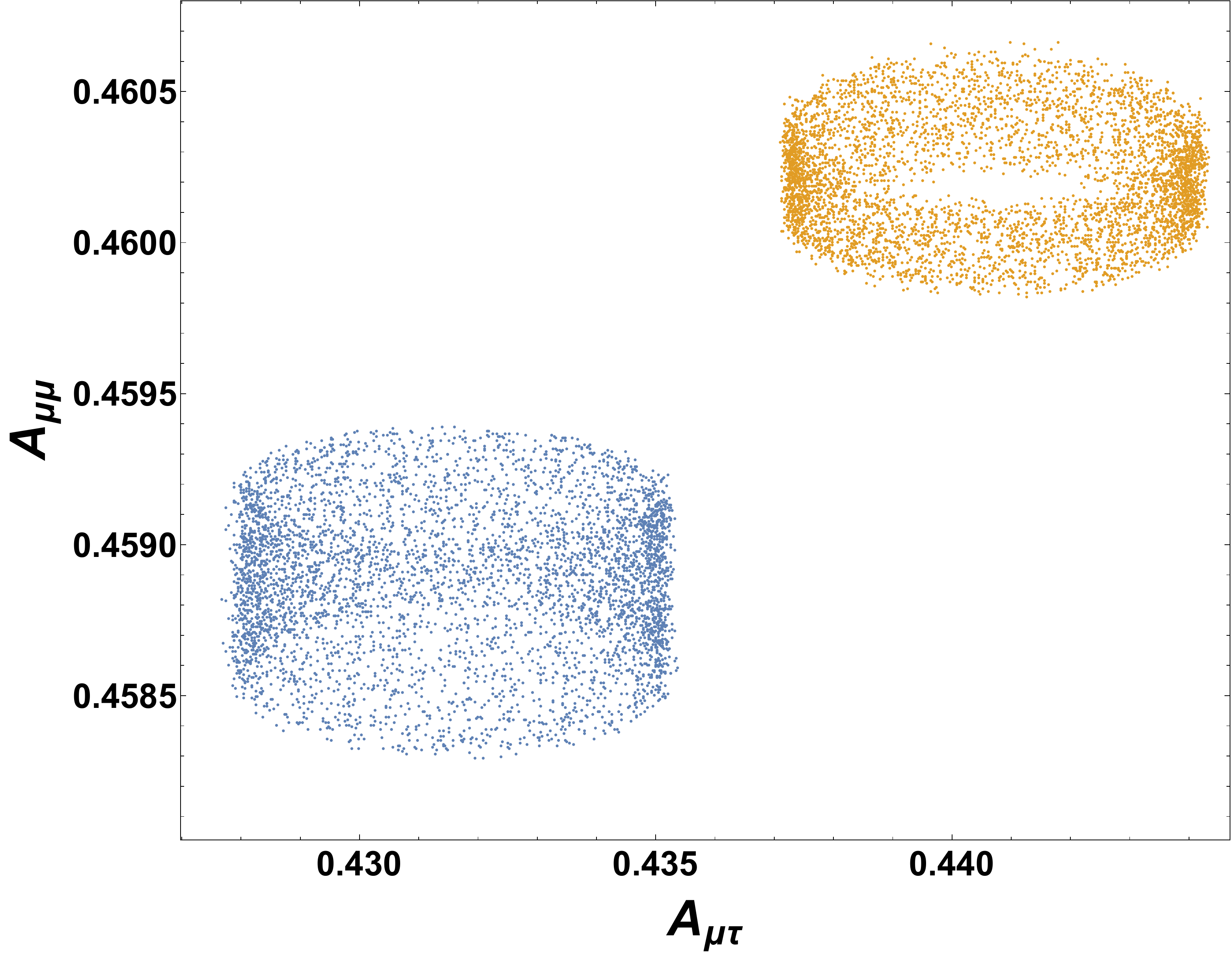}
    \includegraphics[width=7cm,height=7cm]{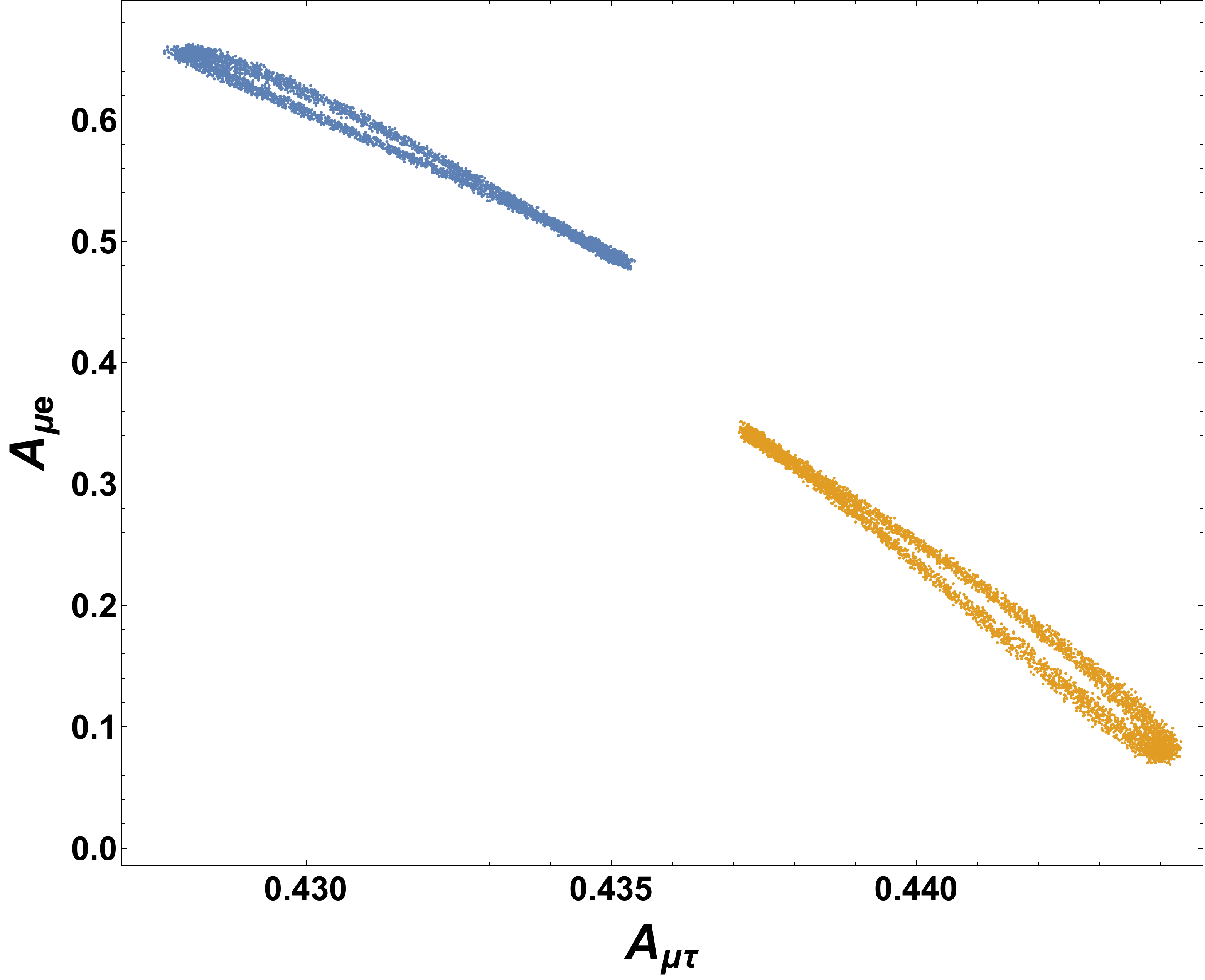}
    \caption{\it Same as Fig.\ref{asystd} but with the optimized flux. Notice the different vertical scales on the left and right panels. }
    \label{asystd_opt}
\end{figure}

The relevant question now is related to the experimental capability to  measure the asymmetries we are considering: in fact, if the CP violating quantities will not be  measured  with a sufficient precision, then we cannot distinguish the deviation from the SM results due to NP. Instead of considering the asymmetries at the probability level, we deal with the experimentally  relevant  integrated  asymmetries built from the number of expected events $N_\beta$ and $\bar N_\beta$:
\begin{eqnarray}
\label{eq:asirate}
A_{\alpha \beta} &=&\frac{N_\beta-\bar N_\beta}{N_\beta+\bar N_\beta}\,,
\end{eqnarray}
where the event rates for the $\nu_\alpha \to \nu_\beta$ and the CP conjugate $\bar\nu_\alpha \to \bar\nu_\beta$ transitions are computed from:
\begin{eqnarray}
\label{eq:asirate2}
N_\beta &=&\int_{E_\nu} dE_\nu \,P_{\alpha \beta}(E_\nu)\,\sigma_\beta(E_\nu)\,\frac{d\phi_\alpha}{dE_\nu}(E_\nu) \,\varepsilon_\beta(E_\nu) \\
\bar N_\beta &=&\int_{E_\nu} dE_\nu \,P_{\bar\alpha \bar\beta}(E_\nu)\,\sigma_{\bar\beta}(E_\nu)\,\frac{d\phi_{\bar\alpha}}{dE_\nu}(E_\nu) \,\varepsilon_\beta(E_\nu)\,,
\end{eqnarray} 
in which $\sigma_{\beta (\bar \beta)}$ is the cross section for producing the lepton $\beta(\bar \beta)$,  
$\varepsilon_{\beta(\bar \beta)}$ the detector efficiency to reveal that lepton and  $\phi_{\alpha(\bar \alpha)}$ the initial neutrino flux at the source.
Since in the SM the only dependence on the CP phase is carried on by $\delta$, the correlations between the pair of asymmetries, for instance $(A_{\mu\tau},A_{\mu e})$ and $(A_{\mu\tau},A_{\mu \mu})$, is maximal and a close curve appears in the related physical planes. If, in addition, we also take into account the experimental errors on angles and mass differences, the curves are scattered as reported in Fig.\ref{asystd}, for the DUNE standard flux and in Fig.\ref{asystd_opt} for the optimized flux. The blue dots are obtained using parameters in the normal hierarchy, while the orange ones are obtained using the inverse hierarchy hypothesis.

The first striking features of the integrated asymmetries is related to the fact that their sign is always positive; in fact, being integrated quantities, they are influences  not only by the relative differences among $\nu$ and $\bar \nu$ probabilities, but also by the  differences among $\nu$ and $\bar \nu$ fluxes and cross sections. As we can observe in Fig.\ref{NSIspectra}, the SM spectra (blue lines) of $\bar\nu$  is always lower, and this helps in justifying the observed signs. 
The other important observation is that, as discussed above, $A_{\mu e}$ is the asymmetry that changes the most with a change of the CP phase. On the other hand, the other two asymmetries $A_{\mu \tau}$ and  $A_{\mu \mu}$ change at a much slower rate. 

%Moreover, the small uncertainties on the measured standard parameters, let the variations of the obtained values very small. 
Eventually, it is worth mentioning that, for each pair of asymmetries, the closed curves corresponding to NH and IH never overlap. This means that, at least in principle, one could be able to solve the neutrino hierarchy problem simply looking at the CP asymmetries. However, in DUNE as well as in other future experiments, the foreseen experimental errors on such asymmetries will probably be too large to allow for such a discrimination, as we will discuss later on in the manuscript.

\subsection{Numerical evaluation of the asymmetries in presence of NSI}

Now we are ready to apply our strategy to check whether other sources of CP violation carried on by NP can be sufficiently distinguished from the SM phase. In order to do that, we first need to evaluate the experimental errors on the SM asymmetries and then recompute them as predicted by the NSI and the $3+1$ sterile models.
From Fig.\ref{asystd} we see that the uncertainties on the standard angles and mass splittings are not playing an important role. 
%Thus, the main source of errors on the integrated asymmetries $\delta A_{\alpha\beta}$ are related to the statistical and the systematic errors. %(we will assume for the rest of the paper that the backgrounds are known and can be subtracted to the total number of events). 
A simple but accurate estimate from error propagation gives:
\begin{eqnarray}
 (\delta A_{\alpha\beta})^2=\frac{4\bar N_\beta^2 (\delta N_\beta)^2+4N_\beta^2 (\delta \bar N_\beta)^2}{(N_\beta+\bar N_\beta)^4}\,,\label{asyerror}
\end{eqnarray}
where $\delta N$ is the uncertainty related to the number of expected events which receives contributions from the systematic error (normalization errors cited in Sect.\ref{DUNE}) and the statistical error. For the $\nu_\mu$ disappearance channel, the first source of uncertainty is always dominating, since the number of events is very large and the statistical error is reduced. On the other hand, in the other two channels both terms are important. In particular, in the $\nu_\tau$ appearance, systematic errors are quite large (due to the poorly known cross section and to the systematics related to the complicated event reconstruction) and the number of events is small. Thus we expect  $\delta A_{\mu\tau}$ to be particularly large. 

\begin{figure}
    \centering
    \includegraphics[width=7cm,height=7cm]{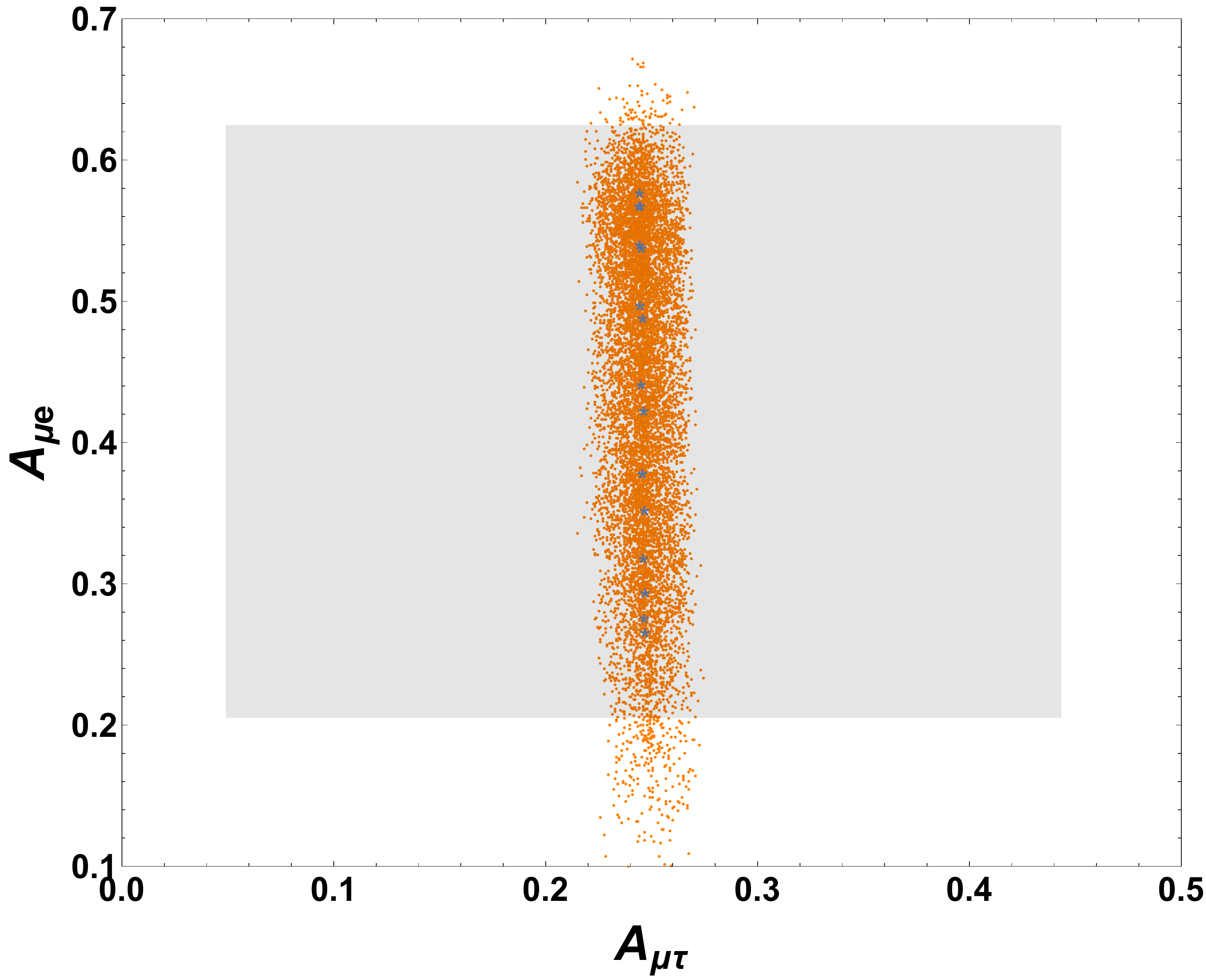}
    \includegraphics[width=7cm,height=7cm]{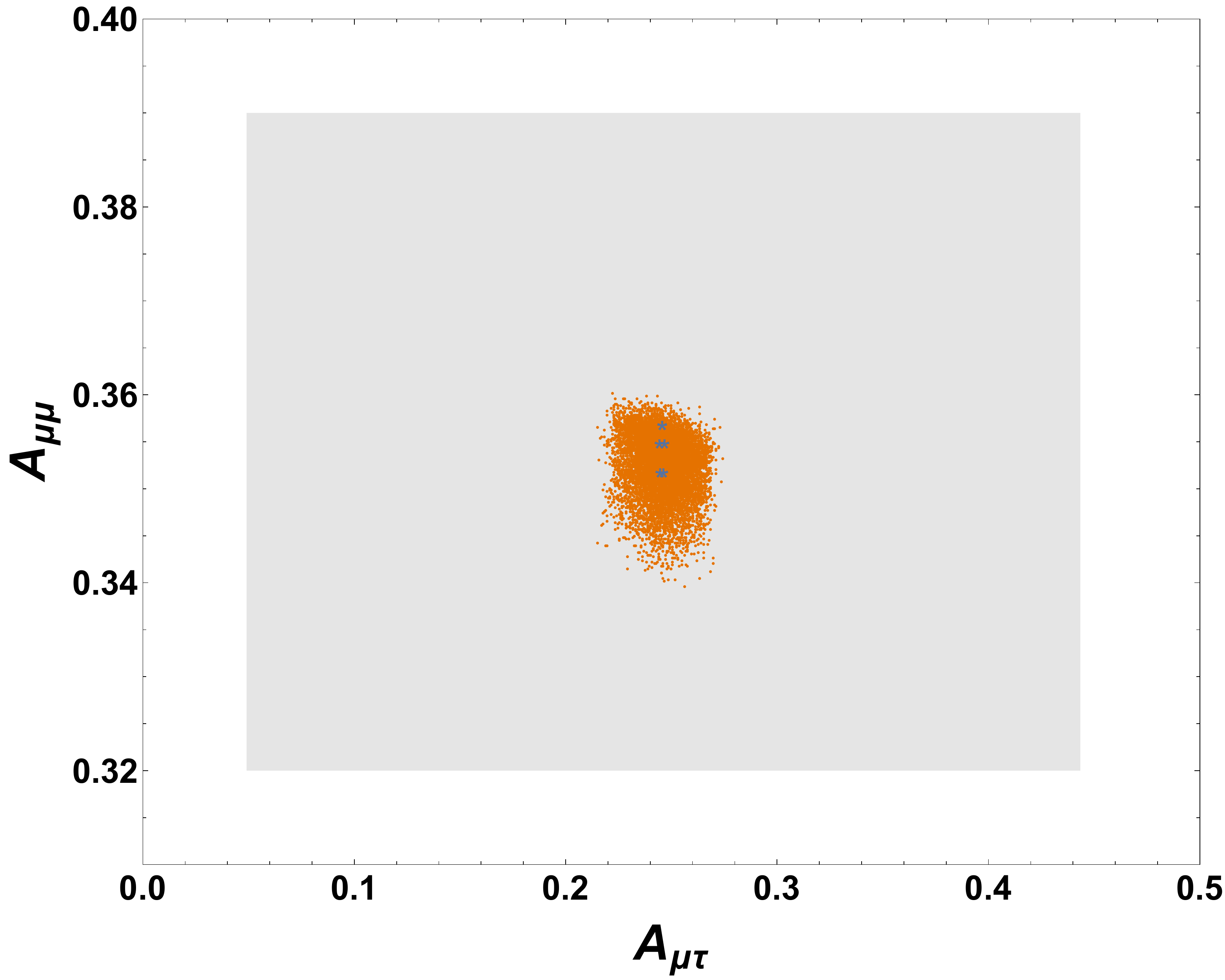}
    \caption{\it Integrated asymmetries in the ($A_{\mu\tau}, A_{\mu e}$) (left plot) and ($A_{\mu\tau}, A_{\mu \mu}$) planes (right plot). Blue stars represent the asymmetries in the SM case while the orange dots are the values obtained in presence of NSI. The grey rectangle shows the 1$\sigma$ error range on the standard asymmetries. For sake of simplicity, we do not report here the error bars on the orange dots. Standard neutrino flux has been employed to compute the number of events.}
    \label{NSIscatter}
\end{figure}

In Fig.\ref{NSIscatter} we show the values of the asymmetries where the effects of NSI are taken into account, computed by using the standard neutrino flux. The blue stars represent the asymmetries in the standard case  (fixing all the standard parameters to their best fits\footnote{We present here only results in Normal Hierarchy, for the Inverted Hierarchy case the conclusions are very similar.} but varying the values of $\delta$), while the orange dots are the results obtained in presence of NSI, computed from the number of events corresponding to random flat extraction of the couplings in the ranges shown in Tab.\ref{tabNSI}. The sides of the grey rectangles represent the maximum 1$\sigma$ error bars on the standard asymmetries at different chosen values of $\delta$ as computed from eq.(\ref{asyerror}). For the sake of illustration, we do not show here the error bars associated to the NSI points because the number of events is not much different from the standard case, thus the error bars in the NSI framework are of the same order of magnitude as the displayed ones. \\
It is clear that $A_{\mu\tau}$ is very sensitive to New Physics. Indeed, the SM asymmetry has almost a fixed value $A_{\mu\tau}\sim 0.245$, as showed in Fig.\ref{asystd}, while the NSI contributions can turn $A_{\mu\tau}$ into the range [0.21,0.27]. However, the error bars are much larger than the produced variation, making this asymmetry at the DUNE conditions not useful for discerning new CP phases. Even though the $A_{\mu e}$ asymmetry  gets very different values in the standard case (in the range [0.28,0.55]),  the inclusion of the NSI is able to even extend the foreseen asymmetry beyond such a range, enough to reach values outside the error bars of the standard asymmetries. The problem in this case is that, as discussed before, we should also take into account the error bars on the orange dots so that, when we include them, also $A_{\mu e}$ cannot give hints of NP at DUNE. 
Finally, for the $A_{\mu\mu}$ the very same analysis done for $A_{\mu\tau}$ applies. 

\begin{figure}
    \centering
    \includegraphics[width=7cm,height=7cm]{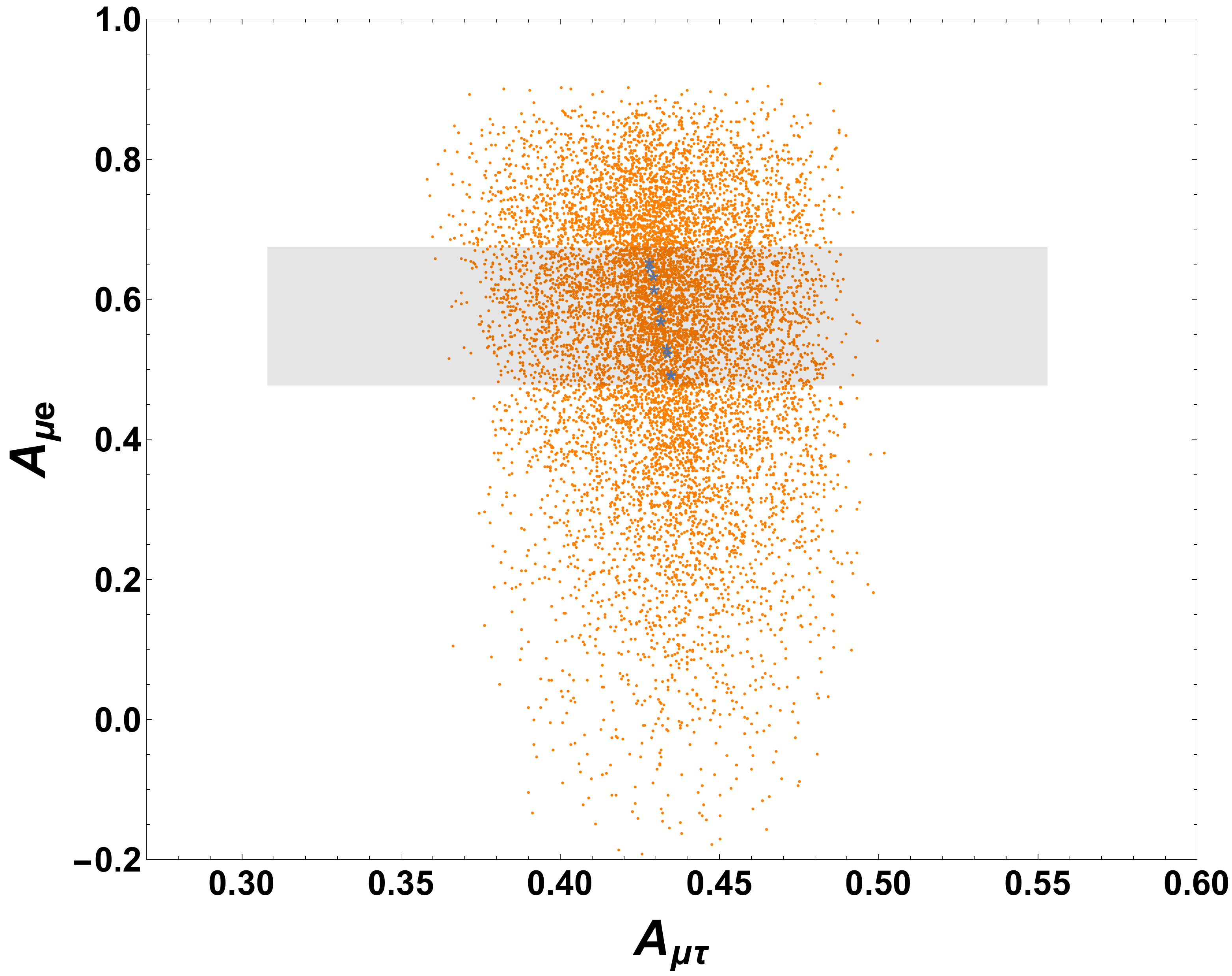}
    \includegraphics[width=7cm,height=7cm]{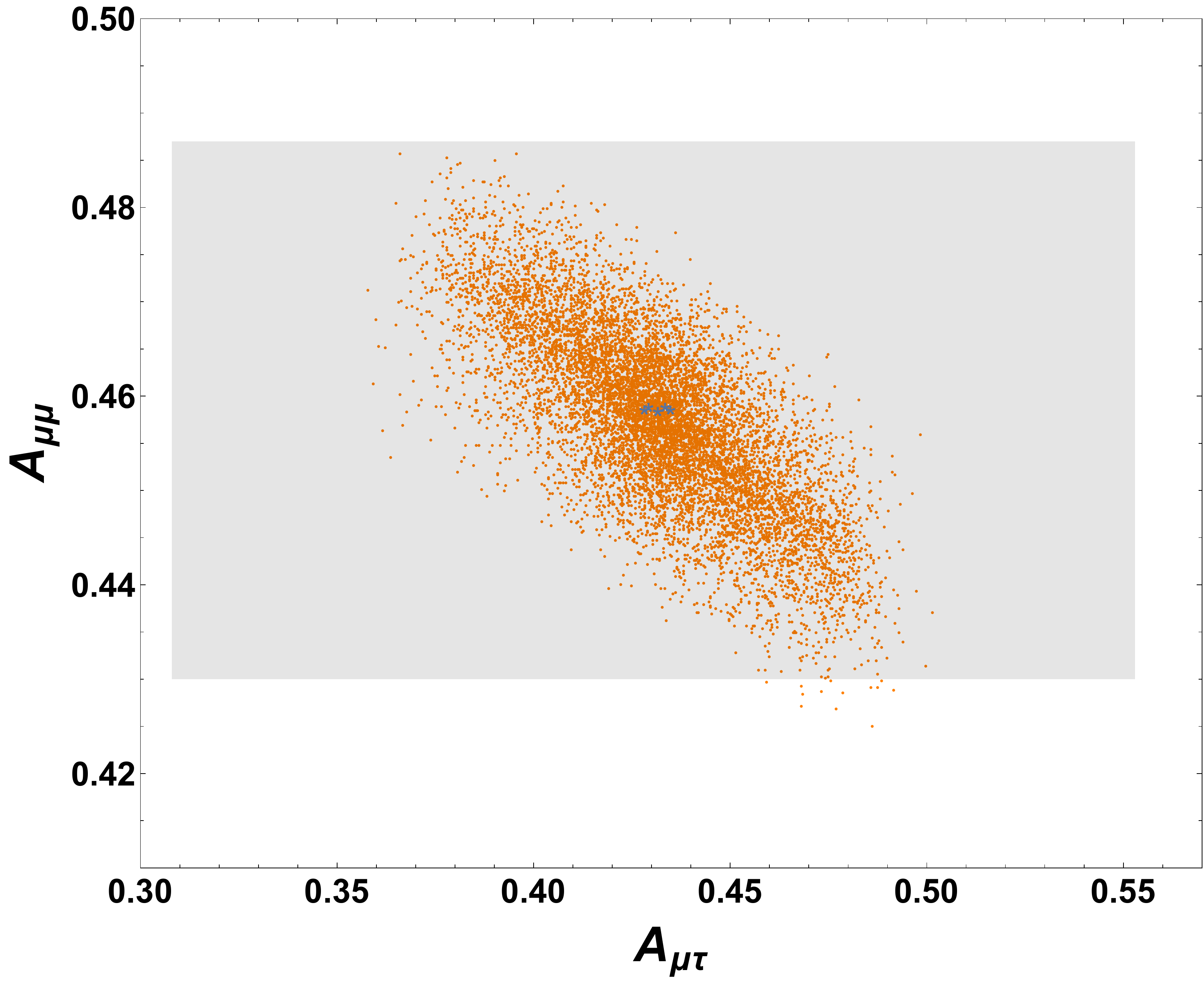}
    \caption{\it Same as Fig.\ref{NSIscatter} but using the optimized flux.}
    \label{NSIscatteropt}
\end{figure}

With a higher energy flux, the results partially differ from what illustrated above (see Fig.\ref{NSIscatteropt}). Even though the larger number of events reduces the error bars, 
%In this case, the broader neutrino flux increases the new physics effects, which can cause a drastic change in the number observed asymmetries.
the $A_{\mu\tau}$ and $A_{\mu\mu}$ with NSI do not change enough in such a way to be clearly distinguished at an acceptable confidence level from the SM case. On the other hand, $A_{\mu e}$ can assume values very different from the SM ones, in particular, a sets of NSI parameters can push it toward negative values. 
%Such small values for this parameter could be a significant hint of new physics. 
Indeed, as it is clear from eqs.(\ref{mue_1}-\ref{mue_3_last}),  NSI corrections to the asymmetries can be comparable to the SM case when  $\varepsilon_{e\mu}$ and $\varepsilon_{e\tau}$ are of $\mathcal{O}(0.1)$. With higher energy fluxes, the appearance transition probabilities are mainly evaluated off peak, making the cosine of $\Delta_{31}$ in eqs.(\ref{mue_2},\ref{mue_3_last}) no longer negligible. Thus NSI corrections become more and more important, causing an opposite sign of the asymmetry with respect to the SM case when $\cos(\delta-\delta_{e\mu,\tau})$ terms become negative.

\subsection{Numerical evaluation of the asymmetries in the $3+1$ sterile neutrino model}

\begin{figure}
    \centering
    \includegraphics[width=7cm,height=7cm]{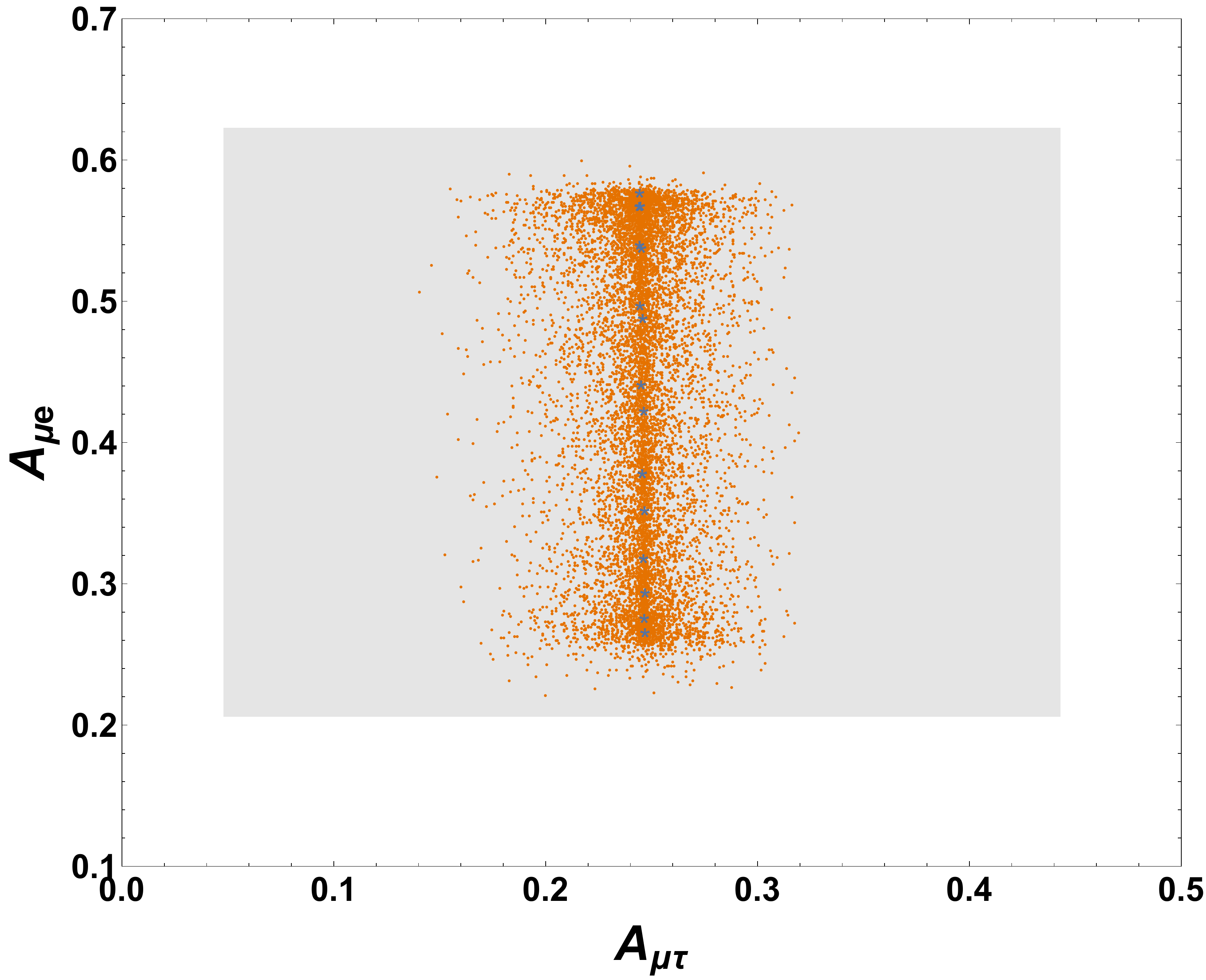}
    \includegraphics[width=7cm,height=7cm]{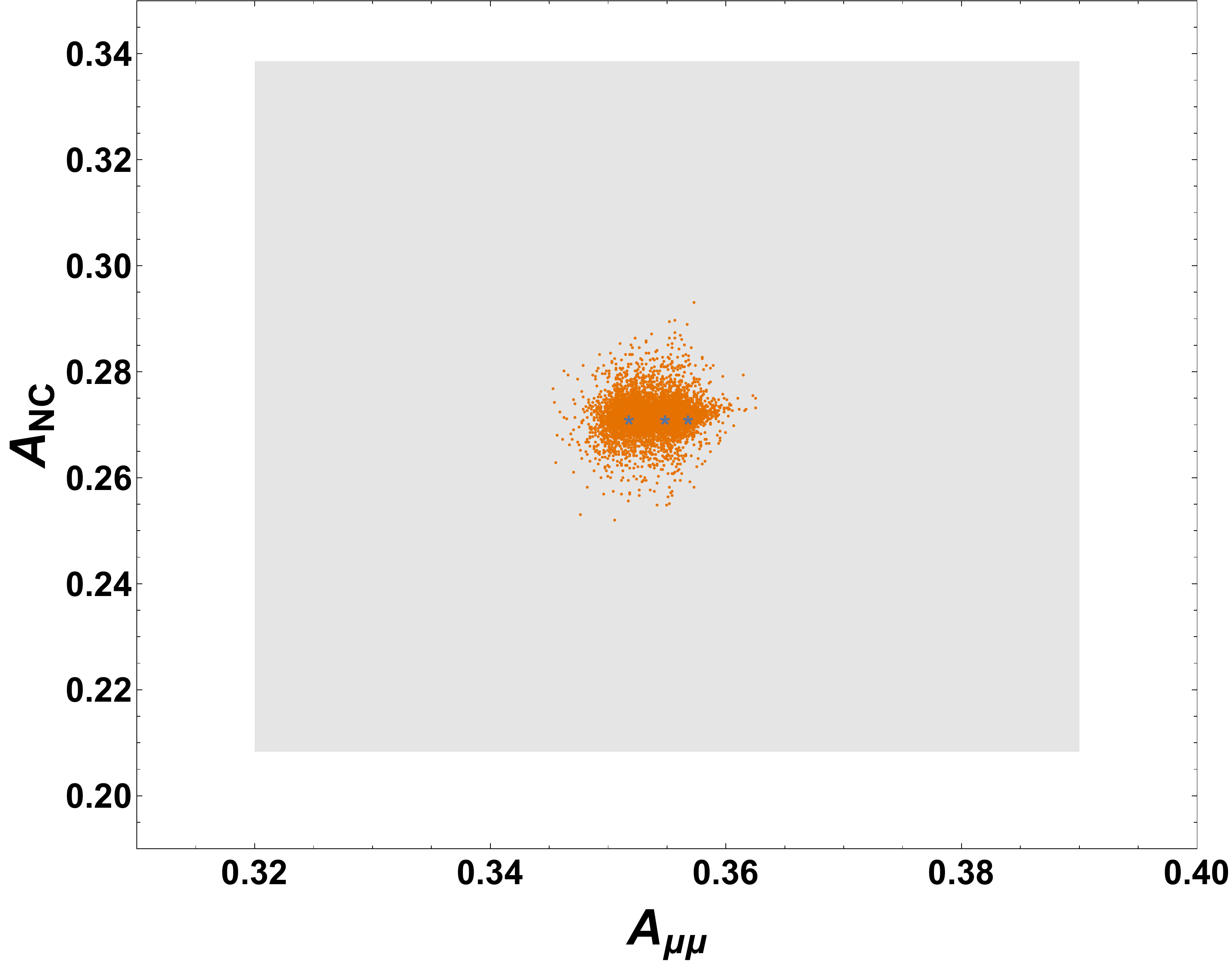}
    \caption{\it Integrated asymmetries in the ($A_{\mu\tau}, A_{\mu e}$) (left plot) and ($A_{\mu\mu}, A_{NC}$) planes (right plot) in the sterile neutrino model. The adopted legend for the symbols is the same as for the other plots. Standard neutrino flux has been employed to compute the number of events.}
    \label{31scatter}
\end{figure}

In Fig.\ref{31scatter}, we report our numerical results for the 3+1 case, obtained for fixed $\Delta m_{41}^2=1$ eV$^2$ and all mixing angles and phases extracted randomly flat in the ranges discussed in Sect.\ref{spectrasect}. Standard neutrino fluxes have been employed.
As previously mentioned, we have four independent asymmetries. Three of them ($A_{\mu e}$, $A_{\mu\mu}$ and $A_{\mu\tau}$) are accessible through the corresponding oscillation channels. The other one, namely $A_{\mu s}$, can be measured looking at the NC events. Indeed, since the NC interactions are flavor independent, the number of events in this channel depends on the sum: 
\begin{eqnarray}
 N_{NC}\propto P(\nu_\mu\to\nu_e)+P(\nu_\mu\to\nu_\mu)+P(\nu_\mu\to\nu_\tau)\,,
\end{eqnarray}
which, from the unitarity relation, corresponds to $1-P(\nu_\mu\to\nu_s)$. Thus, the integrated asymmetry
\begin{eqnarray}
 A_{NC}=\frac{N_{NC}-\bar{N}_{NC}}{N_{NC}+\bar{N}_{NC}}
\end{eqnarray}
is closely related to the $\mu s$ asymmetry. \\
We present our results in the $(A_{\mu \tau},A_{\mu e})$ and $(A_{\mu\mu},A_{NC})$ planes, see Fig.\ref{31scatter}, for the standard flux. The situation is quite clear: even though the analytic corrections to $A_{\mu e}\sim {\cal O}(\lambda)$ and to $A_{\mu \mu, \mu\tau}\sim {\cal O}(\lambda^2)$,
%According to our eq.(\ref{3+1asy}), the largest variation from the SM can be found in $A_{\mu\tau}$. This is mainly due to the weaker bounds on $\theta_{34}$, which plays an important role in the corrections to the probabilities. However, due to 
the relatively large uncertainties do no allow the 3+1 points to spread outside the error bars. 
%From the plots it is also clear that the changes in $A_{\mu e}$ due to the presence of a fourth sterile neutrino are very small. This is because, as discussed before, the new physics corrections to this asymmetry are order of magnitude smaller than the leading order value.

\begin{figure}
    \centering
    \includegraphics[width=7cm,height=7cm]{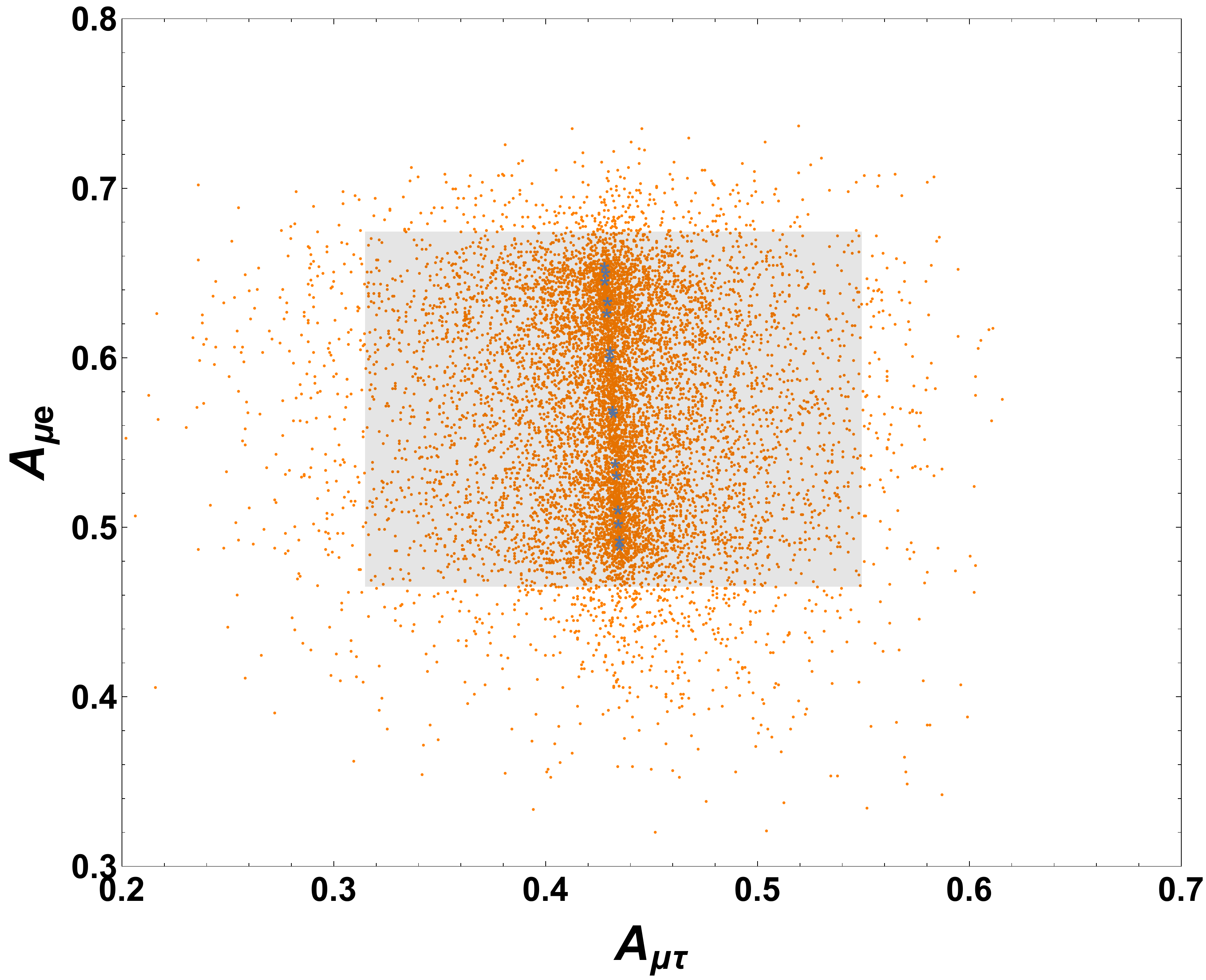}
    \includegraphics[width=7cm,height=7cm]{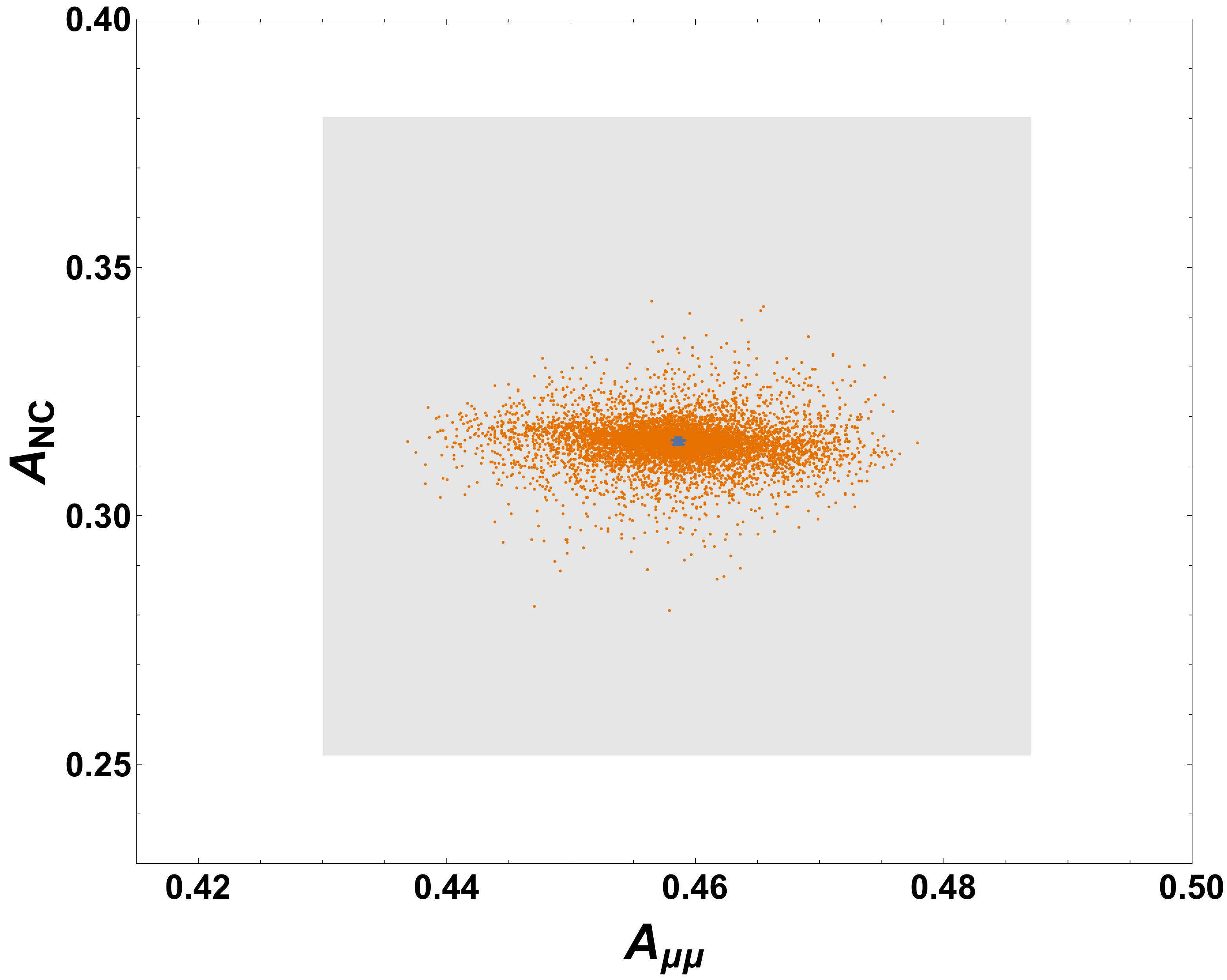}
    \caption{\it Same as Fig.\ref{31scatter} but for the optimized flux.}
    \label{31scatteropt}
\end{figure}

As before, the use of the higher energy flux 
reduces the error bands and increases the number of points outside the SM uncertainties, see Fig.\ref{31scatteropt}. 
Furthermore, as for the NSI case, the asymmetry which vary the most when NP enters into the game is $A_{\mu e}$ since, as shown in  (\ref{3+1asy}), the correction to SM asymmetry is at first order in our perturbative expansion. It is clear from the left panel of Fig.\ref{31scatteropt} that there are some points at more than two sigmas away from the standard values but, differently from the previous case, $A_{\mu e}$ never becomes negative. \\
On the other hand, in the $(A_{\mu\mu},A_{NC})$ plane,  no orange point lies outside the grey rectangle.

\section{Conclusions}
\label{concl}
The search of Physics beyond the Standard Model has become an attractive  research field in the neutrino  sector thanks to the huge experimental efforts in the measurement of the standard oscillation parameters. Beside few exceptions, they are known with a very good precision and the relevant question is now to establish whether New Physics is hidden within the experimental uncertainties. In this respect, the quest for new sources of CP violation beyond the single phase already present in the Standard Model is a pressing one. Contrary to many similar studies in the literature, we avoided to focus on the sensitivity of a given experiment to a particular phase, as many assumptions are usually done in the fit procedure (marginalization over a subset of parameters while other are kept fixed, and so on) which obscure the true sensitivity to CP violation.
Instead, we designed a more objective strategy based on the evaluation within the Standard Model of the integrated CP-odd quantities, the asymmetries $A_{\alpha\beta}$, and compare them with the same observables evaluated in the New Physics scenarios under consideration. In this paper, we performed a perturbative analytic evaluation of all asymmetries accessible at the DUNE experiment under the assumptions that Non-Standard  Interactions and one sterile state affect the standard neutrino  oscillation framework. We then apply our procedure to the more realistic asymmetries built from the expected number of event in DUNE, reaching the conclusion that, for both New Physics scenarios, the $A_{\mu e}$ asymmetry can reach values well beyond the Standard Model expectation, including the foreseen statistics and systematic uncertainties, when an high energy flux is employed. A special mention should be devoted to $A_{\mu \tau}$: while analytic considerations indicates that New Physics sets large corrections compared to the Standard Model results, the uncertainties involved in the evaluation of the number of expected events obscure this important feature. An experimental effort should be carried out to reduce the uncertainties in $\tau$ detection. 

\section*{Appendix A: perturbative expressions of probabilities}

We provide here the perturbative expressions of the probabilities expanded as discussed in Sects.\ref{CPasySM}, \ref{CPasyNSI} and \ref{CPasysterile}, in both NSI and the sterile neutrino models. The standard model probabilities can be obtained putting all new physics parameters to zero.\\
For the NSI case we have:
\begin{eqnarray}
\nonumber P(\nu_\mu\to\nu_e)^{NSI}&=& r^2\sin^2\Delta_{31}+\frac{4}{3}r\alpha\Delta_{31}(\cos\delta\cos\Delta_{31}-\sin\delta\sin\Delta_{31})\sin\Delta_{31}+\frac{4}{9}\alpha^2\Delta_{31}^2+\\
 \nonumber& &+2V_{CC}r\sin\Delta_{31}[\Delta_{31}\cos\Delta_{31}(\varepsilon_{e\mu}\cos(\delta-\delta_{e\mu})-\varepsilon_{e\tau}\cos(\delta-\delta_{e\tau})-r)+
 \\ \nonumber& & +\Delta_{31}\sin\Delta_{31}(\varepsilon_{e\tau}\sin(\delta-\delta_{e\tau})-\varepsilon_{e\mu}\cos(\delta-\delta_{e\mu})+\\
 & & +\sin\Delta_{31}(\varepsilon_{e\mu}\cos(\delta-\delta_{e\mu})+\varepsilon_{e\tau}\cos(\delta-\delta_{e\tau})+r)]
\end{eqnarray}

\begin{eqnarray}
 \nonumber P(\nu_\mu\to\nu_\mu)^{NSI}&=&\cos^2\Delta_{31}+\frac{4}{3} \alpha \Delta_{31}\sin\Delta_{31}\cos\Delta_{31}-\frac{4}{9}\alpha^2\Delta_{31}^2(2-3\sin^2\Delta_{31})+\\
\nonumber & &-\frac{4}{3}r\alpha \cos\delta\cos\Delta_{31}\sin\Delta_{31}+4a^2\sin\Delta_{31}-\frac{4}{3}s\alpha\Delta_{31}\sin\Delta_{31}\cos\Delta_{31}+ \\
 \nonumber & & +V_{CC}\bigg\{-8\Delta_{31}\varepsilon_{\mu\tau}\cos\delta_{\mu\tau}\cos\Delta_{31}\sin\Delta_{31}+\frac{4}{3} \alpha\Delta_{31}^2[ r\cos\delta\cos^2\Delta_{31}+\\
 \nonumber & & -\varepsilon_{e\mu}\cos\delta_{e\mu}\cos^2\Delta_{31}+\varepsilon_{e\tau}\cos\delta_{e\tau}\cos^2\Delta_{31}+4\varepsilon_{\mu\tau}\cos\delta_{\mu\tau}(1-2\sin^2\Delta_{31})]+\\
 \nonumber& &+4a\Delta_{31}\varepsilon_{\tau\tau}\cos\Delta_31\sin\Delta_{31}\}+4r\Delta_{31}\varepsilon_{e\mu}\cos(\delta-\delta_{e\mu})\cos\Delta_{31}\sin\Delta_{31}+\\
 \nonumber& &-4a\varepsilon_{\tau\tau}\sin^2\Delta_{31}+\frac{4}{3}\alpha\Delta_{31}[\varepsilon_{e\tau}\cos\delta_{e\tau}\cos\Delta_{31}\sin\Delta_{31}+\\
 & &-\varepsilon_{e\mu}\cos\delta_{e\mu}\cos\Delta_{31}\sin\Delta_{31}-r\cos\delta\cos\Delta_{31}\sin\Delta_{31}]\bigg\}
\end{eqnarray}

\begin{eqnarray}
 \nonumber P(\nu_\mu\to\nu_\tau)^{NSI}&=&\sin^2\Delta_{31}-\frac{4}{3}\alpha\Delta_{31}\cos\Delta_{31}\sin\Delta_{31}+ \\
 \nonumber & & -\frac{4}{9}\alpha^2\Delta_{31}^2(3\sin^2\Delta_{31}-2)-4a^2\sin^3\Delta_{31}-r^2\sin^2\Delta_{31}+\\
 \nonumber& &-\frac{4}{3}\alpha \Delta_{31}\sin\Delta_{31}(s\cos\Delta_{31}+r\sin\delta'\sin\Delta_{31})+\\
 \nonumber & &+2V_{CC}\bigg[4\Delta_{31}\varepsilon_{\mu\tau}\cos\delta_{\mu\tau}\sin\Delta_{31}\cos\Delta_{31}+r^2\sin\Delta_{31}(\Delta_{31}\cos\Delta_{31}+\\
 \nonumber & & -\sin\Delta_{31})+\frac{4}{27}\alpha^2\Delta_{31}^2\cos\Delta_{31}\Delta_{31} +\frac{2}{3}r\alpha\sin\delta\sin\Delta_{31}(\sin\Delta_{31}-\Delta_{31}\cos\Delta_{31})+\\
 \nonumber & & -\frac{8}{3}\alpha\Delta_{31}^2\varepsilon_{\mu\tau}\cos\delta_{\mu\tau}(1-2\sin^2\Delta_{31})+\frac{2}{3}a\Delta_{31}\varepsilon_{e\mu}\sin\delta_{e\mu}\sin^2\Delta_{31}+\\
 \nonumber & &-2a\varepsilon_{\tau\tau}\sin\Delta_{31}(\Delta_{31}\cos\Delta_{31}-\sin\Delta_{31})+\frac{2}{3}\alpha\Delta_{31}\varepsilon_{e\tau}\sin^2\Delta_{31}(\sin\delta_{e\tau}+\Delta_{31}\cos\delta_{e\tau})+\\
 \nonumber & & +\frac{2}{3}\alpha\Delta_{31}\varepsilon_{e\mu}\sin^2\Delta_{31}(\sin\delta_{e\mu}-\Delta_{31}\cos\delta_{e\mu})+r\Delta_{31}\varepsilon_{e\mu}\cos(\delta-\delta_{e\mu})\sin\Delta_{31}\cos\Delta_{31}+\\
 \nonumber & &-r\varepsilon_{e\mu}\sin^2\Delta_{31}(\cos(\delta-\delta_{e\mu})-\Delta_{31}\sin(\delta-\delta_{e\mu}))+\\
 \nonumber & &+ r\Delta_{31}\varepsilon_{e\tau}\cos(\delta-\delta_{e\mu})\cos\Delta_{31}\sin\Delta_{31}+\\
 & &+r\varepsilon_{e\tau}\sin^2\Delta_{31}(\cos(\delta-\delta_{e\tau})-\Delta_{31}\sin(\delta-\delta_{e\tau}))\bigg] 
\end{eqnarray}\\
For the 3+1 model, we found the following expressions ($\delta'=\delta_2-\delta_1-\delta_3$):

\begin{eqnarray}
\nonumber P(\nu_\mu\to\nu_e)^{3+1}&=& r^2\sin^2\Delta_{31}+\frac{4}{3}r\alpha\Delta_{31}(\cos\delta'\cos\Delta_{31}
-\sin\delta'\sin\Delta_{31})\sin\Delta_{31}+\frac{4}{9}\alpha^2\Delta_{31}^2+\\
\nonumber & & V_{CC}\bigg[-2r^2\sin^2\Delta_{31}-2 r^2\Delta_{31}\cos\Delta_{31}\sin\Delta_{31}+\\
& &+\frac{4}{3}r\alpha\Delta_{31}(\cos\delta'\cos\Delta_{31}-\sin\delta'\sin\Delta_{31})(\sin\Delta_{31}-\Delta_{31}\cos\Delta_{31})\bigg]
\end{eqnarray}

\begin{eqnarray}
 \nonumber P(\nu_\mu\to\nu_\mu)^{3+1}&=&(1-2s_{24}^2)\cos^2\Delta_{31}+\frac{4}{3} \alpha \Delta_{31}\sin\Delta_{31}\cos\Delta_{31}-\frac{4}{9}\alpha^2\Delta_{31}^2(2-3\sin^2\Delta_{31})+\\
\nonumber & &-\frac{4}{3}r\alpha \cos\delta'\cos\Delta_{31}\sin\Delta_{31}+4a^2\sin\Delta_{31}-\frac{4}{3}s\alpha\Delta_{31}\sin\Delta_{31}\cos\Delta_{31}+ \\
\nonumber & &+\frac{4}{27}\alpha\Delta_{31}V_{CC}[9r \cos\delta'\cos\Delta_{31}(\Delta_{31}\cos\Delta_{31}-\sin\Delta_{31}+\\
 & &-2\alpha\Delta_{31}^2\sin\Delta_{31}\cos\Delta_{31}] +4V_{NC}\Delta_{31}s_{24}s_{34}\cos\delta_3\cos\Delta_{31}\sin\Delta_{31}
\end{eqnarray}

\begin{eqnarray}
 \nonumber P(\nu_\mu\to\nu_\tau)^{3+1}&=& \sin^2\Delta_{31}-\frac{4}{3}\alpha\Delta_{31}\cos\Delta_{31}\sin\Delta_{31}+ \\
 \nonumber & & -\frac{4}{9}\alpha^2\Delta_{31}^2(3\sin^2\Delta_{31}-2)-4a^2\sin^3\Delta_{31}-r^2\sin^2\Delta_{31}+\\
 \nonumber& &-\frac{4}{3}\alpha \Delta_{31}\sin\Delta_{31}(s\cos\Delta_{31}+r\sin\delta'\sin\Delta_{31})+(s_{24}^2-s_{34}^2)\sin^2\Delta_{31}+ \\
 \nonumber & & +2s_{24}s_{34}\sin\delta_3\sin\Delta_{31}\cos\Delta_{31} +V_{CC}\bigg[2r^2\sin\Delta_{31}(\Delta_{31}\cos\Delta_{31}-\sin\Delta_{31})+\\
 \nonumber & & +\frac{8}{27}\alpha^2\Delta_{31}^3\cos\Delta_{31}\sin\Delta_{31}-\frac{4}{3}r\alpha\Delta_{31}\sin\delta'\sin\Delta_{31}(\Delta_{31}\cos\Delta_{31}-\sin\Delta_{31})\bigg]+\\
  & & -4V_{NC}s_{24}s_{34}\cos\delta_3\cos\Delta_{31}\sin\Delta_{31}
\end{eqnarray}

\begin{eqnarray}
 P(\nu_\mu\to\nu_s)^{3+1}=s_{24}^2(2-\sin^2\Delta_{31})+s_{34}^2\sin^2\Delta_{31}-2s_{24}s_{34}\sin\delta_3\cos\Delta_{31}\sin\Delta_{31}
\end{eqnarray}

\section*{Appendix B: probabilities in the 3+1 model for not-averaged $\Delta m_{41}^2$}

In this appendix, we will provide the perturbative expressions for the asymmetries in the 3+1 model in the case  the oscillations driven by $\Delta m_{41}^2$ cannot be averaged out. 
The leading order of $A_{\mu e}$ is unchanged, thus there are no corrections to the SM at the chosen perturbative order.

For the other asymmetries, we give here a first order expansion in the matter potentials $V_{CC}$ and $V_{NC}$. Thus, we put the corrections in the following form:

\begin{equation}
    A_{\alpha\beta}^{3+1}=(A_{\alpha\beta}^{3+1})_0+V_{CC} (A_{\alpha\beta}^{3+1})_{CC}+V_{NC} (A_{\alpha\beta}^{3+1})_{NC}+\mathcal{O}(V^2,\lambda^3)\,.
\end{equation}
For the $\mu\tau$ asymmetry we have:
\begin{eqnarray}
 (A_{\mu \tau}^{3+1})_0 &=& A_{\mu\tau}^{SM_0}+4 s_{24}s_{34} \sin\delta_3(\cot\Delta_{31}\sin^2\Delta_{41}+\sin\Delta_{41}\cos\Delta_{41}) \nonumber \\
 (A_{\mu \tau}^{3+1})_{CC} &=& A_{\mu\tau}^{SM_1} \\
(A_{\mu \tau}^{3+1})_{NC} &=& s_{24}^2 b_{\mu\tau}^{s_{24}^2}+s_{34}^2 b_{\mu\tau}^{s_{34}^2}+s_{24}s_{34} b_{\mu\tau}^{s_{24}s_{34}}\nonumber
\end{eqnarray}
where $\Delta_{41} = \Delta m^2_{41} L/E_\nu$ and:
\begin{eqnarray}
 b_{\mu\tau}^{s_{24}^2} &=& \frac{2\Delta_{41}-2\Delta_{31}(\sin^2\Delta_{31}+\cot\Delta_{31}\sin\Delta_{41}\cos\Delta_{41})}{(\Delta_{41}/\Delta_{31})(\Delta_{31}-\Delta_{41})} \\
 b_{\mu\tau}^{s_{34}^2} &=& b_{\mu\tau}^{s_{24}^2}
 \\
 b_{\mu\tau}^{s_{24}s_{34}} \nonumber &=&-8\cos\delta_3(\cot\Delta_{31}\cos^2\Delta_{41}+\sin\Delta_{41}\cos\Delta_{41})+\\
 \nonumber & & +2\cos\delta_{3}\Delta_{31}\frac{2\sin^2\Delta_{41}-\cot\Delta_{31}\sin\Delta_{41}\cos\Delta_{41}}{(\Delta_{41}/\Delta_{31})(\Delta_{31}-\Delta_{41})}+\\
 & & +4\cos\delta_{3}\Delta_{41}\frac{1-2\sin^2\Delta_{41}+\cot\Delta_{31}\sin\Delta_{41}\cos\Delta_{41}}{(\Delta_{41}/\Delta_{31})(\Delta_{31}-\Delta_{41})}
\end{eqnarray}
For the $\mu\mu$ asymmetry: 
\begin{eqnarray}
 (A_{\mu \mu}^{3+1})_0 &=& 0 \nonumber \\
 (A_{\mu \mu}^{3+1})_{CC} &=& A_{\mu\mu}^{SM_1} \\
(A_{\mu \mu}^{3+1})_{NC} &=& s_{24}^2 b_{\mu\mu}^{s_{24}^2}+s_{24}s_{34} b_{\mu\mu}^{s_{24}s_{34}}\nonumber\,
\end{eqnarray}
where
\begin{eqnarray}
 \nonumber b_{\mu\mu}^{s_{24}^2} &=&-4\Delta_{31}[\tan\Delta_{31}(1-2\sin^2\Delta_{41})-2\sin\Delta_{41}\cos\Delta_{41}]+\\
 & &\frac{4(\Delta_{31}-2\Delta_{41})}{(\Delta_{41}/\Delta_{31})(\Delta_{31}-\Delta_{41})}(\tan\Delta_{31}\sin\Delta_{41}\cos\Delta_{41}-\sin^2\Delta_{41})
 \\
 b_{\mu\mu}^{s_{24}s_{34}} &=&4\Delta_{31}\cos\delta_3\tan\Delta_{31}+4\cos\delta_3 \Delta_{31}\frac{\sin^2\Delta_{41}+\tan\Delta_{31}\sin\Delta_{41}\cos\Delta_{41}}{(\Delta_{41}/\Delta_{31})(\Delta_{41}-\Delta_{31})}
\end{eqnarray}
Finally, for the $\mu s$ asymmetry we found:
\begin{eqnarray}
 (A_{\mu s}^{3+1})_0 &=& \frac{4s_{24}s_{34}\sin\delta_3(\sin\Delta_{41}\cos\Delta_{41}\sin^2\Delta_{31}-\sin\Delta_{31}\cos\Delta_{31}\sin^2\Delta_{41})}{d_0} \nonumber \\
 (A_{\mu s}^{3+1})_{CC} &=& 0 \\
(A_{\mu s}^{3+1})_{NC} &=& \frac{s_{24}^3 s_{34} b_{\mu s}^{s_{24}^3s_{34}}+s_{34}^3s_{24}b_{\mu s}^{s_{34}^3s_{24}}+s_{24}^2s_{34}^2 b_{\mu s}^{s_{24}^2s_{34}^2} }{s_{24}^3 s_{34} d_{\mu s}^{s_{24}^3s_{34}}+s_{34}^3s_{24}d_{\mu s}^{s_{34}^3s_{24}}+s_{24}^2s_{34}^2 d_{\mu s}^{s_{24}^2s_{34}^2} }\nonumber\,
\end{eqnarray}
where
\begin{eqnarray}
\nonumber d_0&=&(s_{34}^2-s_{24}^2)\sin^2\Delta_{31}+\\
 \nonumber & &+2 s_{24}^2(\sin^2\Delta_{31}+2\sin^2\Delta_{41}\cos^2\Delta_{31}+2\sin\Delta_{31}\cos\Delta_{31}\sin\Delta_{41}\cos\Delta_{41})+\\
 & & +2s_{24}s_{34}\cos\delta_3 \sin\Delta_{31}[\sin\Delta_{31}(1-2\sin^2\Delta_{41})-2\cos\Delta_{31}\sin\Delta_{41}\cos\Delta_{41}]
\end{eqnarray}
and the $b_{\mu s}$ and $d_{\mu s}$ are complicated functions which, in the limit $\Delta_{41}\gg\Delta_{31}$ can be reduced to
\begin{eqnarray}
 \nonumber b_{\mu s}^{s_{24}^3s_{34}}& = & 0 \\
 \nonumber b_{\mu s}^{s_{34}^3s_{24}}& = & 2 \Delta_{31} \cos\delta_3 \sin^2\Delta_{31} (1-2\sin^2\Delta_{41}) \\
 \nonumber b_{\mu s}^{s_{34}^2s_{24}^2}& = & 4 \Delta_{31} \sin\Delta_{31}\cos\Delta_{41}\sin\Delta_{41} (4\sin^2\Delta_{41}-3) \\
 d_{\mu s}^{s_{24}^3s_{34}}& = & 4 \cos\delta_3\sin\Delta_{41}\cos\Delta_{41}(\sin^2\Delta_{31}-4\sin^2\Delta_{41})\\
 \nonumber d_{\mu s}^{s_{34}^3s_{24}}& = & -2 \cos\delta_3 \sin^2\Delta_{31}\sin\Delta_{41}\cos\Delta_{41} \\
 \nonumber d_{\mu s}^{s_{34}^2s_{24}^2}& = &  \sin\Delta_{31}(4\sin^2\Delta_{41}+8\sin^2\Delta_{41}\cos^2\Delta_{41}\cos^2\delta_3-\sin^2\Delta_{31})
\end{eqnarray}

\end{document}